\documentclass[twocolumn]{aastex631}

\usepackage{graphicx}	
\usepackage{amsmath}	
\usepackage{amssymb}	
\usepackage{color}
\usepackage{algorithm}
\usepackage{booktabs}
\usepackage[noend]{algpseudocode}
\usepackage{pifont}
\usepackage{soul}
\usepackage{float}
\usepackage{multibib}
\newcites{Appendix}{Appendix References}
\usepackage{natbib}
\usepackage[flushleft]{threeparttablex}
\usepackage{mathrsfs}
\usepackage{tikz}
\usepackage{esint}
\usepackage[draft]{todonotes}
\usepackage{lipsum}
\usepackage{longtable}
\usepackage{pifont}
\usepackage{savesym}
\savesymbol{tablenum}
\usepackage{siunitx}
\restoresymbol{SIX}{tablenum}
\usepackage{placeins}
\usepackage{bm}

\newcommand{\appropto}{\mathrel{\vcenter{
  \offinterlineskip\halign{\hfil$##$\cr
    \propto\cr\noalign{\kern2pt}\sim\cr\noalign{\kern-2pt}}}}}

\newcommand{\Exp}[2]{\left\langle{#1}\right\rangle_{#2}}

\renewcommand{\d}[1]{\ensuremath{\operatorname{d}\!{#1}}}
\newcommand{\dthree}[1]{\ensuremath{\operatorname{d}^3\!{#1}}}

\DeclareMathOperator\D{\mathcal{D}}
\DeclareMathOperator\V{\mathcal{V}}
\newcommand{\M}{\mathcal{M}}

\newcommand\kperp{k_{\perp}}

\DeclareMathOperator\s{\text{s}}
\DeclareMathOperator\f{\text{f}}

\newcommand\ekin{E_{\rm kin}}

\newcommand\emag{E_{\rm mag}}

\DeclareMathAlphabet\mathbfcal{OMS}{cmsy}{b}{n}

\renewcommand{\Re}{{\rm{Re}}}

\renewcommand{\v}{\mathit{v}}
\renewcommand{\b}{\mathit{b}}

\newcommand\kpar{k_{\parallel}}

\renewcommand{\u}{\bm{u}}
\newcommand{\us}{\bm{u}_s}
\newcommand{\uc}{\bm{u}_c}
\renewcommand{\k}{\bm{k}}
\newcommand{\x}{\bm{x}}
\newcommand{\K}{\bm{K}}
\renewcommand{\P}{\mathcal{P}_{u}}
\newcommand{\bom}{\bm{\omega}}
\newcommand{\At}{\mathbb{A}}

\newcommand{\St}{\mathbb{S}}
\newcommand{\It}{\mathbb{I}}
\newcommand{\T}{\mathcal{T}}
\newcommand{\bnabla}{\bm{\nabla}}
\newcommand{\ie}{{i.e.,}}

\renewcommand{\bell}{\bm{\ell}}

\begin{document}

\correspondingauthor{\\
$^{\dagger}$James R. Beattie: \href{mailto:james.beattie@princeton.edu}{james.beattie@princeton.edu};\\ 
$^\star$ these authors made equal contributions and should be deemed as joint first authors for the publication.}

\title{So long Kolmogorov: the forward and backward turbulence cascades in a supernovae-driven, multiphase interstellar medium}

\author[0000-0001-9199-7771]{James R. Beattie$^{\dagger,\star}$}
\affiliation{Department of Astrophysical Sciences, Peyton Hall, Princeton University, Princeton, NJ 08544, USA}
\affiliation{Canadian Institute for Theoretical Astrophysics, University of Toronto, 60 St. George Street, Toronto, ON M5S 3H8, Canada}
\affiliation{Department of Astronomy and Astrophysics, University of California, Santa Cruz, 1156 High Street, Santa Cruz, CA 96054}

\author[0000-0001-7364-4964]{Anne Noer Kolborg$^{\star}$}
\affiliation{Department of Astronomy and Astrophysics, University of California, Santa Cruz, 1156 High Street, Santa Cruz, CA 96054}

\author[0000-0003-2558-3102]{Enrico Ramirez-Ruiz}
\affiliation{Department of Astronomy and Astrophysics, University of California, Santa Cruz, 1156 High Street, Santa Cruz, CA 96054}

\author[0000-0002-0706-2306]{Christoph Federrath}
\affiliation{Research School of Astronomy and Astrophysics, Australian National University, Canberra, ACT 2611, Australia}

\begin{abstract}
    The interstellar medium (ISM) of disk galaxies is turbulent, and yet the fundamental nature of ISM turbulence, the energy cascade, is not understood in detail. In this study, we use high-resolution simulations of a hydrodynamical, gravitationally stratified, supernova (SNe)-driven, multiphase ISM to probe the nature of a galactic turbulence cascade. Through the use of velocity flux transfer functions split into interactions between compressible $\uc$ and incompressible $\us$ modes, we show that there exists a large-to-small-scale cascade in both $\uc$ and $\us$ when mediated by an additional $\us$ mode. But the $\us$ cascade is highly non-local. Moreover, there is a $\uc$ mediated component of the $\us$ cascade that proceeds in the opposite direction -- an inverse cascade from small-to-large scales. The cascade feeds flux into scales well beyond the scale height, energizing the winds and fueling the direct cascades. Both the strongly non-local and the inverse $\us$ cascades happen on scales that have a power law $\us$ energy spectrum, highlighting how degenerate the spectrum is to the true underlying physical processes. We directly show that the inverse cascade comes from $\us$ modes interacting with expanding SNe remnants (SNRs) and that $\us$ modes are generated to leading order via baroclinic, highly corrugated cooling layers between warm $(T\lesssim 10^4\,\rm{K})$ and hot $(T\gg10^4\,\rm{K})$ gas in these SNRs. Finally, we outline a complete phenomenology for SNe-driven turbulence in a galactic disk, estimate a $10^{-16}\,\rm{G}$ Biermann field generated from SNR cooling layers, and highlight the strong deviations that SNe-driven turbulence has from the conventional Kolmogorov model.
\end{abstract}


\section{Introduction} \label{sec:intro}
    The interstellar medium (ISM) of our Galaxy is turbulent, boasting hydrodynamical Reynolds numbers, $\Re = UL/\nu$, between $\sim 10^2$ in the hot ($T > 10^6\,\rm{K}$) ionized medium, to $\sim 10^{9}$ in the cold ($T = 10\,\rm{K}$) molecular medium \citep{Low1999,Krumholz2015_star_formation_text,Ferriere2020_reynolds_numbers_for_ism}, where $U$ is characteristic velocity, $L$ the system size scale, and $\nu$ the kinematic viscosity. Turbulence plays a diverse set of roles in the ISM, from regulating the star formation rate \citep{Klessen2000,Krumholz2005,Hennebelle2008,Hennebelle2014_SN_driven_turb,Federrath2012,Burkhart2018,Hennebelle2024_inefficient_SF} and modifying the initial mass function of stars \citep{Nam2021,Mathew2023_IMF_driving_modes}, to mixing metals \citep{Krumholz2018_metallicity_SF,Kolborg2022_metal_mixing_1,Kolborg2023_metal_mixing_2,Sharda2024_galactic_metal_mixing} and generating and maintaining dynamically relevant magnetic fields \citep{Schekochihin2002_saturation_evolution,Federrath2011_mach_dynamo,Schober2012_saturation_Re_Rm_dependence,Schober2015_saturation_of_turbulent_dynamo,Xu2016_dynamo,Seta2022_multiphase_dynamo,Kriel2022_kinematic_dynamo_scales,Beattie2023_growth_or_decay,Beattie2023_bulk_viscosity,Kriel2025_SSD}. However, the exact origin of the turbulence is still an open question in ISM physics \citep[e.g.,][]{Krumholz2016_source_of_turb}.
    
    There is sufficient energy from supernova (SNe) alone to drive the turbulence to a steady state within our own Galaxy \citep{Beck1996_galactic_magnetism,Elmegreen2004_ISM_turbulence_review,Hennebelle2014_SN_driven_turb,Padoan2016_supernova_driving,Girichidis2016_SLICC,Chamandy2020_SN_driven_turbulence,Bacchini2020_supernova_drives_turb}, and maintain the multiphase structure \citep{McKee1977_SN_ISM,Fielding2020_fractal_cooling_layer,Guo2024_high_res_SN}, but the details of exactly how SNe, detonated on small scales, $\ell$ can facilitate an energy cascade from large-to-small scales, $L > \ell$, are quantitatively unknown. The large-to-small ``direct" cascade, derived through \citet{Kolmogorov1941_dissipation}'s 4/5 law, is one of the few exact results in incompressible, homogeneous, isotropic turbulence. It reads,
    \begin{align}\label{eq:energy_flux_exact}
        \Exp{\delta u^3}{\ell} = -\frac{4}{5}\varepsilon\ell \iff \bnabla_{\bm{\ell}} \cdot \Exp{\delta u^2 \delta\u}{\ell} = - 4\varepsilon,
    \end{align}
    where $\Exp{\hdots}{\ell}$ is the average over $\ell$, $\delta \u = \u({\rm{r}} +\bell) - \u({\rm{r}) }= \u' - \u$ is the velocity increment, and $\varepsilon$ is the velocity flux, which on the system scale is simply $\varepsilon \sim U^3/L$. We have also used the differential version of the expression as derived in \citet{Monin1975_mechanics_of_turbulence} and recited in \citet{Banerjee2017_exact_relations}, where the left-hand side of this equation is the kinetic energy flux on the scale $\ell$, and because the right-hand side is negative, this implies that the flux acts like a converging ($\bnabla\cdot\u < 0$) flow, moving down from large $L$ to small $\ell$. It has further been shown for incompressible turbulence, \citet{Kolmogorov1941}-style turbulence the transfer of flux from large to small scales is dominated by local flux transfers between $k$ modes \citep{1988PhFl...31.2747D,1990PhFlA...2..413D,1992FlDyR..10..499S,2007PhFl...19h5112D,Aluie2009_hydro_local_transfers,2016PhRvE..93f1102Y}.
    
    Indeed, \autoref{eq:energy_flux_exact} was extended to homogeneous, compressible turbulence by \citet{Ferrand2020_exact_relations_compressible}, 
    \begin{align}\label{eq:energy_flux_exact_comp}
         \bnabla_{\bm{\ell}} \cdot\Exp{\overline{\delta\rho}\delta u^2 \delta\u}{\ell} - \frac{1}{2}\Exp{(\rho\theta' + \rho'\theta)\delta u^2}{\ell} = - 4\varepsilon,
    \end{align}
    where $\rho$ is the mass density, $\theta = \bnabla\cdot\u$ and $\overline{\delta\rho} = (\rho' + \rho) / 2$. Clearly, this means that $\varepsilon$ can now change sign, based on whether or not the energy flux from the more classical cascade $\bnabla_{\bm{\ell}} \cdot\Exp{\overline{\delta\rho}\delta u^2 \delta\u}{\ell}$, is larger or smaller than the energy flux coupled to the dilatations and compressions, $\Exp{(\rho\theta' + \rho'\theta)\delta u^2}{\ell}/2$. One of the key goals of this paper is to understand how the flow of energy works in supernova-driven turbulence, which we will not explore through these exact relations, but instead velocity flux transfer functions that allow us to directly probe the sign of $\varepsilon$.
    
    Certainly, as we showed in \autoref{eq:energy_flux_exact} and \autoref{eq:energy_flux_exact_comp}, the energy spectrum alone is insufficient to understand the fundamental nature of a turbulent plasma. This is already well-known, and we have to seek statistical tools that go beyond simply measuring Fourier amplitudes \citep[see recent review][]{Burkhart2021_stats_turb_review}. To make the reason why this is true clear, consider the following case. \citet{Kolmogorov1941}-type turbulence has $\varepsilon = \rm{const}.$ energy flux exchange between the $k$ modes, and the energy is transported down a cascade with $u^2(k) \sim k^{-5/3}$, where $\varepsilon > 0$, as shown by \autoref{eq:energy_flux_exact}. However, in two-dimensional turbulence (or quasi-two-dimensional, which may be realized in a strongly stratified or magnetized limit)  has the same $u^2(k) \sim k^{-5/3}$ spectrum as in \citet{Kolmogorov1941}, but with $\varepsilon < 0$ with a cascade moving from small $\ell$ to large $\ell$ \citep[e.g.,][]{Kraichnan1967_two_dimensional_turbulence,Boffetta2012_two_d_review}, immediately showing that the spectrum alone is a degenerate statistic for understanding the nature of turbulence. \citet{Padoan2016_supernova_driving} showed that indeed the energy spectrum from SNe-driven turbulence in a triply-periodic box looks correct, in that the spectrum looks like the spectrum we get from local turbulent boxes, even with $u^2(k) \sim k^{-5/3}$ (or a bit shallower). But, as we highlight above, this does not mean that \citet{Kolmogorov1941} is at play.
    
    \citet{Kolmogorov1941}-type turbulence is regularly invoked (or at least broadly imagined) for the ISM in our own Galaxy \citep[e.g.,][]{Armstrong1995_power_law,Elmegreen2004_ISM_turbulence_review,Falceta2014_turbulence_in_ism,Hopkins2021_testing_gev_cr_models,Nandakumar2023_large_scale_cascade}, but what if supernova-driven turbulence is different? It could be \citet{Batchelor1959_passive_tracer}-like instead, where low-$k$ turbulent modes couple $\varepsilon$ non-locally to all scales \citep[e.g.,][]{Adkins2018_phase_space_turbulence}, and the self-similarity of the energy spectrum is not a repercussion from a local cascade at all? Of course, \citet{Burgers1948}-type turbulence is also regularly invoked \citep{Federrath2013_universality,Krumholz2015_star_formation_text,Federrath2021,Cernetic2024_GPU_supersonic_turb,Beattie2025_nature_astro}, which should not have a cascade at all because it is simply the spectrum one gets from Fourier transforming a velocity discontinuity and dumping $\varepsilon$ on all scales. The key point is that $\varepsilon$ is what we need to measure as a function of $k$ to distinguish between these different turbulent models.
    
    For decades, turbulent boxes have been used as local models of the ISM \citep[e.g., ][ and many more]{Stone1998_MHD_dissipation,Burkhart2020_CATS,Federrath2021,Hu2022_superdiffusion_CRs,Kempski2023_b_field_reversals,Fielding2022_ISM_plasmoids,Beattie2025_nature_astro,Kriel2025_SSD}, usually employing Fourier space driving, where momentum is directly injected on low $k$ modes (large scales), and in the absence of kinetic and magnetic helicity \citep[e.g., ][]{Waleffe1992_triads,Alexakis2017_helically_decomposed,Plunian2020_inverse_cascade}, a self-consistent direct cascade is formed, albeit over a limited range of scales (see \citealt{Beattie2025_nature_astro} for the currently highest resolution turbulence box in the world, with $\Re \gtrsim 10^6$, resolving a few orders of magnitude of length scales within the turbulent cascade). However, if the ISM is energized by SNe on the small scales, it is not clear how well these local simulations faithfully represent the ISM. We aim to address this question throughout the study, starting with the spectrum, energy fluxes, and then finishing with the generation of vorticity.

    In this study, we directly explore how the velocity flux, $\varepsilon \sim U^3/L$ and turbulent cascade $\ekin(k) \sim k^{-\alpha}$, work in stratified supernova-driven turbulence, with a time-dependent chemical network that provides a multiphase ISM. To do this, we use shell-to-shell velocity flux transfer functions, \citep[motivated by e.g.,][]{Grete2017_shell_models_for_CMHD,Grete2020_as_a_matter_of_state,Grete2021_as_a_matter_of_tension,Grete2023_as_a_matter_of_dynamical_range} generalized to calculate compressible and incompressible mode interactions. Previous works have performed detailed and important analyses of supernova-driven energetics \citep{Martizzi2016,Li2020_SNE_energetics,Mohapatra2024_SNe_turbulence_ellipticals}. However, no study has performed them with a $k$ mode-by-mode focus, making our study rather unique and highly illuminating. We neglect magnetic fields, galactic rotation and the cold phases of the ISM plasma, to focus solely on the impact of the supernova-driven turbulence on the large-scale ISM dynamics with a previously utilized setup \citep{Martizzi2016,Kolborg2022_metal_mixing_1,Kolborg2023_metal_mixing_2} at much higher grid resolutions of up to $1,\!024^3$. This allows us to fully characterize the effect of SNe-driven turbulence at a level of detail that has never been done before.

    This study is organized as follows. In \autoref{sec:sims} we discuss the gravito-hydrodynamical fluid model, initial and boundary conditions, SNe-driving prescription, cooling function and phase structure, our key dimensionless parameters, and reaching stationarity in a SNe-driven ISM. In \autoref{sec:velocity_structure} we define the Helmholtz decomposition that we perform on our velocity field, and discuss the qualitative behavior of each component. In \autoref{sec:power_spectra} we define the velocity power spectrum and show the results from the cylindrically integrated velocity spectrum, separated into compressible and incompressible modes and cylindrical coordinates. In \autoref{sec:transfer_functions} we define and discuss results of Helmholtz decomposed shell-to-shell velocity flux transfer functions, building on the methods from \citet{Grete2017_shell_models_for_CMHD,Grete2020_as_a_matter_of_state,Grete2021_as_a_matter_of_tension,Grete2023_as_a_matter_of_dynamical_range}, focusing on like-mode interactions, the emergence of an inverse cascade, and time variability in the standard incompressible, \citet{Kolmogorov1941}-type, cascade. In \autoref{sec:vorticity} we compute each of the source terms in the vorticity evolution equation, show that the baroclinic term dominates the generation of vorticity in this type of turbulence, and identify the main sources as fractal cooling layers embedded inside of expanding SNRs. We use our simulations to estimate how strong the Biermann magnetic field would be from the cooling layer, which is $\sim 10^{-16}\,\rm{G}$. Finally, in \autoref{sec:summary}, we define an entire phenomenology for SNe-driven turbulence and how it may work in the disk of a galaxy, from the initial adiabatic expansion of the SNR to the velocity flux of all different mode combinations. We focus on and highlight how compressible and incompressible modes both have different but vitally important roles in this type of turbulence, and further emphasize how vastly different SNe-driven turbulence ends up being compared to Kolmogorov. 
    

\section{Numerical Simulations \& Methods}\label{sec:sims}
    We model the ISM in a section of a disk galaxy, and the simulation parameters are chosen such that the resulting galaxy parameters are similar to those of the present-day Milky Way. In the following, we summarize the fluid model, the chosen model parameters, and some of the basic statistics from the simulation, including the phase structure.

\begin{table*}
    \caption{Main simulation parameters and derived quantities.}
    \hspace{-6em}
    \begin{tabular}{lccccccccc}
    \hline
    \hline
    Galaxy & $\Exp{\rho}{\V}$ & $L$ & $\ell_0$ & $t_0$ & $\Exp{u^2}{\V}^{1/2}$ & $\M$ & $\ell_{\rm cor,\parallel}$ & $\ell_{\rm cor,\perp}$ &  $N_{\text{grid}}^3$ \\
     model & [g cm$^{-3}$] & [pc] & [pc] & [Myr] & [km/s] & & [pc] & [pc]  & \\ 
    (1) & (2) & (3) & (4) & (5) & (6) & (7) & (8) & (9) & (10) \\ 
    \hline
    \texttt{MW\_1024  }  & \SI{7.85e-25}{}  & 1\!,000  & $85 \pm 6$& $2.9 \pm 0.7$  & 28 $\pm$ 6  & 1.75 $\pm$ 0.05 & $(35 \pm 3) \times 10^1$ & $(29 \pm 2) \times 10^1$ & 1\!,024$^3$ \\
    \texttt{MW\_512  }  &  \SI{7.84e-25}{}  & 1\!,000  & $86 \pm 5$ & $2.3 \pm 0.5$ &  37 $\pm$ 8 & 1.82 $\pm$ 0.02 & $(35 \pm 2) \times 10^1$ & $(29 \pm 2) \times 10^1$ & 512$^3$ \\
    \texttt{MW\_256  }  &   \SI{7.83e-25}{} & 1\!,000  & $88 \pm 5$ & $2.7 \pm 0.6$ & 32 $\pm$ 7  & 1.80 $\pm$ 0.05 & $(36 \pm 2) \times 10^1$ & $(32 \pm 2) \times 10^1$  & 256$^3$ \\
    \hline
    \hline
    \end{tabular}
    \begin{tablenotes}[para]
        \textit{\textbf{Notes.}} Column (1): Simulation label. Column (2): volume-weighted mean gas density. Column (3): the length of the cubic simulation domain. Column (4): the effective scale height of the gaseous disk fit by an empirical model in the steady state. Column (5): the turbulent turnover time at the gaseous scale height, column (4), as shown in \autoref{eq:turnover_time}. Column (6): the root-mean-squared (rms) turbulent (rest-frame) velocity. Column (7): the volume-weighted turbulent Mach number, as shown in \autoref{eq:mach_number}. Column (8): the correlation scale of the total velocity fluctuations parallel to $\bnabla\phi$, computed from the power spectrum. Column (9): the same as column (8) but for the total velocity fluctuations perpendicular to $\bnabla\phi$. Column (10): the total number of cells used in the grid discretisation of the simulation. All statistics in the table are averaged over 10$t_0$ in the statistically steady state shown in \autoref{fig:mach}. 
    \end{tablenotes}
    \label{tab:sims}
\end{table*}

%


\subsection{Multiphase, supernova-driven, gravito-hydrodynamical fluid model}
    We model sections of a galaxy disk based on the setup of \citet{Martizzi2016}. For our ISM simulations, we solve the three-dimensional, compressible Euler equations in a static gravitational field with mass, momentum and energy sources from stochastic supernova events ($\sim$ point sources of extreme mass, momentum and energy) using the \textsc{ramses} code \citep{Teyssier2002_ramses} employing the Monotonic Upstream-centered Scheme for Conservation Laws (MUSCL) scheme and HLLC Riemann solver. The model is
    \begin{align}
    \dfrac{\partial \rho}{\partial t} + \bnabla \cdot \left( \rho \u \right) = {}& \dot{n}_{\rm SNe} M_{\rm ej}, \label{eq:mass_conservation} \\
    \dfrac{\partial \rho \u}{\partial t} + \bnabla \cdot \left( \rho \u \otimes \u + P \mathbb{I} \right) = & - \rho \bnabla \phi
    +\dot{n}_{\rm SNe} \bm{p}_{\rm SNe}(Z, n_{\rm H}), & \label{eq:momentum_conservation} \\
    \dfrac{\partial \rho e}{\partial t} + 
    \bnabla \cdot \left[\rho \left( e + P \right) \u \right] =& - n_{\rm H}^2 \Lambda - \nonumber\\
    \rho \u \cdot \bnabla \phi
     + \dot{n}_{\rm SNe}\bigg[E_{\rm th, SNe}(&Z, n_{\rm H}) + \frac{p^2_{\rm SNe}(Z, n_{\rm H})}{2 (M_{\rm ej} + M_{\rm swept})} \bigg], \label{eq:energy_conservation}  \\
    e = & \epsilon + \frac{u^2}{2}, \,\, P = \frac{2}{3} \rho \epsilon, \label{eq:EOS}
    \end{align}
    where $\otimes$ is the tensor product, such that $\u\otimes\u = u_i u_j$. $\u$ is the gas velocity, $\rho$ is the gas density. $\bnabla\phi$ describes the static gravitational potential, $\partial_t \bnabla\phi = 0$, the details of this potential are described in \autoref{ssec:phi}. $P$ is the scalar pressure, and $\mathbb{I}$ is the unit tensor, $\delta_{ij}$. $\rho e$ is the total energy composed of both the kinetic energy of the gas, $\rho u^2/2$ and the internal energy, $\epsilon$. $\dot{n}_{\rm SNe}$ is the volumetric rate of SNe, $\bm{p}_{\rm SNe}$, $M_{\rm ej}$, and $E_{\rm th, SNe}$ are the radial momentum, the mass of the ejecta, and the thermal energy of each SNe. $M_{\rm swept}$ is the mass of ISM material swept up by the SNe shock wave. $Z$ is the metallicity of the ambient medium. $\Lambda$ is the cooling function, which encompasses both the heating and cooling terms; the underlying physical model is discussed in \autoref{ssec:cooling}. Finally, $n_{\rm H}$ is the number density of hydrogen (atomic and ionized). 

    A variety of metals are injected into the medium from the SNe, see Appendix in \citet{Kolborg2022_metal_mixing_1} for more details. Each metal, $Z_i$, follows the passive scalar transport equation, 
    \begin{align}\label{eq:metal_transport}
        \frac{\partial Z_i}{\partial t } + \u \cdot\bnabla Z_i = \mathcal{S}_{\rm SNe},
    \end{align}
    where $\mathcal{S}_{\rm SNe}$ is the source of metals from SNe. For further, detailed discussion of, e.g., $Z_i$ yields, we refer to \citet{Kolborg2022_metal_mixing_1}. Further details of the SNe driving are discussed in \autoref{ssec:SNe}. The $Z$ composition of the medium is utilized by $\Lambda$, which then contributes to determining the local thermodynamic properties of the plasma.
    
    We choose a cubic simulation domain of side length, $L = 1,\!000\,\rm{pc}$, with periodic boundary conditions on the four sides perpendicular to the disk midplane and outflow boundaries on the top and bottom boundaries. The domain is discretized using a regular Cartesian grid of up to $N_{\rm grid} = 1024$ cells for each $L$. For second-order spatial reconstruction methods such as the one used to solve our model, the numerical diffusive effects influence $\approx 10\d{x}$ \citep{Malvadi2023_numerical_diss,Beattie2023_bulk_viscosity}, where $\d{x} = L/N_{\rm grid}$, which means that at $N_{\rm grid} = 1024$ we properly resolve roughly $10\,\rm{pc} \lesssim \ell \lesssim 1,\!000\,\rm{pc}$ in our simulations. This study includes three simulations with exactly the same fluid parameters, but with resolutions of $N_{\rm grid} = 256$ to $N_{\rm grid} = 1024$ for convergence tests. The parameters of each of the runs are summarized in \autoref{tab:sims}.
    
\subsubsection{Gravitational potential} \label{ssec:phi}
    The simulations employ a static gravitational potential, $\phi(z)$, with stellar disk of scale height $z_0$ and surface density $\Sigma_*$ and a spherical dark matter halo of density $\rho_{\rm halo}$,
    \begin{equation}\label{eq:grav_pot}
        \phi(z) = 2 \pi G \Sigma_*\left( \sqrt{z^2 + z_0^2} - z_0 \right) + \frac{2\pi G \rho_{\rm halo}}{3}z^2
    \end{equation}
    with the accelerations $2 \pi G \Sigma_*z/\sqrt{z^2 + z_0^2}$ due to the disk component and $(4/3)\pi G \rho_{\rm halo}z$ due to the halo  \citep{KuijkenGilmore}. The stellar scale height, $z_0$, and the halo density $\rho_{\rm halo}$ are chosen to match the Milky Way model in \citet[][; see Table~1, MW model]{Martizzi2016}. Likewise, we fix $\Sigma_*$ to $\Sigma_{\rm gas}$ by the gas fraction $f_{\rm gas} = \Sigma_{\rm gas}/\Sigma_* = 0.088$, which is chosen to mimic that of the present-day Milky Way within the Solar neighborhood \citep[MW; ][]{McKee_solarneigh}. With these parameters, the gravitational potential has a scale height, $z_{\rm eff} = 180\,\rm{pc}$ \citep{Kolborg2023_metal_mixing_2}.

\subsubsection{Initial conditions} \label{ssec:code&conditions}
    The simulation is initialized at $t = t_{{\rm init}}$ in hydrostatic equilibrium, where we use $\rho(z,t_{{\rm init}}) = \rho_0\exp\left\{ -f \phi(z) \right\}$ and $P(z,t_{{\rm init}})/k_{\rm B} = T_0\rho(z,t_{{\rm init}})/(\mu m_{\rm H})$, where $k_B$ is the Boltzmann constant, $\mu = 0.6$ is the mean molecular weight, $m_{\rm H}$ is the mass of hydrogen, and $\rho_0 = \SI{2.1e-24}{g\,cm^{-3}}$ is chosen such that the gas surface density, $\Sigma_{\rm gas} = \SI{5}{{\rm M_\odot} / pc^2}$, is similar to that of the present-day solar neighborhood value (\citealp{McKee_solarneigh}, \citealp[see also][]{Martizzi2016}), and similarly for the initial metallicity of the simulation $Z = Z_{\odot}$. The constant in the exponential $\rho$ atmosphere is $f = m_{{\rm H}}\mu/(k_BT_0)$, where $T_0 = 12,\!891\,\rm{K}$ is the constant initial temperature. In addition, through $T_0$ and $\rho_0$, we define a midplane pressure of $P_0/k_B = \SI{1.6e4}{K\,cm^{-3}} =  T_0\rho_0/(\mu m_{\rm H})$. The gas velocity is initialized $|\u| = 0$, without initial velocity perturbations.
    
\subsubsection{Supernova driving} \label{ssec:SNe}
    Energy, mass and momentum are injected into the medium via core collapse SNe detonations. The injection is modeled using the sub-grid model in \citet{Martizzi2015}. The model takes into account ambient $\rho$ and $Z$ in the local medium where a SNe detonates and partitions the total energy, $E_{\rm SNe} = 10^{51}\,\rm{erg}$ into (radial) momentum, $\bm{p}_{\rm SNe}$ and thermal energy, $E_{\rm th, SNe}$. Each SNe ejects $M_{\rm ej } = \SI{6.8}{{\rm M_\odot}}$ new material, including $Z_i$, into the ISM, and as the shock wave expands through the medium it continuously sweeps material $M_{\rm swept}$. $M_{\rm ej}$, $\bm{p}_{\rm SNe}$ and $E_{\rm th, SNe}$ are deposited over a region of size $\ell_{\rm inj} = 2 \d{x}$, fixed for all $N_{\rm grid}$. This means that the SNe are detonated on progressively smaller physical scales as we go to higher $N_{\rm grid}$. This is, of course, a degree of freedom, in that we could have fixed $\ell_{\rm inj}$ to a physical scale, but in reality the SNe will be detonated on scales well below any of the resolved scales ($\d{x}\sim1\;\rm{pc}$ at the highest resolutions), so this is indeed well justified, but does mean our energy injection scale shifts as we go to higher $N_{\rm grid}$. We perform convergence studies on the spectra in \autoref{app:convergence} to ensure that nothing peculiar happens to the turbulence between the different $N_{\rm grid}$. The volumetric rate of SNe, $\dot{n}_{\rm SNe}$ is given,
    \begin{equation}
        \dot{n}_{\rm SNe} = \frac{\dot{\Sigma}_*}{2 z_{\rm eff} 100\,{\rm M_\odot}},
    \end{equation}
    where $\dot{\Sigma}_* \propto \Sigma_{\rm gas}^{1.4}$ is the surface density of star formation \citep{Kennicutt1998_SFR, Kennicutt2007_obsSFR}. This results in one SNe occurring for every $100\,{\rm M_\odot}$ stars formed. 
    
    The positions of a SNe is chosen at random within a fixed volume. Parallel to the gravitational potential, the distribution function or SNe positions is,
    \begin{align}
        p(z) = \left\{\begin{matrix}
        1/(2z_{\rm eff}), & |z| \leq z_{\rm eff}, \\
         0, & |z| > z_{\rm eff}, \\ 
        \end{matrix}\right.
    \end{align}
    \ie the SNe have an equal probability of happening anywhere within $z \pm z_{\rm eff}$ of the disk midplane and zero probability outside this region. In the perpendicular direction (\ie in the disk plane) the SNe positions follow a uniform distribution. This is a very simple prescription, but we intend to explore other schemes in future works.
    
    \subsubsection{Heating and cooling} \label{ssec:cooling}
    We employ a heating and cooling model to capture the multiphase nature of the ISM \citep{McKee1977_SN_ISM,Wolfire1995_isothermal_ISM}. The model is based on the microphysical heating and cooling prescriptions studied in \citet{Theuns1998_Cooling} and \citet{SutherlandDopita}. The cooling function $\Lambda$ has both cooling $\mathcal{C}$ and heating $\mathcal{H}$ terms, 
    \begin{align}
        \Lambda(n_{\rm H},Z) = \mathcal{H}(n_{\rm H}) - \mathcal{C}(n_{\rm H},Z).
    \end{align}
    The model solves a time-dependent chemical network consisting of HI, HII, HeI, HeII, HeIII and free electrons. The solution to this network forms the basis for computing $\mathcal{H}$ and $\mathcal{C}$. Our heating term, $\mathcal{H}$ is due solely to photoheating, which is the excess energy in free electrons that have been ejected from HI, HeI and HeII atoms,
    \begin{equation}\label{eq:heating}
        \mathcal{H} = \left( n_{\rm HI} \epsilon_{\gamma \rm HI} + n_{\rm HeI}  \epsilon_{\gamma \rm HeI} + n_{\rm HeII} \epsilon_{\gamma \rm HeII} \right) / n_{\rm H}^2,
    \end{equation}
    where the photo-heating rates, $\epsilon_{\gamma}$, are those reported in Table~B4 of \citet{Theuns1998_Cooling} and depend on the rate of photoionization by the background power-law UV spectrum, $J(\nu)$ (see Equation~B11 of \citealt{Theuns1998_Cooling}). The total cooling rate, $\mathcal{C}$, is given by 
    \begin{equation}\label{eq:cooling}
        \mathcal{C} = \sum_{i=1}^{10} c_i(T, n_{\rm H}) + c_{\rm metal}(T, n_{\rm H}, Z),    
    \end{equation}
    where we sum over each of five physical processes: collisional ionization (HI, HeI and HeII); recombination (HII, HeII, HeIII); dielectronic recombination (HeII); collisional excitation (HI and HeII); and, Bremsstrahlung (HII, HeII and HeIII). We use the rates from Appendix~B1 of \citet{Theuns1998_Cooling}. However, our cooling model neglects the inverse Compton cooling term, which is included in the original \citet{Theuns1998_Cooling} model. Furthermore, we supplement the \citet{Theuns1998_Cooling} model with cooling due to metal line emission, $c_{\rm metal}$, following the model from \citet{SutherlandDopita}. $Z$ is sourced by the injection of metals from SNe, as we indicated in \autoref{eq:metal_transport}. As discussed in \citet{Karpov2020_metallicity_patterns}, SNe injection will not create a solar abundance pattern, and in reality, one needs many additional metal production channels, which we do not take into account in our model.

    \begin{figure}
        \centering
        \includegraphics[width=\linewidth]{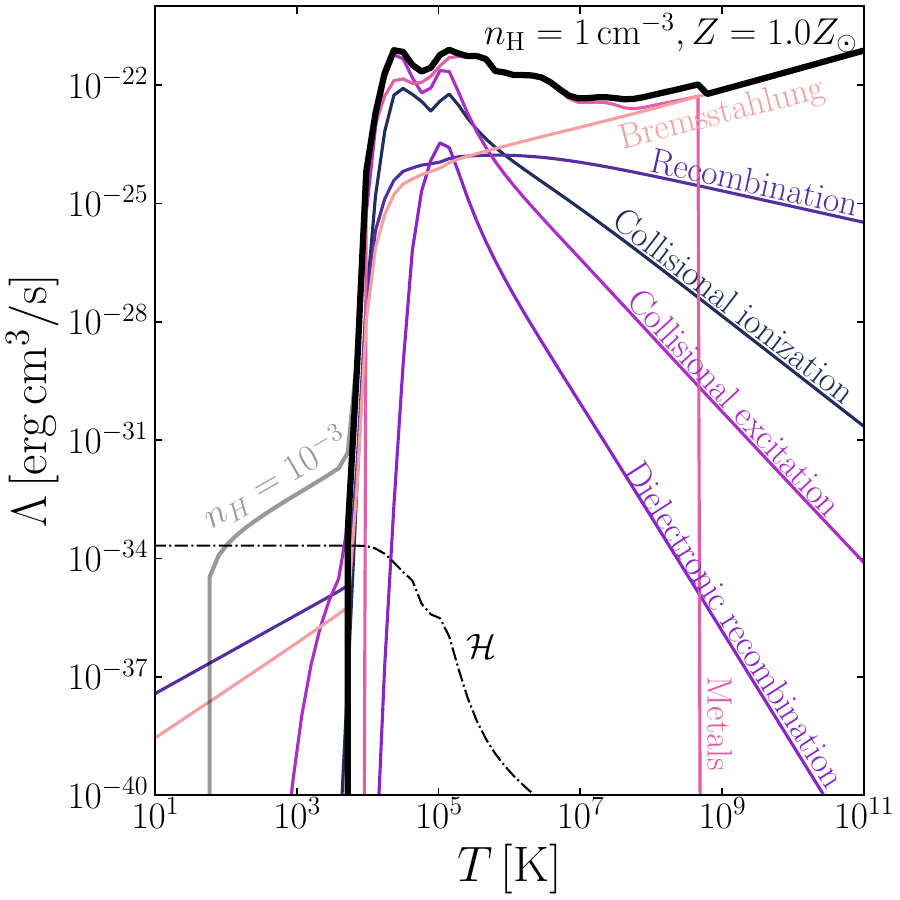}
        \caption{The heating and cooling curve we employ in our simulation. The black solid line is the total cooling, $\Lambda$, at $n_{\rm H} = 1\,\rm{cm}^{-3}$ and $Z = Z_\odot$. Each component of the total cooling is plotted for the same density and metallicity in colored lines, labeled by the associated physical process. The gray line is the total cooling at the indicated lower density, showing that the gas density only significantly influences the cooling at $T \lesssim 10^4\,\rm{K}$. The black, dash-dotted line is the total photoheating, $\mathcal{H}$, at gas density $n_{\rm H} = 1\,\rm{cm}^{-3}$. }
        \label{fig:cooling}
    \end{figure}
    
    The total heating and cooling implemented in the code is shown in \autoref{fig:cooling} at fixed density $n_{\rm H} = 1\,\rm{cm}^{-3}$ and solar metallicity. The thick black line is the total net-cooling, $\Lambda$, while the dash-dotted black line is photoheating, $\mathcal{H}$. The colored lines indicate the contributions to the cooling by each process modeled as indicated along the respective lines. At temperatures $10^4\,{\rm K} \lesssim T \lesssim 10^8\,{\rm K}$ cooling is dominated by metal line emission for $Z \approx Z_{\odot}$ and at $Z < Z_{\rm \odot}$ collisional excitation and ionization dominate the cooling when $T \lesssim 10^6$. However, for $T \gtrsim 10^8\,\rm{K}$ (and for $T \sim 10^7 \,\rm{K}$, for low $Z$) bremsstrahlung dominates the cooling function. Changes in $n_{\rm H}$ only result in significant changes to the total cooling at temperatures $T < 10^4\,\rm{K}$ as shown by the gray line for $n_{\rm H} = 10^{-3}\;\rm{cm}^{-3}$. In general, most of the cooling terms are truncated at $T \approx 10^4\,\rm{K}$; therefore, the simulations do not self-consistently form a cold (or WNM) phase of the ISM via condensation, and any gas at $T \lesssim 10^4\,\rm{K}$ is from the adiabatic expansion of the disk under the impact of the SNe. Additional cooling terms are unlikely to lead to significantly different results in our simulations because the cold gas would be poorly resolved, since we already expect numerical diffusion effects to influence $\sim 10\;\rm{pc}$, which is approximately the size-scale of the largest cold clumps (that can individually have $\Re \sim 10^{10}$, but would be resolved with $\Re \sim 1$ in our simulations; \citealt{Ferriere2020_reynolds_numbers_for_ism}) in bistable simulations of turbulence boxes \citep{Fielding2022_ISM_plasmoids}. Hence, we focus on simulating the large-scale warm (WIM) and hot (HIM) phases of the ISM.
    
\begin{figure}
        \centering
        \includegraphics[width = \linewidth]{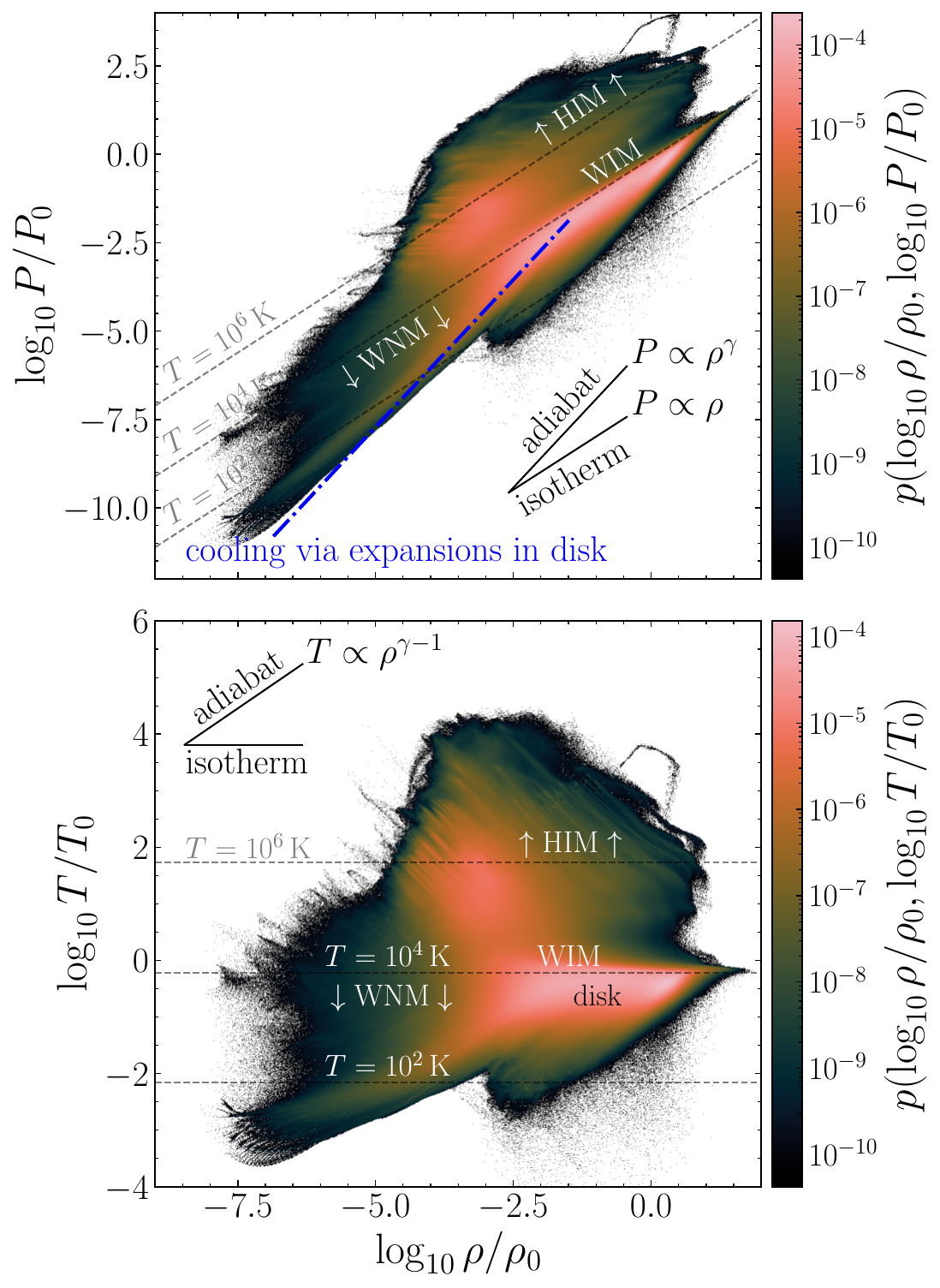}
        \caption{Two-dimensional, time-averaged pressure-density (top panel) and temperature-density (bottom panel) probability distribution functions, illustrating the presence of a multiphase ISM with a hot $T \geq 10^6\,\rm{K}$ (HIM) and volume-filling warm phases $T \sim 10^4\,\rm{K}$ (WIM) and $T \lesssim 10^4\,\rm{K}$ (WNM), annotated on each of the panels. All parameters have been normalized by their initial midplane values, $\rho_0 = 2.1 \times 10^{-24}\,\rm{g}\,\rm{cm}^{-3}$, $P_0/k_B = \SI{1.6e4}{K\,cm^{-3}}$ and $T_0 = 12,\!891\,\rm{K}$. We plot example adiabats $P \propto \rho^{\gamma}$ and isotherms $P \propto \rho$ in both panels. Due to the lack of explicit cooling to $T \leq 10^4\,\rm{K}$ temperatures (see \autoref{fig:cooling}), the $T \lesssim 10^4\,\rm{K}$ gas is reached only through the adiabatic cooling from expanding gas in the disk, highlighted with a blue adiabat from $T \sim 10^4\,\rm{K}$ to $T \sim 10^2\,\rm{K}$ in the top panel. The distributions have been averaged over all realizations in the steady state (see \autoref{ssec:steady_state_units}).}
        \label{fig:phase}
    \end{figure}

    To more clearly understand how $\Lambda$ translates into different ISM phases in our simulation, in \autoref{fig:phase} we plot time-averaged phase diagrams of $P/P_0$ (top panel) and $T/T_0$ (bottom panel) as a function of $\rho/\rho_0$. All parameters have been normalized by their initial midplane condition values (see \autoref{ssec:code&conditions} and \autoref{tab:sims}), and we plot isotherms to help guide the eye. As expected, the impact of the drastic drop in cooling efficiency for $T \lesssim 10^4\,\rm{K}$ is translated into both the $P/P_0$-$\rho/\rho_0$ and $T/T_0$-$\rho/\rho_0$ distribution functions by the build-up of probability density (i.e., the volume filling factor of the gas) around the $T = 10^4\,\rm{K}$ isotherm. The most efficient mechanism for gas to cool below this $T \lesssim 10^4\,\rm{K}$ is through adiabatic expansion caused by expanding gas in the disk, and we observe directly that the trajectory from the $T = 10^4\,\rm{K}$ to the $T = 10^2\,\rm{K}$ isotherm follows an adiabat, $P \propto \rho^{\gamma}$, which we plot in blue (where we use $\gamma = 5/3$ for a monoatomic gas). Indeed, we observe small volumes of the ISM (small probability densities) getting to $T \lesssim 10^2\,\rm{K}$ via this mechanism. 

\subsection{Characteristic scales, \\ dimensionless numbers and stationarity}\label{ssec:steady_state_units}
    In this section, we describe the characteristic velocities, length scales, and time scales of the simulation. We first calculate the (instantaneous) global average turbulent Mach number,
    \begin{equation}\label{eq:mach_number}
        \M = \Exp{\left(\frac{u}{c_s} \right)^2}{\V}^{1/2},
    \end{equation}
    given by the ratio between the dispersion of the local nonthermal, turbulent (rest-frame) velocities $\Exp{u^2}{\V}^{1/2}$ and thermal velocity dispersion, $\Exp{c_s^2}{\V}^{1/2}$. Furthermore, it is useful to consider the measured scale height of the gaseous disk (rather than $z_{\rm eff}$, which is the analytical steady state, not including SNe contributions) $\ell_0$, such that $\rho(z) \propto \exp\left( - |z|/\ell_0 \right)$. With our cooling function, as well as the SNe detonation and turbulence, it makes the analytical equilibrium profile and $\ell_0$ challenging to derive; hence, we resort to numerical means for computing this scale. We directly fit an exponential model to the steady state data and measure $\ell_0$, and then average it over time. From $\left\langle u^2 \right\rangle^{1/2}_{\V}$ and $\ell_0$ we define the characteristic time scale of the simulation, the turbulent turnover time at $\ell_0$,
    \begin{equation}\label{eq:turnover_time}
        t_0 = \frac{\ell_0}{\langle u^2 \rangle_{\mathcal{V}}^{1/2}}.
    \end{equation}
    We will use $\M$, $\ell_0$ and $t_0$ regularly throughout the study. This is, in many ways, analogous to the non-dimensionalizations used regularly in turbulence box simulations \citep[e.g.,][]{Beattie2022energy_balance}. We do, however, note that the $\ell_0$ we compute above is not the outer scale of the turbulence, which we compute independently below in \autoref{sec:power_spectra}, directly from the velocity spectrum. 

    Because we are using a second-order spatial reconstruction method, we can directly estimate the hydrodynamical Reynolds number utilizing numerical dissipation relations in \citet{Malvadi2023_numerical_diss}. For our $N_{\rm grid}$ we get $\Re \sim 10^4$ on the outer scale, which maps to the $\Re$ of the transonic, volume-filling WIM $(T \sim 10^4\,\rm{K})$ in our simulations. At these $\Re$, we do resolve the HIM $(\Re \sim 10^2)$, which is relatively viscous because of the high temperatures, $T \sim 10^6\;\rm{K}$, and low electron densities ($n_e \sim 10^{-3}$), leading from the fact that ${\Re} \propto n_eT^{-5/2}$ \citep{Ferriere2020_reynolds_numbers_for_ism}. However, for the WIM and WNM, which are the volume-filling phases in our simulation, this is a few orders of magnitude away from a realistic $\Re$, with $\Re\sim 10^7$. However, at $\Re \sim 10^4$ we will still have a turbulent warm phase that ought to have at least some of the cascade resolved (indeed we see this is the case in \autoref{sec:transfer_functions}), at least when we compare to turbulent box simulations with the same order spatial reconstruction method \citep[e.g.,][]{Federrath2013_universality}. Based on the required $\Re$, to get a completely resolved WIM we would need to go to grid resolutions of $N_{\rm grid} \gtrsim 10,\!000$ (as in \citealt{Federrath2021} and \citealt{Beattie2025_nature_astro}) and above, which would be an exciting direction for future work.

    The early time in the simulation is characterized by relaxation from the initial conditions -- as the gas cools it collapses in $\bnabla\phi$; simultaneously, feedback from SNe turns on injecting mass, momentum and energy into the disk \citep{Kolborg2022_metal_mixing_1,Kolborg2023_metal_mixing_2}. These injections drive turbulence in the ISM and contribute to pressure support of the gas (via $\bnabla u^2$) in $\bnabla\phi$. Eventually the disk settles into a statistically steady-state where 
    \begin{align}\label{eq:equilibrium}
        \left\langle\bnabla \cdot \left( \rho \u \otimes \u + P \mathbb{I} \right)\right\rangle_{\V} =& - \left\langle\rho \bnabla \phi\right\rangle_{\V} \nonumber \\ 
        &+ \left\langle\dot{n}_{\rm SNe} \bm{p}_{\rm SNe}(Z, n_H)\right\rangle_{\V},
    \end{align}
    in which a stationary turbulent cascade forms, e.g., $\varepsilon_{\rm injection} = \varepsilon_{\rm dissipation}$ \citep{Li2020_SNE_energetics}, where $\varepsilon_{\rm injection}$ is the flux from the SNe injection, and $\varepsilon_{\rm dissipation}$ is the dissipated flux, which is both a function of \autoref{eq:heating} and the numerical dissipation from the discretisation, which we discuss in detail in \autoref{sec:power_spectra} and \autoref{sec:transfer_functions}.
    
\begin{figure}
        \centering
        \includegraphics[width = 0.45 \textwidth]{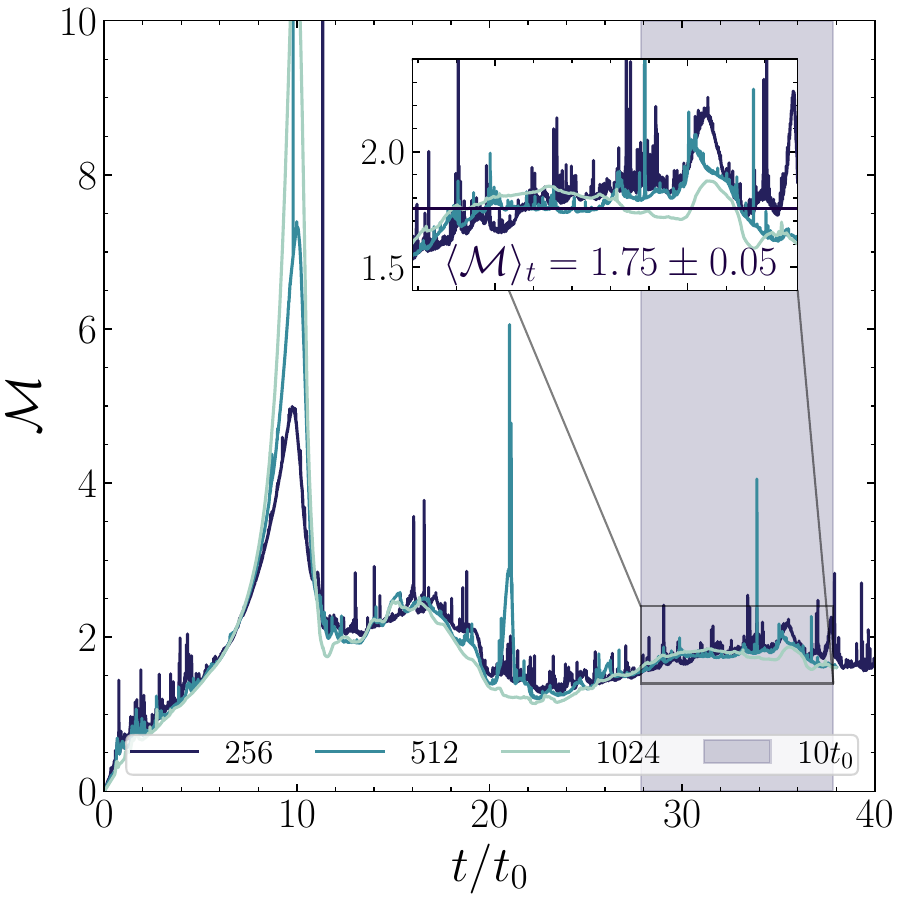}
        \caption{Evolution of the global, volume-weighted turbulent Mach number, $\M$ (\autoref{eq:mach_number}), plotted as a function of the turbulent turnover time, $t_0 = 2.9 \pm 0.7\,\rm{Myr}$ (\autoref{eq:turnover_time}), with each color indicating a different linear grid resolution. The turbulence reaches a steady state for $t \gtrsim 27\,t_0$. The shaded region marks the steady state over which all of the results in this study are taken, unless explicitly stated otherwise, for which the ISM is at $\M \approx 1.8$, \ie transonic, consistent with radio observations of the Milky Way's ISM \citep{Gaesnsler_2011_trans_ISM}.}
        \label{fig:mach}
    \end{figure}

    To determine when the stationary state is reached we plot the evolution of the volume-weighted $\M$ in \autoref{fig:mach}. Volume-averaged over the entire ISM, we find $\M = 1.37 \pm 0.04$ in the steady state (and $\Exp{u^2}{\V}^{1/2} = 28\pm6\,\rm{km\,s^{-1}}$), consistent with the $\M$ derived from observations of the volume-filling, warm-ionized medium in the ISM \citep[$\M \approx 2$][]{Gaesnsler_2011_trans_ISM,Ferriere2020_reynolds_numbers_for_ism,Gerrard2024_nuclear_wind_Mach}. Compared to Fourier driven turbulent boxes \citep[e.g.,][]{Federrath2013_universality,Beattie2022_spdf}, $\M$ has many more intermittent events, associated with sudden intense detonations of SNe \citep{Kolborg2022_metal_mixing_1}. However, the intensity of these fluctuations decreases as we go from $N_{\rm grid} = 256$ to $N_{\rm grid} = 1,\!024$, most likely due to the reduction of the volume-filling factor for the initial SNe seeds. In \autoref{sssec:variability}, we will later explore the effect (the time variability, not $N_{\rm grid}$) that this has on the $k$ space statistics, including the velocity flux in the velocity cascade. Using our exponential profile model, in this state we find $\ell_0 = 85 \pm 6 \,\rm{pc}$ for $N_{\rm grid} = 1,\!024$, allowing us to derive $t_0 = 2.9 \pm 0.7\,\rm{Myr}$. We list $\ell_0$, $t_0$ and for $\Exp{u^2}{\V}^{1/2}$ all grid resolutions in \autoref{tab:sims}, showing good agreement up to 1$\sigma$.
    
    We identify a range of $t/t_0$ wherein the time-averaged $\M$ no longer changes significantly for different time windows, $\Exp{\M(t) - \M(t+\Delta t)}{\Delta t} \approx 0$. This occurs at approximately $t/t_0 \gtrsim 27$, after the initial transient stage associated with reaching a new quasi-equilibrium, \autoref{eq:equilibrium}. Throughout this study, we will average over realizations in $27 \lesssim t/t_0 \lesssim 37$ ($10t_0$ in the statistically steady state), indicated with the gray band in \autoref{fig:mach}. Compared to momentum driven turbulence in Fourier space, it takes roughly an order of magnitude longer in $t/t_0$ to reach a steady state \citep{Beattie2022_spdf,Beattie2022energy_balance}. Of course, this could be reduced by finding better initial conditions that capture \autoref{eq:equilibrium} more accurately, specifically the enhancement in the scale-height from $\left\langle\dot{n}_{\rm SNe} \bm{p}_{\rm SNe}(Z, n_{\rm H})\right\rangle_{\V}$.
    
\begin{figure*}
    \centering
    \includegraphics[width = \linewidth]{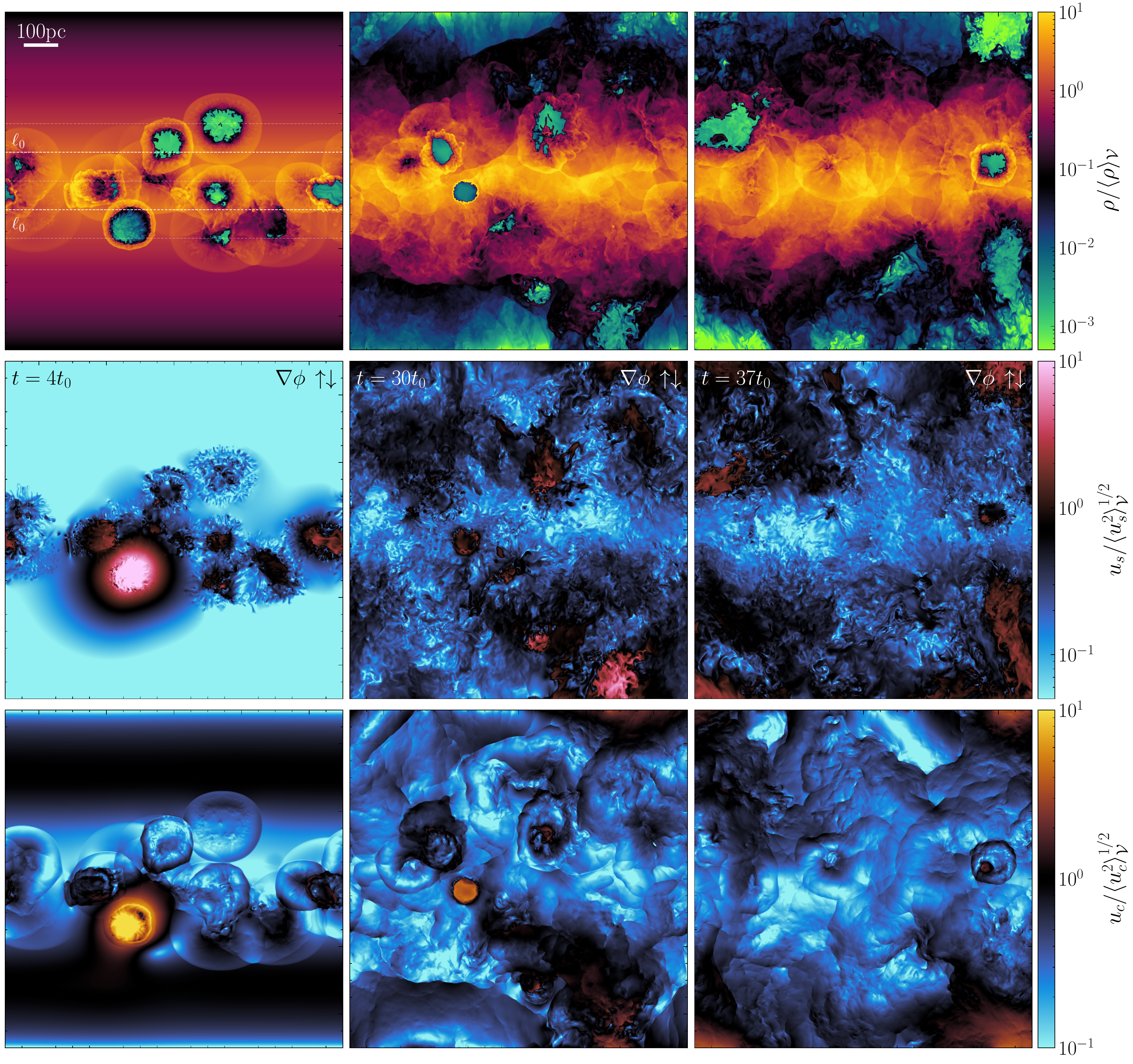}
    \caption{Two-dimensional slices parallel to $\bnabla\phi$, through the center of the simulation domain, for the mass density, $\rho$ (top row), incompressible (solenoidal) velocity, $u_s$ ($\bnabla\cdot\us = 0$; middle row) and compressible velocity, $u_c$ ($|\bnabla\times\uc| = 0$; bottom row) at three different time realizations through the evolution of the simulation. All quantities are normalized by either the volume average in the case of the mass density $\Exp{\rho}{\V}$, or the volume-weighted root-mean-squared (rms) for the velocities. The time of each snapshot is indicated in the panel on the middle row. The first snapshot is taken at the close to the beginning of the simulation $t \approx 4t_0$, the second just before the stationary state $t \approx 30t_0$ and the final in the stationary state, $t \approx 37t_0$ (see the full evolution of the turbulent Mach in \autoref{fig:mach}, which explicitly shows a steady state is reached within $t \sim 20t_0$). We show calculated gaseous scale height, $\ell_0$, with dashed white lines in the top left panel, as well as the midplane and $2\ell_0$ with more transparent white lines.  Fluctuations in $\rho$ and $u_c$ are spatially correlated, with $u_c$ generation corresponding to the expanding shock fronts ($\bnabla\cdot\u<0$) and the dilating regions inside of the remnants ($\bnabla\cdot\u>0$). Whilst $u_s$ is more closely correlated with the post-shock regions and internal SNe structures, where the SNs generate intense regions of vorticity as they expand, eventually contributing to the vortical winds coming out of the disk.}
    \label{fig:map}
\end{figure*}


\section{Velocity structure \\ \& Helmholtz Decomposition} \label{sec:velocity_structure}
     Critical to our analysis of the turbulent cascade is the nature of the velocity modes. We decompose $\u$ into compressible $\uc$ ($|\bnabla \times \uc| = 0$) and incompressible $\us$ ($\bnabla \cdot \us=0$) modes, using a Helmholtz decomposition of the field, such that,
     \begin{align}\label{eq:H_decomp}
        \u = \uc + \us, \quad \text{and} \quad \uc \cdot \us = 0.
    \end{align}
    We do this in Fourier space, where
    \begin{align}
        \uc(\bell) = \int\d{\bm{k}}\;\frac{\bm{k}\cdot\Tilde{\u}(\bm{k})}{k^2}\bm{k} \exp\left\{2\pi i \bm{k}\cdot\bell\right\},
    \end{align}
    is the inverse Fourier transform of the Fourier transformed velocity $\Tilde{\u}(\bm{k})$ projected along the wave vector $\k = 2\pi/\bell$, where $\Tilde{\u}(\bm{k})$ is defined,
    \begin{align}
        \Tilde{\u}(\k) = \frac{1}{N^3_{\rm grid}} \int\d{\bell}\; \u(\x)\exp\left\{-2\pi i \k\cdot\bell \right\}. \label{eq:Fourier}
    \end{align}
    and \autoref{eq:H_decomp} is used to calculate $\us$. In \autoref{fig:map} we plot the mean normalized gas density, $\rho/\Exp{\rho}{\V}$ (top row), and the rms normalized magnitudes, $u_s/\Exp{u_s^2}{\V}^{1/2}$ (middle row) and $u_c/\Exp{u_c^2}{\V}^{1/2}$ (bottom row), components of velocity. In the left-hand column we show these fields at a time shortly after the beginning of the simulation, $t/t_0 = 4$, where the simulation is yet to form a time-stationary state (\ie \autoref{eq:equilibrium}). During this time it is straightforward to pick out the regions influenced by individual SNR, which are spheres of large density gradients around low-density interiors. It is also clear to see that the interiors of SNRs are closely associated with $\us$ modes, while the SNe shock fronts are associated with $\uc$ modes. In the middle and right-hand column we show two-dimensional slices in the steady state (see \autoref{fig:mach}), $t/t_0 = 30$ and $t/t_0 = 37$, respectively. The density has collapsed in the $\bnabla\phi$ direction, leading to a steeper gradient parallel to $\bnabla\phi$, resulting in the disk being supported by both thermal, $\bnabla P$, and turbulent, $\bnabla u^2$, pressure gradients. The effects of multiple, interacting SNRs become apparent for these $t/t_0$, and turbulent velocity modes have developed over the entire simulation domain (see the middle row). Having gained a qualitative understanding of the velocity structures in our simulated ISM we now turn our attention to quantitative methods to dissect the energy spectrum and $k$ space velocity flux transfer, beginning with the velocity power spectrum. 

    \begin{figure}
        \centering
        \includegraphics[width = \linewidth]{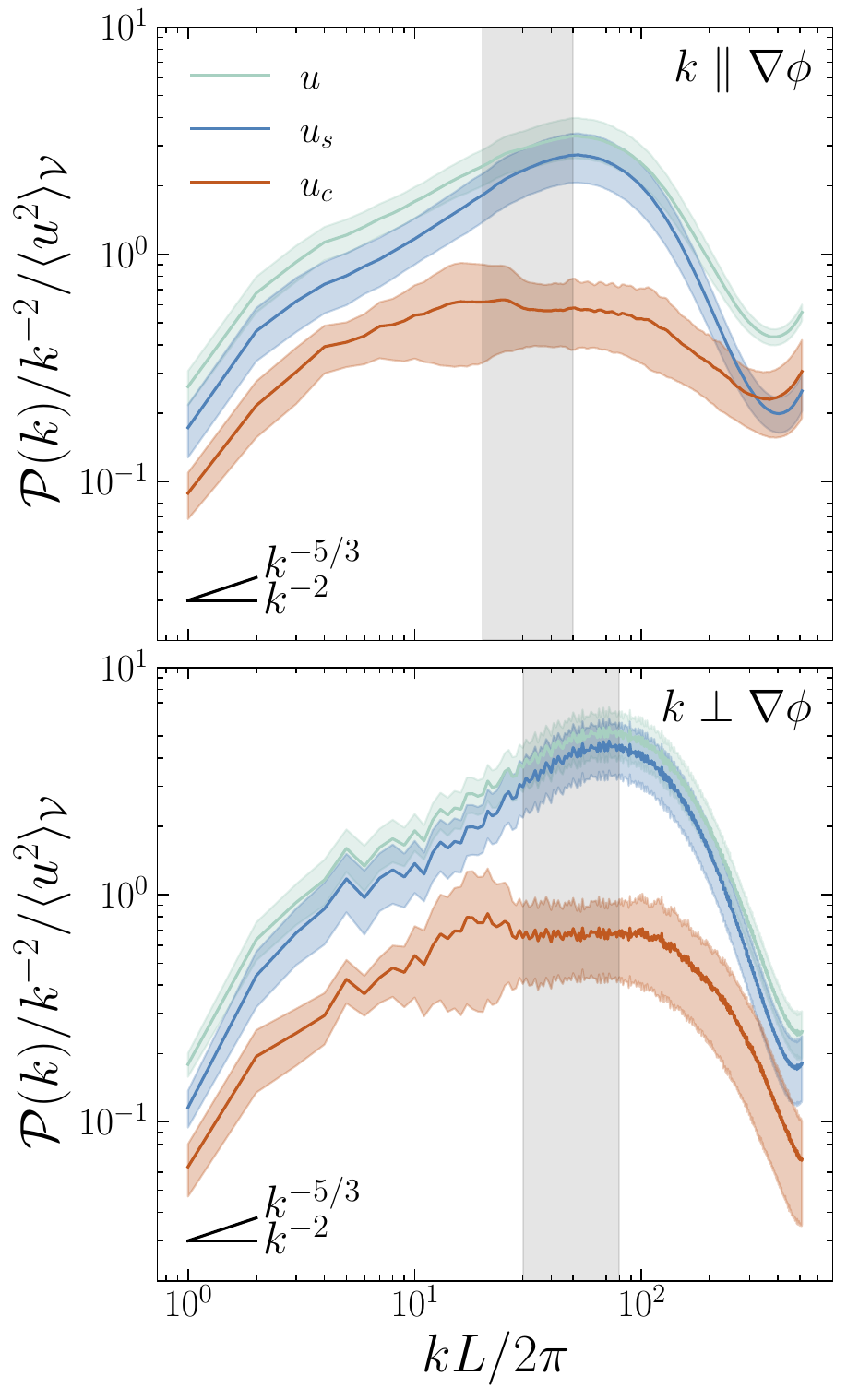}
        \caption{Cylindrically integrated, compensated velocity power spectra, $\P(k)$, as a function of wavenumbers parallel to the static gravitational potential $\bnabla\phi$, \autoref{eq:grav_pot} ($\kpar = |k_z|$; top row) and perpendicular to it ($\kperp = \sqrt{k_x^2 + k_y^2}$; bottom row). Each color represents either the total velocity $\u$ (green), the incompressible velocity component of the velocity $\us$ (blue) or the compressible component of the velocity $\uc$ (orange). All $\P(k)$ are compensated by the power in the total fluctuations $\Exp{u^2}{\V}$ (\autoref{eq:parseval}), and $k^{-2}$ for \citet{Burgers1948}-type turbulence. The $\us$ spectra approximately follow $\sim k^{-3/2}$ scaling and $\uc$ spectra a $\sim k^{-2}$ scaling, the \citet{Burgers1948} expectation for Fourier transforms of velocity discontinuities (all exponents are reported in \autoref{tab:powerspec}). A shaded gray region is shown where the inertial range, $\varepsilon = \rm{const.}$, is very approximately true (see \autoref{app:cross}).} 
        \label{fig:spectra}
    \end{figure}

    \begin{table}[h!]
    \caption{Velocity power spectrum parameters for \texttt{MW\_1024}}
    \hspace{-3.5em}
    \begin{tabular}{ccccc}
    \hline\hline
    & $\alpha_{\parallel}$ & $ \alpha_{\perp} $ &  $\ell_{\mathrm{cor}, \parallel} / \ell_0$ & $ \ell_{\mathrm{cor}, \perp} / \ell_0$ \\ 
    (1) & (2) & (3) &  (4) & (5) \\ \hline
    \hline
    $\u$  & $1.63 \pm 0.01$ & $1.58 \pm 0.03$ & $ 4.1 \pm  0.3$ & $ 3.5 \pm 0.3$ \\ 
    $\us$ & $1.51 \pm 0.02$ & $1.51 \pm 0.03$ & $ 4.1 \pm 0.3$ &  $ 3.3 \pm 0.2$ \\
    $\uc$ & $2.08 \pm 0.02$ & $1.98 \pm 0.02$ & $ 4.5 \pm 0.3$ &  $ 4.0 \pm 0.3$ \\
    \hline\hline
    \end{tabular}
    \begin{tablenotes}[para]
    \textit{\textbf{Notes.}} Column (1): the velocity component, total, $\u$, incompressible, $\us$ and compressible, $\uc$. Column (2) and (3): the best fitting power law exponent to the power spectrum, $\mathcal{P}_u(k) \propto k^{-\alpha}$, along the parallel $\alpha_{\parallel}$ and perpendicular $\alpha_{\perp}$ directions to $\bnabla\phi$. Column (4) and (5): the correlation length scale, \autoref{eq:lcorr}, along the parallel, $\ell_{\mathrm{cor}, \parallel}$, and perpendicular, $\ell_{\mathrm{cor}, \perp}$, directions in units of the gaseous scale height, $\ell_0$, where $\ell_0 \approx 85\;\rm{pc}$ for $N_{\rm grid} = 1,\!024$.
    \end{tablenotes}
    \label{tab:powerspec}
\end{table}

\section{The velocity spectra of supernova-driven turbulence}\label{sec:power_spectra}
    We calculate the turbulent velocity\footnote{See \autoref{app:spect_comparison} for a comparison and discussion of the power spectra for the variables $\bm{u}$, $\rho^{1/3}\bm{u}$, $\rho^{1/2}\bm{u}$.} power spectrum
    \begin{align}
        \P(\k) = |\Tilde{\u}(\k)\cdot\Tilde{\u}^{\dagger}(\k)|,
    \end{align}
    where $\square^{\dagger}$ denotes the complex conjugate. Due to the acceleration from $\bnabla\phi$, there is a global anisotropy in the velocity field which we must account for in the analysis of the spectra. Therefore we integrate the spectra into cylindrical wave numbers, which are defined as follows. Following the geometry of the galaxy disk we define two length scales, $\ell_{\parallel}$ and $\ell_{\perp}$. The first, parallel to $\bnabla\phi$, $\ell_{\parallel}^{-1} \sim \kpar = \vert k_z \vert$, and the second, perpendicular to $\bnabla\phi$ (parallel to the extent of the disk) $\ell_{\perp}^{-1} \sim \kperp= \sqrt{k_x^2 + k_y^2}$. In our analysis we integrate the three-dimensional spectra into a two-dimensional function of these scales,
    \begin{align}
        \P(\kpar,\kperp) = \int\d{\Omega_{\kperp}}\;\P(\k) \, 2\pi \kperp,   
    \end{align}
    where $\d{\Omega_{\kperp}}$ is the angle-integration at fixed radial shells of $\kperp$. We further integrate the $\P(\kpar,\kperp)$ along each dimension to achieve two one-dimensional spectra,
    \begin{equation}
        \P(k_j) = \int\d{k_i}\;\P(k_i, k_j), 
    \end{equation}
    where all of our integrals conserve the total power in the field such that 
    \begin{align}\label{eq:parseval}
        \Exp{u^2}{\V} = \int\d{k}\; \P(k),
    \end{align}
     \ie Parseval's theorem. In \autoref{fig:spectra} we plot the integrated spectra of the total velocity $\u$ and both the $\us$ and $\uc$ components for the highest-resolution run ($\texttt{MW\_1024}$). We normalize all spectra by $\Exp{u^2}{\V}$. The shaded regions around each spectrum denote the $1 \sigma$ fluctuations from the time-average. For $\u$ and $\us$ the variation is generally quite small, while $\uc$ shows more substantial variation. As discussed in \autoref{sec:velocity_structure}, $\uc$ is closely associated with SNe shock fronts and the large time variability in $\mathcal{P}_{u_c}(k)$ is likely tied directly to the stochastic nature of the SNe energy injections.
    
    We identify a single inertial range for the spectra in each dimension, this is indicated by the shaded gray regions in each panel and is bracketed by $20 \lesssim k_\parallel \lesssim 50$ and $30 \lesssim k_\perp \lesssim 80$ for the parallel and perpendicular directions, respectively (see \autoref{app:cross} for further discussion of the inertial range, but to summarize, we use the transfer functions from \autoref{sec:transfer_functions} to find where in $k$ space is the $\varepsilon \approx \rm{const.}$, \ie the Kolmogorov definition). We calculate the correlation (outer) scale of turbulence along each dimension of the volume, 
    \begin{equation} \label{eq:lcorr}
        \frac{\ell_{\mathrm{cor}, i}}{L} = \frac{\displaystyle\int_0^\infty \d{k_i}\; (kL/2\pi)^{-1} \P(k_i)}{\displaystyle\int_0^\infty\d{k_i} \; \P(k_i)}.
    \end{equation}
    In \autoref{tab:powerspec} we summarize $\ell_{\rm cor}$ for all the velocity components. For $\u$ we find $\ell_{\rm{cor}, \parallel} \approx \SI{350}{pc} \approx 4.1 \ell_0$ and $\ell_{\rm{cor}, \perp} \approx \SI{295}{pc} \approx 3.5 \ell_0$. Hence, in general, correlated turbulent motions exist on significantly larger scales than the gaseous scale height of the disk. The $\us$ modes exhibit slightly shorter $\ell_{\rm cor}$ than $\uc$, albeit the differences are minor. The $\ell_{\rm cor}$s we list here are not significantly influenced by resolution -- the values for the lower resolution runs are included in \autoref{app:convergence}. Generally, $\ell_{\rm cor, \parallel} > \ell_{\rm cor, \perp}$ suggesting that the winds blowing out of the disk support correlated $\u$, $\uc$ and $\us$ at larger physical scales than in $\ell_{\rm cor, \perp} $, larger than the classical $\sim$ argument where the outer scale of the turbulence is set by the gaseous scale height \citep[e.g.,][]{Beattie2022_ion_alfven_fluctuations}. We return to this point in \autoref{subsec:shell2shell}. In general, this means that we should consider the galactic winds and the ISM in the disk as being correlated, communicating through the turbulence, and that we should differentiate between driving scales $\ell_{\rm inj}$ ($\approx 2\;\rm{pc}$ for this simulation) and correlation scales, $\ell_{\rm cor}$ ($\approx 350\;\rm{pc}$), \ie a spectrum with energy peaked and correlated at low $k$ does not mean the driving scale is also on low $k$, as inferred in \citet{Bialy2020_driving,Colman2022_large_scale_driving_in_ISM}.
    
    Next, we fit a power law, $\P(k) \propto k^{-\alpha}$ to the spectra over the inertial range based on where the velocity flux is approximately constant. The exact values with uncertainties are reported in \autoref{tab:powerspec}. For the $\u$ spectrum we find $\alpha_\parallel = 1.63 \pm 0.01$ and $\alpha_\perp = 1.58 \pm 0.03$, where the uncertainties are reported as $1\sigma$ errors on the fit. The $\us$ spectrum is slightly shallower with indexes of $\alpha \approx 1.5$ in both directions, while the $\uc$ spectra are steeper showing $\alpha \approx 2$. The $\uc$ spectrum is consistent with \citet{Burgers1948}-type turbulence with $\mathcal{P}_{u_c}(k) \sim k^{-2}$, whilst the $\u$ spectrum is marginally consistent with \citet{Kolmogorov1941}-type turbulence (which predicts a slope of $\alpha = 5/3 \approx 1.67$). Curiously, the $\us$ spectrum deviates from \citet{Kolmogorov1941}, with a scaling close to the MHD \citet{Iroshnikov_1965_IK_turb}-\citet{Kraichnan1965_IKturb} spectrum $\alpha = 3/2$, with the shallower than \citet{Kolmogorov1941} slopes found for magnetized supernova-driven turbulence \citet{Gent2021_supernova_turbulence_and_dynamo}. However, as we highlight later in \autoref{sec:transfer_functions}, the flux between the different $\us$ modes can be mediated by $\uc$, resulting in effects that strongly deviate from the classical \citet{Kolmogorov1941} picture of turbulence (e.g., inverse cascades, negative energy fluxes). \citet{Connor2025} recently has shown that the $-3/2$ spectrum extends from the galactic disk into the winds. Our $\alpha$ values are broadly consistent with previous simulations of SNe detonations in a periodic box \citep{Padoan2016_supernova_driving}, even though we have cylindrically integrated spectra and stratification.
    
    It is noteworthy to further highlight that the $\alpha$ values that best fit each $\u$ are quite similar to those found in \citet{Beattie2025_nature_astro}. In the study, \citet{Beattie2025_nature_astro} modeled supersonic magnetohydrodynamic turbulence using Fourier space driving on large scales at extremely high resolution grids up to $10,\!080^3$. They also performed the decomposition of the $\P(k)$ into $\uc$ and $\us$ modes, showing very similar spectral scalings as presented in this study (e.g., $\alpha \approx 3/2$ for incompressible and $\alpha \approx 2$ for compressible). The agreement between the values found here and those in \citet{Beattie2025_nature_astro} is suggestive that standard turbulent boxes, the workhorses of numerical turbulent experiments, may capture, to a certain degree, more realistic, global set-ups such as the one studied here, at least at the level of the spectrum, and at least for the compressible modes.

    \begin{figure}
        \centering
        \includegraphics[width = \linewidth]{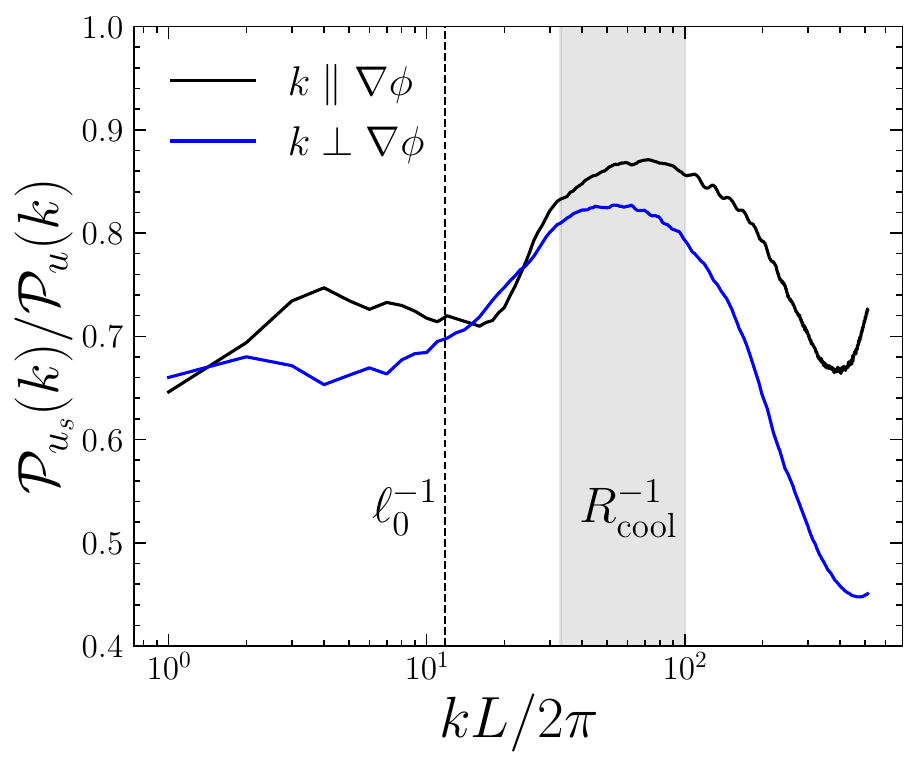}
        \caption{The ratio between the power spectrum of incompressible velocity modes $\mathcal{P}_{u_{s}}(k)$ and the total velocity modes $\P(k)$, $\mathcal{P}_{u_{s}}(k)/\P(k)$, colored by $\kperp$ (blue) and $\kpar$ (black). The gaseous scale height $\ell_0$ is indicated on the plot with the vertical black dashed line. The ratio shows that all scales in the ISM are dominated by $\us$ modes, with a further significant enhancement on scales $\ell \approx R_{\rm cool}$, where $R_{\rm cool} \approx (10-30)\,\rm{pc}$, \autoref{eq:cooling_radius}, is the cooling radius of the SNe \citep{Cioffi1988_SNe_cooling_radius,Blondin1998_SNe_cooling_radius}.} 
        \label{fig:spectra_ratio}
    \end{figure}
    
    For all $\u$ components, the slope of the spectra (and even the normalization) along each direction of $\bnabla\phi$ is consistent with the same value to within $\sim 1\sigma$. This suggests that the SNe-driven turbulence in our simulation develops into fairly isotropic turbulence $(\kpar \sim \kperp \sim |\k|)$ on small enough $\ell$ (or high enough $k$), despite the presence of the stratification from $\bnabla\phi$\footnote{Given the nature of the geometry of the disk splitting scales into parallel $\ell_{\parallel}$ and perpendicular $\ell_{\perp}$ along the $\bnabla\phi$, stratified turbulence theory \citep[e.g.,][]{Maffioli2017_StratTurb} is a natural framework to consider for explaining the measured spectra. However, our simulations, and potentially in general supernova-driven turbulence, fail to meet the basic assumptions that the stratified models have. Specifically, the components of the velocity field modeled here are very similar along and across $\bnabla\phi$, $u_{\parallel} \sim u_\perp$, whereas stratified turbulence theory assumes $u_{\parallel} \ll u_\perp$. Hence, even though we have a disk structure the turbulence is rather isotropic and is not strongly stratified. Regardless, we retain the $\ell_{\parallel}$ and $\ell_{\perp}$ decomposition of the domain demanded upon us by $\bnabla\phi$, which will certainly be an appropriate framework for understanding longer box setups, where the stratification will play a more significant role, as in, e.g., \citet{Kim2017_tigeress}.}. Such universality is regularly invoked for turbulence at the small scales \citep[e.g.,][]{Nazarenko2011_universality} and has previously been realized for a number of different global geometries \citep{Schumacher2014_universality}. This is potentially part of the reason why the spectral exponents end up being quite similar to the local models.

    Interesting to understand is the ratio between the power in the $\us$ and $\uc$ modes as a function of $k$, which allows us to understand what kinds of modes dominate each scale in the ISM \citep{Federrath2010_solendoidal_versus_compressive,Padoan2016_supernova_driving,Pan2016_SNII_comp_ratio,Li2020_SNE_energetics}. We plot $\mathcal{P}_{u_{s}}(k)/\P(k)$ in \autoref{fig:spectra_ratio}, for both of the two directions, indicated with the color. The turbulence is dominated by incompressible modes $\mathcal{P}_{u_{s}}(k)/\P(k) \gtrsim 0.6$, which usually is the case, and has been found before for supernova-driven turbulence \citep{Padoan2016_supernova_driving,Pan2016_SNII_comp_ratio,Gent2021_supernova_turbulence_and_dynamo}, and even for turbulent boxes driven with purely compressible modes \citep[top right panel Figure~8 in][]{Federrath2013_universality}. On $k > \ell_0^{-1}$, \ie smaller $\ell$ than the gaseous scale height, the $\us$ modes become highly energized $\mathcal{P}_{u_{s}}(k)/\P(k) \approx 0.9$, even more so than in purely incompressible-driven turbulent boxes, both isothermal \citep[top left panel Figure~8 in][]{Federrath2013_universality} and bistable \citep{Kobayashi2022_multiphase_turbulence_boxes}. For a radiatively cooled SNe, the cooling radius (for which $t_{\rm cool} = t_{\rm age}$ of the SNe) is,
    \begin{align}\label{eq:cooling_radius}
        R_{\rm cool}\approx 14\,\mathrm{pc}\left(\frac{E_{\rm SN}}{10^{51}\,\rm{erg}}\right)^{0.29}\left(\frac{n_0}{1\,\rm{cm}^{-3}}\right)^{-0.43},
    \end{align}
    at $Z = Z_\odot$, where $E_{\rm SN}$ energy of the SNe explosion, and $n_0$ is the ambient hydrogen density \citep{Cioffi1988_SNe_cooling_radius,Blondin1998_SNe_cooling_radius}. Considering that $n_0 \approx (0.1 - 1)\, \rm{cm}^{-3}$ in the warm phase of our simulations, $R_{\rm cool} \approx (10 - 30)\,\rm{pc}$. We annotate this on \autoref{fig:spectra_ratio} showing that the peak incompressible mode generation occurs on $\approx R_{\rm cool}^{-1}$. 
    
    We will address the most likely reason for the intense surge of incompressible mode power at $ R_{\rm cool}^{-1}$ in \autoref{sec:vorticity}. Regardless of the source, the key result from \autoref{fig:spectra_ratio} is that even in the presence of purely compressible driving, the velocity modes are dominated by the $\us$ component, even more so in supernova-driven turbulence than in turbulent boxes.
    
    The $\P(k)$ provides some first-order understanding of the nature of SNe-driven turbulence in the ISM (e.g., Fourier amplitudes of the velocity $\iff$ the distribution of the second moment of velocity across $k$ space, which one can directly model), but the spectrum alone is insufficient to differentiate between different types of turbulence. Motivated strongly by the work of \citet{Grete2017_shell_models_for_CMHD}, to gain a deeper understanding of the type of turbulence energy interactions, we turn our attention to the spectral energy transfer functions. 

\section{Spectral transfer functions}\label{sec:transfer_functions}
    With the goal of understanding the underlying turbulent cascade mechanisms and nature of the velocity flux, we utilize shell-to-shell velocity transfer functions based on \citet[][and for an incompressible plasma, see \citealt{Alexakis2005_shell_to_shell}]{Grete2017_shell_models_for_CMHD}. This method allows us to both probe $\varepsilon$ between sets of $k$-mode shells, $K$ and $Q$, and associate it with a particular mechanism directly from the momentum equation of the fluid, \autoref{eq:momentum_conservation}. Because we are also interested in interactions between $\uc$ and $\us$, and how they might differ, we take the \citet{Grete2017_shell_models_for_CMHD} method a step further and decompose the transfers into incompressible and compressible mode interactions, which we detail in the following subsection.
    
    \subsection{Helmholtz-decomposed transfer functions}
    The turbulent cascade arises from the quadratic nonlinearity in \autoref{eq:momentum_conservation}, $\u\cdot\bnabla\otimes\u$ \citep{Kraichnan1971_triads,Waleffe1992_triads,Alexakis2005_shell_to_shell}, therefore, to keep this analysis as simple as possible, we focus solely on this  $\u\cdot\bnabla\otimes\u$ for the kinetic energy transfer functions, where the nonlinearity becomes $\u\cdot\u\cdot\bnabla\otimes\u = \u\otimes\u:\bnabla\otimes\u = u_iu_j\partial_ju_i$,\footnote{Note that this can be directly related to \autoref{eq:energy_flux_exact} through the triple product $\partial_i (u_i u_j u_j) = u_j u_j \partial_i u_i + 2u_iu_j\partial_i u_j$ using the symmetry of the $u_iu_j$ tensor. For incompressible turbulence $\partial_i (u_i u_j u_j)=2u_iu_j\partial_i u_j$, so the exact relations we report in \autoref{sec:intro} are proportional to the incompressible transfer functions that we derive in this section, as expected.} which can be filtered to construct the transfer function,
    \begin{align}
        \T_{uu}^u(Q,K | P) = - \int\dthree{\bell}\; \u^K \otimes \u^P : \bnabla \otimes \u^Q,
    \end{align}
     to indicate the transfer of velocity flux from shell $Q$ to $K$, mediated by $\u^P$, where, e.g., $\u^K(\bell)$ is the real space value of $\u$ integrated over the cylindrical integrated $k$-mode shell $K$ (see \autoref{ssec:shells} for more details on the shell definitions). In general, the transfer function probes the interaction between three velocity modes, $\k^Q + \k^P + \k^K= 0$. However, following \citet{Mininni2005_transfer_functions} and \citet{Grete2017_shell_models_for_CMHD}, we do not require knowing the localization of the $\u^P$ mode, and hence we sum over all mediating $P$ mode shells (simply giving the total velocity), which is therefore
    \begin{align}\label{eq:advective_transfer_function}
        \T_{uu}^u(Q,K) = - \int\dthree{\bell}\; \u^K \otimes \u : \bnabla \otimes \u^Q.
    \end{align}
     This means that our transfer functions are not able to localize the interaction (mediated by $\u^P$), but can localize the transfer of velocity flux from $\u^Q$ to $\u^K$ \citep{Mininni2005_transfer_functions,Alexakis2005_shell_to_shell,Grete2017_shell_models_for_CMHD}.

\begin{figure}
        \centering
        \includegraphics[width=\linewidth]{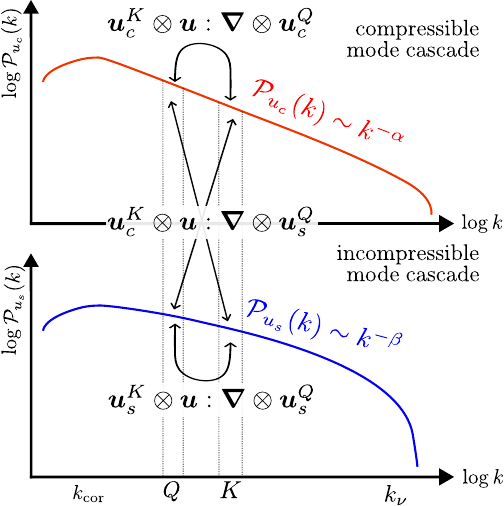}
        \caption{A visual summary of the flux transfers described in \autoref{sec:transfer_functions}. $\us \xrightarrow{\uc,\,\us} \uc$ and $\uc \xrightarrow{\uc,\,\us} \us$ transfers (top and bottom expressions) define velocity fluxes between modes within each of the cascades, and $\uc \xrightarrow{\uc,\,\us} \us$ and $\us \xrightarrow{\uc,\,\us} \uc$ transfers (middle expression) defines fluxes between cascades (shown in \autoref{app:mixed-modes}).}
        \label{fig:transfer_schematic}
    \end{figure}

    In this study, we are interested in the interplay between the compressible and incompressible modes, which we can separate in the velocity via the Helmholtz decomposition as in \autoref{eq:H_decomp}. By applying it to \autoref{eq:advective_transfer_function}, we get
    \begin{align}\label{eq:decomp_TF}
        \T_{uu}^{c+s}&(Q,K) = \\ 
        &- \int\dthree{\bell}\; (\u^K_c + \us^K) \otimes (\uc + \us) : \bnabla \otimes (\u^Q_c + \us^Q), \nonumber 
    \end{align}
    hence we can calculate transfers between $\uc$ and $\us$, mediated by either $\us \cdot \bnabla$ or $\uc \cdot \bnabla$. We can write transfer functions between like-modes as follows,
    \begin{align}
        & \quad\quad\quad\quad \uc^Q \xrightarrow{\uc} \uc^K \nonumber \\
        \T_{cc}^c(Q,K) &= - \int\dthree{\bell}\; \uc^K\otimes \uc : \bnabla \otimes \u^Q_c, \label{eq:Tccc}\\[1em]
        & \quad\quad\quad\quad \uc^Q \xrightarrow{\us} \uc^K \nonumber \\
        \T_{cc}^s(Q,K) &= - \int\dthree{\bell}\; \uc^K\otimes \us : \bnabla \otimes \u^Q_c,\label{eq:Tccs}\\[1em]
        & \quad\quad\quad\quad \us^Q \xrightarrow{\uc} \us^K \nonumber \\
        \T_{ss}^c(Q,K) &= - \int\dthree{\bell}\; \us^K\otimes \uc : \bnabla \otimes \u^Q_s, \label{eq:Tscs}\\[1em]
        & \quad\quad\quad\quad \us^Q \xrightarrow{\us} \us^K \nonumber \\
        \T_{ss}^s(Q,K) &= - \int\dthree{\bell}\; \us^K\otimes \us : \bnabla \otimes \u^Q_s, \label{eq:Tsss}
    \end{align}
    where we provide both the full transfer function definition and shorthand notation, e.g., $\u_a \xrightarrow{\u_c} \u_b$, to emphasize that the notation $\T_{ab}^c(Q,K)$ means $a$-type modes in shell $Q$ donate ($\T >0$) energy to $b$-type modes in shell $K$ via mediation by any $c$-type modes. For example, the $\T_{ss}^c(Q,K)$ transfer function (or concisely written, $\us \xrightarrow{\uc} \us$) describes the local flux from incompressible $\us$ modes in shell $Q$ to incompressible $\us$ modes in shell $K$, mediated by a compressible mode interaction $\uc\cdot\bnabla$ from an unspecified shell. We provide a schematic for the transfers in \autoref{fig:transfer_schematic}. We will regularly use the $\u_a \xrightarrow{\u_c} \u_b$ and $\T_{ab}^c$ notation throughout the rest of the study. With these transfer functions, we can probe not only the location, extent, and directionality of the turbulent cascade in our SNe-driven turbulence simulations, but also the nature of $\uc$ and $\us$ mode interactions. 

    A convenient property of any transfer function is antisymmetry, \ie $\T(Q,K) = - \T(K,Q)$ \citep{Mininni2005_transfer_functions,Alexakis2005_shell_to_shell,Grete2017_shell_models_for_CMHD}. We address the preservation of the antisymmetry property for the $\T(Q,K)$ in \autoref{app:antisymmetry}, highlighting three important aspects. The first is that there is a violation in the antisymmetry with the outflow boundary conditions in our simulation, as expected. We show later that this ends up being quite negligible. However, important for our unique Helmholtz decomposed transfer functions is that the antisymmetric property for $\us$ and $\uc$ mediated $\T(Q,K)$ are different. Indeed, we show that $\us$ mediated transfers are antisymmetric with themselves, but $\uc$ mediated transfers are not, and require an additional transfer mediated by $\bnabla\cdot\uc$ (see \autoref{app_eq:anti_sym_transfers} for more details).
    
    There also exists a set of transfer functions for flux transfers between mixed modes ($\us \xrightarrow{\uc,\,\us} \uc$ and $\uc \xrightarrow{\uc,\,\us} \us$), these interactions carry less velocity flux and therefore we limit our analysis in the main text to just the like-mode transfers. However, the flux between mixed modes transfers is still generally interesting because it describes the process of turning $\uc$ modes from detonating SNe directly into $\us$ modes, which define the classical turbulence cascade; for this reason, we include the mixed-mode transfers and the discussion of them in \autoref{app:mixed-modes}.

    \subsection{Shell definition}\label{ssec:shells}
        We define our shells in the cylindrical coordinate system (see \autoref{sec:power_spectra}). For perpendicular scales, $\ell\perp\bnabla\phi$, it is  
            \begin{align}
                    \u_{\perp}^K(\bell) &= \int\d{\k}\; \delta^2(k_{\perp} - K)\Tilde{\u}(\k) \exp\left\{2\pi i\k \cdot \bell\right\}, \label{eq:shellquant_perp}
            \end{align}
        and for parallel scales it is,
            \begin{align}
                    \u_{\parallel}^K(\bell) &= \int\d{\k}\; \delta(|k_{\parallel}| - K)\Tilde{\u}(\k) \exp\left\{2\pi i\k \cdot \bell\right\}, \label{eq:shellquant_parallel}
            \end{align}
        without loss of generality. In the \autoref{eq:shellquant_perp} $\delta^2(k_{\perp} - K)$ is the two-dimensional delta function that selects the $K$ shell from the cylindrical $k_{\perp}$ coordinate, and the same for $\delta(|k_{\parallel}| - K)$ but for the one-dimensional delta function, since the $k_{\parallel}$ is one-dimensional. Each set of $K$ and $Q$ shells are chosen to be logarithmically spaced, such that the shell edges are given by,
        \begin{align}
            \left\{ Q_i \right\} = \left\{ K_i \right\} &= \left\{ 2^{ (i -1) / 4 + 2} \right\}, \\ & i = 0,\; \dots,\; 4 \frac{\ln(N_{\rm grid}/8)}{\ln(2)} + 1,
        \end{align}
        where $N_{\rm grid}$ is the number of resolution elements per linear dimension (\ie $N_{\rm grid} = 1024$ for the $1024^3$ simulation). This allows us to extract turbulent eddies that are local in log space, rather than specific wave modes, which are local in linear space, as discussed in \citet{Grete2017_shell_models_for_CMHD}. Because we compute the transfer functions for both $\ell_{\parallel}$ and $\ell_{\perp}$, we have a total of 20 transfer functions, considering all of the $\u \rightarrow \u$,  $\uc \rightarrow \uc$,  $\us \rightarrow \us$ and $\uc \leftrightarrow \us$ velocity flux transfers.

    \begin{figure*}
        \centering
        \includegraphics[width = \textwidth]{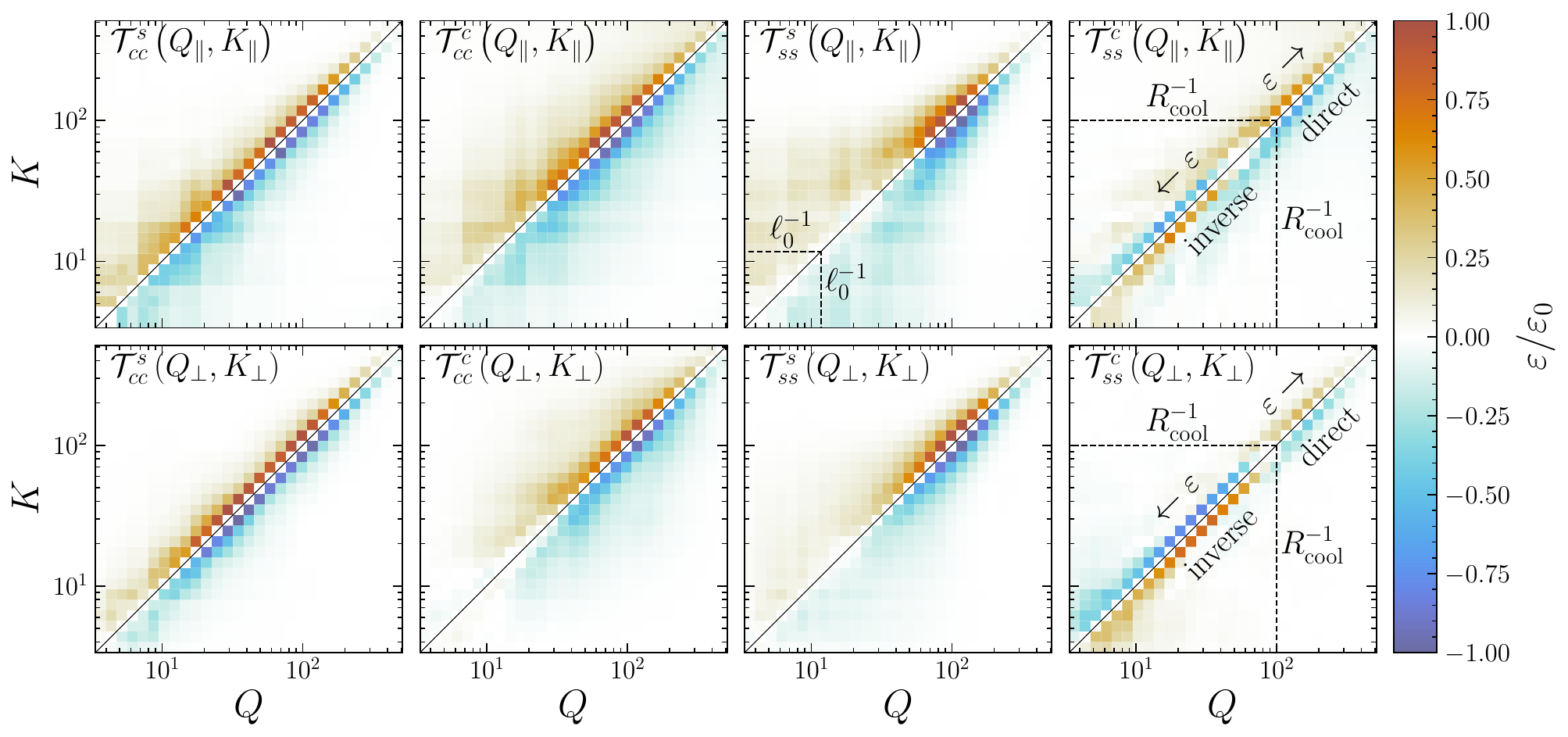}
        \caption{Time averaged shell-to-shell transfer of kinetic energy between logarithmic shells, with transfers parallel to the gravitational potential ($\kpar$; top row) and perpendicular to it ($\kperp$; bottom row). All transfers are between like-modes ($\uc \leftrightarrow \uc$ or $\us \leftrightarrow \us$) of velocity, \autoref{eq:Tccc}-\autoref{eq:Tsss}, colored by the velocity flux $\varepsilon$, normalized to the maximum, absolute value, $\varepsilon_0$, in each transfer. The black horizontal and vertical lines in the $\T_{ss}^s(Q_\parallel, K_\parallel)$ panel indicate $k$ of the gaseous disk scale height, $k_0 \sim \ell_0^{-1}$ (\autoref{ssec:code&conditions}), and the lines in the $\T_{ss}^c(Q, K)$ panels indicate the approximate cooling radius, $R_{\rm cool}^{-1}$ of the SNe, \autoref{eq:cooling_radius}. Most compressible $\uc$ and incompressible $\us$ modes exhibit direct turbulent cascades (positive velocity flux from large to small scales). However, $\us \rightarrow \us$ mediated by $\uc$ interactions exhibits an inverse (negative velocity flux) cascade from small to large scales. This takes energy from below the disk scales $k > k_0$, out to larger scales, $k < k_0$, potentially then fueling the direct cascades in the other transfer functions (and e.g., the winds out of the disk). Shell-to-shell transfers between mixed-modes are included in \autoref{app:mixed-modes}.}
        \label{fig:shell2shell}
    \end{figure*}
            
    \subsection{Shell-to-shell energy transfer} \label{subsec:shell2shell}
       In \autoref{fig:shell2shell} we visualize the shell-to-shell energy transfer between velocity like-modes ($\uc \rightarrow \uc$ or $\us \rightarrow \us$ fluxes). The top row of the figure contains the transfers in $\kpar$, whilst the bottom row includes the transfers in $\kperp$. Each column contains a particular three mode transfer type, which is indicated in the top left hand corner of the panel. As with the spectrum in \autoref{sec:power_spectra}, there is not a large difference between $\kpar$ and $\kperp$, so we will discuss the trends found for the flux transfer across both the directions at the same time. When expanding out each transfer function in terms of its triple product one finds a surface flux term, $\sim \oiint_{\partial\mathcal{V}} u_i u_j u_i\d{\partial\mathcal{V}_j}$ that disappears in the triply-periodic case. This does not strictly disappear for our simulations, but one can probe how large the effect is by looking for deviations away from $\T(Q,K) = - \T(K,Q)$ in the $\us$ mediated transfers (see \autoref{app:antisymmetry}). In \autoref{fig:shell2shell} we see perfect antisymmetry in all $\us$ mediated transfers, hence we conclude $\oiint_{\partial\mathcal{V}} u_i u_j u_i\d{\partial\mathcal{V}_j} \approx 0$, preserving the $\T(Q,K) = - \T(K,Q)$ property.

       \subsubsection{Compressible to compressible mode flux}
       We begin with the velocity flux $\uc \rightarrow \uc$, corresponding to either $\T^c_{cc}$ or $\T^s_{cc}$ (1$^{\rm st}$ column in \autoref{fig:shell2shell}). Both of these transfers exhibit direct energy cascades from large to small scales, regardless of the mediating mode or the direction. $\T_{cc}^s$ supports a highly localized direct cascade which extends over most of the $k$-modes in both $\kpar$ and $\kperp$. This may potentially be from the post-shock regions (energetically dominated by $\us$; see Figure~13 in \citealt{Hew2023_lagrangian_stats}) scattering $\uc$ modes down a cascade. This is quite peculiar since the $\uc$ power spectrum is $u_c^2(k) \sim k^{-2}$ (see \autoref{fig:spectra}), which is usually interpreted as \citet{Burgers1948}. But \citet{Burgers1948} is highly non-local, in that $k^{-2}$ comes from Fourier transforming a single velocity discontinuity (a single real-space structure transforms to all $k$-modes). Clearly this is not what is happening in the $\T^s_{cc}$ transfers, which has the most extended and local cascade out of all of the transfers functions. Based on this result, we can confidently say that there is a $\uc$ mode cascade, but it is mediated by interacting with $\us$ modes. 
       
       Now we turn our focus to the three-mode interaction $\T_{cc}^c$ (2$^{\rm nd}$ column in \autoref{fig:shell2shell}). This transfer function shows a direct cascade over a large range of $k$ modes that is much less localized than $\T_{cc}^s$ (much more velocity flux in the off-diagonal components). This is potentially associated with \citet{Burgers1948}-type of turbulence (\ie not a cascade, just non-local dumping of energy), indicating that the velocity flux imprint from the velocity discontinuities is contained within the $\T_{cc}^c$ transfer but not within $\T_{cc}^s$. The sources of mediating $\uc$ modes in the $\T_{cc}^c$ transfers are from SNe detonations. Hence, we may consider $\T_{cc}^c$ as a probe of the velocity flux from the interactions of the expanding shells with the $\uc \rightarrow \uc$ cascade, somewhat visualized in the bottom row of \autoref{fig:map}. Since both the $\T_{cc}^c$ and $\T_{cc}^s$ mode interactions transfer velocity flux between $\uc \rightarrow \uc$ modes, both of these interactions contribute to the spectrum in $u_c^2(k)$, shown in \autoref{fig:spectra}. Based on the total energy fluxes in and out of each $Q$ and $K$ (the coss scale transfer) in \autoref{app:cross}, the $\T_{cc}^c$ velocity flux is significantly larger than $\T_{cc}^s$, so  $\T_{cc}^c$ is a much faster\footnote{We say ``faster" in that $\varepsilon$ and the cascade rate, $t^{-1}_{\rm nl}$ are related by $\varepsilon \propto t_{\rm nl}^{-3}$ (assuming Kolmogorov turbulence, $u \sim (\varepsilon\ell)^{1/3}$, and $t^{-1}_{\rm nl}\sim u/\ell$), hence larger $\varepsilon$ means a faster cascade, which intuitively makes sense.} cascade (energy moves faster to high $k$ with $\T_{cc}^c$). Regardless of the absolute values of the flux, what we have shown here by separating out $\T_{cc}^s$ from $\T_{cc}^c$, is that even though the $u_c^2(k) \sim k^{-2}$ spectrum is regularly associated with the \citet{Burgers1948}-type phenomenology \citep[e.g.,][]{Federrath2013_universality}, there is a real $\uc \rightarrow \uc$ cascade component of this transfer that is highly local in $\uc$, when mediated by $\us$. 

       \subsubsection{Incompressible to incompressible mode flux}
       Now to the incompressible $\us \rightarrow \us$ transfers, \ie $\T^{s}_{ss}$ and $\T^{c}_{ss}$ (3$^{\rm rd}$ column in \autoref{fig:shell2shell}). The $\T^{s}_{ss}$ transfers are the classical Kolmogorov-type transfer, where \autoref{eq:energy_flux_exact} should hold. Interestingly, these transfers are the most non-local out of all of the energy fluxes, with all $kL/2\pi \lesssim 50$ (well above to below the gaseous scale height) being dominated by non-local low velocity flux transfers across the modes. It is only when we get to $k \gtrsim 30$ modes, \ie $\ell \lesssim 30\,\rm{pc}$, where we see something that starts to resemble a more local cascade, but these scales have a significant numerical viscous component, as discussed in \autoref{ssec:steady_state_units}. This is interesting because, as we showed in \autoref{fig:spectra}, the $u_s^2(k) \sim k^{-3/2}$ spectrum is self-similar over a large range of scales (e.g.,$4 \lesssim k_{\perp} \lesssim 80$ and same for $k_{\parallel}$), which includes most of the scales here that are dominated by non-local transfers. Indeed, this would suggest that the self-similar region of the spectrum is not at all generated from local transfers, \'a la Kolmogorov. In fact, it seems like the compressible modes, in general, are undergoing much more local cascades, which is completely at odds with the conventional wisdom for supersonic turbulence.
       
       Strikingly, the $\T_{ss}^c$ (4$^{\rm th}$ column in \autoref{fig:shell2shell}) supports both an inverse and a direct cascade, split by the inverse cooling radius scale, $R^{-1}_{\rm cool}$, as shown in \autoref{eq:cooling_radius}. The flux for the inverse cascade is significantly stronger than the forward one (energy is transported more efficiently up, i.e., $\varepsilon < 0$, than down, i.e., $\varepsilon > 0$). The inverse cascade extends over a significant range of $k$ modes, extending well beyond $\ell_0^{-1}$, and into the galactic winds. Like $\T_{ss}^s$, a small region of net zero $\varepsilon$ exists between the inverse and the direct cascades, although this region is much more localized than is seen for $\T_{ss}^s$. In \autoref{ssec:inverse_cascade} we examine more closely the mechanisms for establishing and supporting the inverse transfer of energy on $R_{\rm cool}^{-1}$, but already we can assume that this is not a standard incompressible inverse cascade, which is mediated by helical $\us$ modes \citep{Plunian2020_inverse_cascade}. Indeed, this is one of the key results in this paper -- the compressible mode mediated $\uc \rightarrow \uc$ transfer functions support a direct $\varepsilon > 0$ and inverse $\varepsilon < 0$ cascade. In both the $\ell_{\parallel}$ and $\ell_{\perp}$ directions in $\T_{ss}^c$ there is also significant energy loss at $k \lesssim k_0$ (red lobes in $\T_{ss}^c$ at low $k$, which indicate energy coming out of those modes). This is potentially evidence for feeding the other cascade with the inverse fluxes. We explore this in a little more detail in \autoref{app:mixed-modes}.

    \begin{figure}
        \centering
        \includegraphics[height = 0.8\textheight]{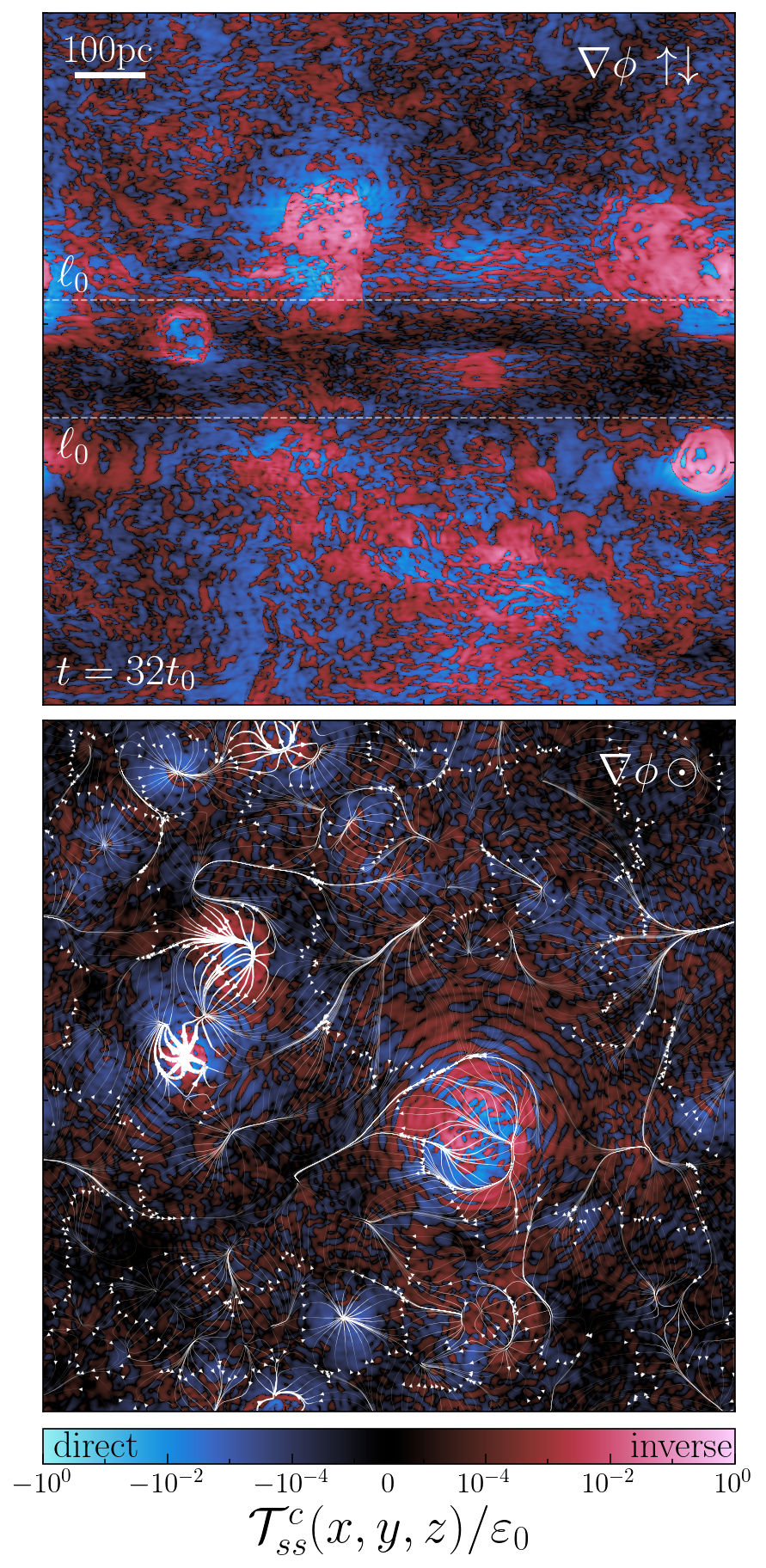}
        \caption{Two-dimensional slices of $\T_{ss}^c(x,y,z)$ (the velocity flux between $\us$ modes mediated by $\uc$ modes; \autoref{eq:energy_flux_field}), Fourier filtered over $11 \leq \kperp \leq 45$ with $Q_\perp < K_\perp$, covering the range of modes where the inverse cascade is observed in the $\T_{ss}^c$ transfer function (see bottom right hand panel of \autoref{fig:shell2shell}). For $\T_{ss}^c>0$ (red) the region of space is undergoing an inverse cascade, and for $\T_{ss}^c < 0$ (blue), a direct cascade. In the top panel the slice is parallel to $\bnabla\phi$ through the center of the domain, with the measured gaseous scale height $\ell_0$ annotated in a similar way as in \autoref{fig:map}. In the bottom panel we slice perpendicular to $\bnabla \phi$, through the center of the galactic disk.}
        \label{fig:Tscs_w_U}
    \end{figure}

       Let us summarize these last two paragraphs, because this is a point that needs to be emphasized. In the self-similar range of the incompressible power spectrum ($\mathcal{P}_{u_{s }}(k) \sim k^{-\alpha}$; shown in \autoref{fig:spectra}), the velocity flux transfers are direct but strongly non-local when we consider $\us \xrightarrow{\us}\us$ transfers, and inverse but local when we consider $\us \xrightarrow{\uc} \us$ transfers on scales larger than the inverse SNe cooling radius. These are the two types of transfers that make up the incompressible mode cascade. This is completely unlike the \citet{Kolmogorov1941} phenomenology for hydrodynamical turbulence, which is local and direct. Because the $\us$ cascade is energetically dominant, and because it is a mixture of $\varepsilon > 0$ and $\varepsilon < 0$, this will act to reduce the total $\varepsilon$, potentially softening the cascade and creating deviations from the strongly nonlinear, local \citet{Kolmogorov1941} case, which we show is evident in $\mathcal{P}_{u_s}(k)$ (\autoref{sec:power_spectra}). Finally, these flux results do not resemble local box turbulence simulations from \citet{Grete2017_shell_models_for_CMHD}, where there is no evidence for inverse cascades and where the $\uc$ mediated transfers were shown to be more non-local than the classical advective ones.  However, note that \citet{Grete2017_shell_models_for_CMHD} did not perform the Helmholtz decomposed transfers, and has additional magnetic fields, so indeed, the fluxes we probe here $u_i^Qu_{j,c}\partial_ju_i^K$ are different than what \citet{Grete2017_shell_models_for_CMHD} investigated $\sim u^Q_iu^K_i\partial_ju_{j,c}$ (we show the relation between these transfer functions in \autoref{app_eq:anti_sym_transfers}).

 \subsection{Unraveling the inverse cascade mechanism} \label{ssec:inverse_cascade}
    Now we intend to provide direct evidence for the mechanism responsible for driving the inverse cascade seen in the rightmost panels of \autoref{fig:shell2shell}. In \autoref{fig:Tscs_w_U} we show two-dimensional slices of the $\T_{ss}^c(Q_\perp, K_\perp)$ transfer function before we integrate it, \ie the spatial slices of 
    \begin{align}\label{eq:energy_flux_field}
        \T_{ss}^c(x,y,z) = - \us^{K_{\perp}}\otimes\uc:\bnabla\otimes\us^{Q_{\perp}},
    \end{align}
    from a single time realization of the simulation, normalized by the mean velocity flux $\varepsilon_0$, and filtered at $11 \leq \kperp \leq 45$ with $Q_\perp < K_\perp$, to ensure that we pick the structures inside of the inverse cascade that we showed in \autoref{fig:shell2shell}. The top panel shows the energy transfer in a $\parallel \bnabla\phi$ slice, and the bottom panel shows a $\perp \bnabla\phi$ slice. $\T_{ss}^c(x,y,z) > 0$ (red) corresponds to regions that participate in the inverse cascade, whilst $\T_{ss}^c(x,y,z) < 0$ (blue) corresponds to the regions participating in the direct cascade, as indicated in annotations on the colormap. We filter and add a slice of the $\uc$ vector field streamlines to the bottom panel, with the streamlines weighted by $|\bnabla\cdot\uc|$ to expose the correlation between $\T_{ss}^c(x,y,z)$ and the convergence $\bnabla\cdot\u < 0$ and $\bnabla\cdot\u > 0$ divergence of $\uc$. This might be the first plot that directly visualizes the direction of a turbulent cascade as a function of space.
    
    Immediately, one can observe a plethora of detailed fluctuations in $\T_{ss}^c(x,y,z)$. From the top panel, we see the winds coming out of the disk have regions with significant $\T_{ss}^c(x,y,z) > 0$, where the inverse cascade is the strongest. Indeed, in most places in the winds we see inverse cascade, indicating that the $\uc$ modes contribute to energizing low-$k$ vortical modes in the galactic winds. Active SNe detonations are also bright red, indicating that the inverse cascade is strongly correlated with SNe detonation events. This is also demonstrated in the bottom panel, where the $\uc$ modes are mostly diverging out of red regions and mostly converging into blue regions. 
    
    As we suggested in \autoref{sec:transfer_functions}, this demonstrates that the $\uc$ modes from SNe detonations drive $\us$ modes to lower $k$ modes by advecting, $\uc\cdot\bnabla$, and stretching, $\bnabla\otimes\us^{Q_{\perp}}$, them around the SNe shells. This is qualitatively similar to the FRB model proposed by \citet{Thompson2023_FRB_emission}, where Alfv\'en modes are stretched by an expanding outflow around a magnetar (see Thompson's figure~2). Note that this is not a $u_i^Ku_i^Q\partial_k u_k^P$ transfer of velocity flux where the compressible mode is mediating the transfer via divergence (from a transfer function perspective, that would be symmetric in energy flux). It is an advective process, where $\us$ modes move to lower $k$ modes by interacting (or scattering) with $\uc\cdot\bnabla$, \ie this is really an inverse cascade and not simply an expansion, which would be probed by a $u_i^Ku_i^Q\partial_k u_k^P$ type transfer function. This is a completely new mechanism for inverse cascade, that does not have anything obviously associated with the helicity of the $\us$ modes, as required in inverse cascades in three-dimensional incompressible turbulence \citep[][see \autoref{ssec:inverse_cascade} for further details about the global helicity in these simulations]{Plunian2020_inverse_cascade}, but a more detailed analysis of the local helicity of the $\us$ modes is required to discuss the detailed deviations.

    \begin{figure*}
        \centering
        \includegraphics[width=\linewidth]{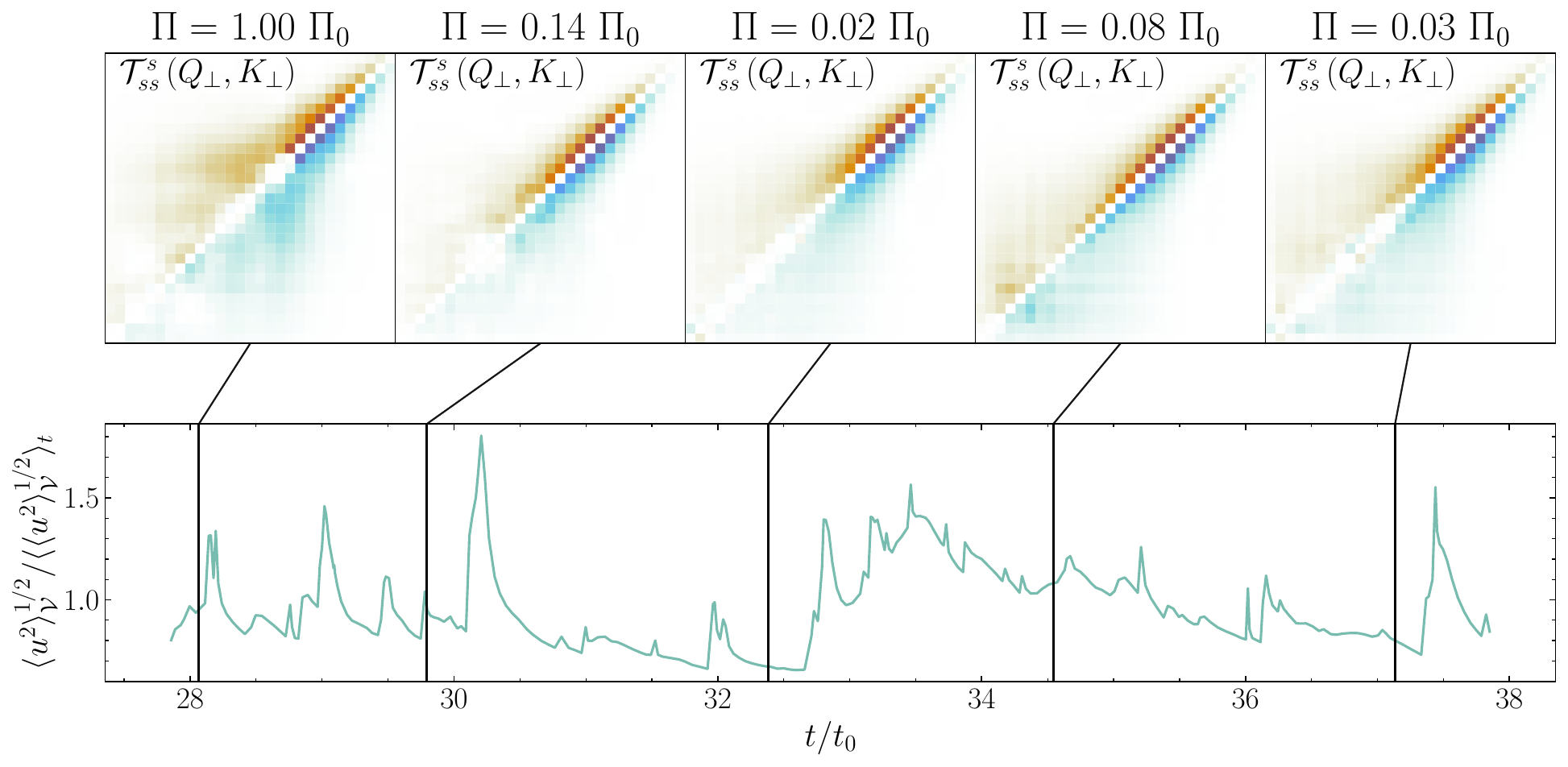}
        \caption{\textbf{Top:} The shell-to-shell energy transfers $\us \rightarrow \us$ modes, mediated by $\us$ modes, $\T_{ss}^s$ (\autoref{eq:Tsss}; \ie the transfer function that would capture the classical, incompressible, \citet{Kolmogorov1941}-type cascade), for the five time realizations averaged over in the transfer function analysis. The extent (how many $k$ modes in the cascade), direction of cascade (whether it is inverse or direct) and the total velocity flux $\Pi/\Pi_0$, changes significantly between different realizations in the stationary state, indicating that the entire ISM cascade fluctuates significantly in time. \textbf{Bottom:} The time evolution of $\langle u^2 \rangle_\mathcal{V}$,normalized by the time-averaged $\langle u^2 \rangle_\mathcal{V}$, within the stationary state of the simulation. }
        \label{fig:vel_disp_snapshotsID}
    \end{figure*}

    We close this subsection with an interesting observation that the $\us\xrightarrow{\uc}\us$ transfers are rather suppressed within the disk, which is best visualized in the top panel of \autoref{fig:Tscs_w_U}, where we have added $|\ell_0|$ contours. This just means that the $\us\xrightarrow{\uc}\us$ transfer is rather low-volume filling factor in the galactic disk, mostly concentrated close to SNRs and in the galactic winds. An interesting future endeavor would be to use the diagnostics we have developed in this section to understand how inhomogeneous other types of energy fluxes are, e.g., those in \citet{Grete2017_shell_models_for_CMHD} throughout a galaxy (one can imagine extending our preliminary analysis by computing poloidal, toroidal and radial $\varepsilon$ profiles for many different types of velocity fluxes). Of course, this will not only result in a better understanding of how turbulence works in our own Galaxy, but will be critical for understanding how heating and cooling work in turbulent plasmas, which are directly related to the velocity flux \citep[e.g.,][]{Howes2024_fundamental_turb_parameters,Mohapatra2024_SNe_turbulence_ellipticals,Sampson2025_CRMHD_heating}.

    \subsection{Time variability of the Kolmogorov cascade} \label{sssec:variability}
        As observed in \autoref{ssec:steady_state_units} (specifically, the strong time variability in \autoref{fig:mach}), our ISM is subject to considerable time variability, at least in $\Exp{u^2}{\V}^{1/2}$ (the square root total energy in the fluctuations integrated over $\P(k)$; \autoref{eq:parseval}). Hence, before concluding this section we examine the nature of the variability in $\T(Q,K)$, which should translate to variability in both slope and extent of the cascade. \autoref{fig:vel_disp_snapshotsID} shows the $\T_{ss}^s(Q_\perp, K_\perp)$ shell-to-shell transfer in the bottom panel (as we have written previously, what ought to be considered the closest to the standard \citet{Kolmogorov1941} cascade probe, since all three interacting modes are $\us$) for each snapshot included in the averages (top row), as well as, $\langle u^2 \rangle_{\mathcal{V}}^{1/2}$ (non-dimensionalized by its time average in the steady state) as a function of $t/t_0$. We probe only the steady state of evolution, derived previously from \autoref{fig:mach}. The black vertical lines indicate where each $\T(Q,K)$ was extracted from in the time series.
    
        From the $\T_{ss}^s(Q_\perp, K_\perp)$ transfer functions, it is immediately apparent that the nature and extent of the cascade change significantly over the span of a few $t_0 \sim 2.9\;\rm{Myr}$. For example, in the first realization, the left-most $\T_{ss}^s(Q_\perp, K_\perp)$ shows a strongly truncated cascade, most likely from strong SNe detonating and disrupting the cascade (which happens just before the spike in $\Exp{u^2}{\V}^{1/2}$; \citealt{Kolborg2023_metal_mixing_2}). Compare this to the $\T_{ss}^s(Q_\perp, K_\perp)$ second from the right, which has an extended cascade starting all the way from the lowest $k$ modes during a low variability period in the evolution of $\Exp{u^2}{\V}^{1/2}$. 
        
        It is not only the extent of the cascade that changes in time (\ie which modes are participating in the cascade) but also the strength of the non-local fluxes (\ie the off-diagonal fluxes). In the first realization, one observes strong non-local energy transfers, which disappear almost completely by the next realization. To complicate matters even further, in the last realization, there appears to be a very weak and very local inverse cascade on some low-$k$ modes, even for this completely incompressible transfer, $\T_{ss}^s$. This suggests that even the $\uc\xrightarrow{\uc}\uc$ mode interactions within our Galaxy may develop into complex velocity flux behaviors (inverse versus direct, local versus non-local) that vary significantly in time. Specifically in more active (e.g., starburst) galaxies, the cascade may become strongly truncated on the large scales (even if one measures a power law in $\P(k)$ on those scales). Indeed, all of these features are simply not captured by the \citet{Kolmogorov1941} phenomenology at all. Of course, we explore here only the hydrodynamics, and for a magnetized ISM this picture may only become more complicated. We leave the magnetized case for a future endeavor, but what we learn from our study is there are still a lot of unknowns to be discovered and made more precise, even in the hydrodynamical case.

    \begin{figure*}
        \centering
        \includegraphics[width=\linewidth]{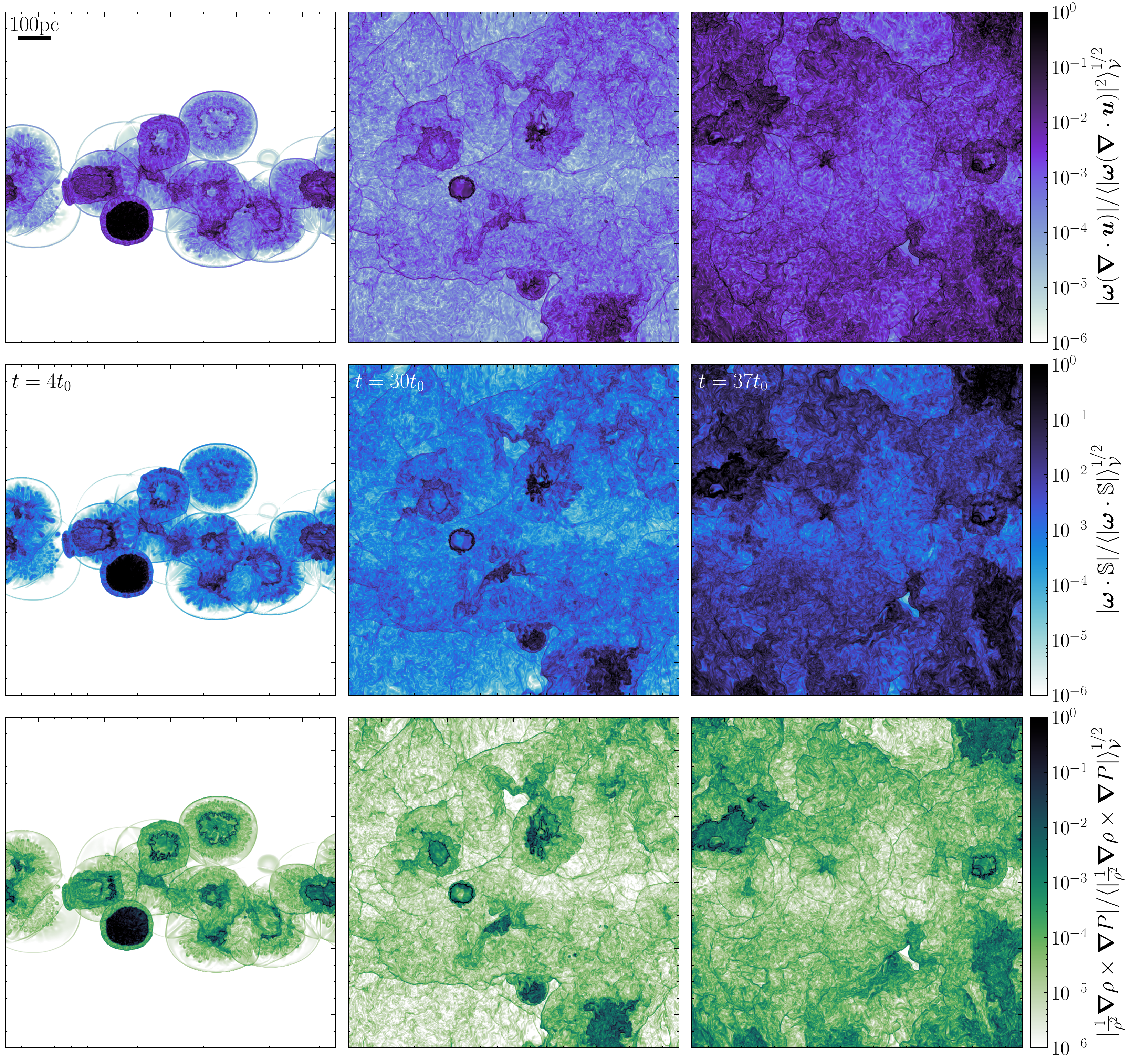}
        \caption{Similar to \autoref{fig:map}, but each row corresponds to slices of the terms responsible for generating vorticity modes, $\partial_t \bom$ (\autoref{eq:vorticity}), probing the generation of modes that participate in the incompressible turbulence cascade. \textbf{Top row:} vortex compression, $\sim\bom(\bnabla\cdot\u)$. \textbf{Middle row:} vortex stretching, $\sim \bom\cdot\St$, where $\St$ is the rate of strain tensor (see \citealt{Beattie2023_bulk_viscosity} for why we use $\St$ instead of the full $\bnabla\otimes\u$). \textbf{Bottom row:} baroclinicity, $\sim \bnabla \rho \times \bnabla P$ the only battery term, in that it generates $\bom$ without any initial seed $\bom$. Each panel is normalized by the instantaneous rms value of the respective field, therefore these panels do not give information about the absolute values of each term (see \autoref{fig:vort_eq_timeline} for the rms amplitudes of each term), but emphasize the spacial structure of each individual component. All of the $\bom$ generation terms are strongly correlated, mostly sensitive to the post-shock regions inside of the SNRs, and the SNe shock fronts themselves. However, in the $\bnabla \rho \times \bnabla P$ term, the most notable sources of $\bom$ generation are the fractal cooling layers \citep[e.g.,][]{Fielding2020_fractal_cooling_layer} between the cold (exterior; $T \sim 10^4\,\rm{K}$) and hot (interior; $T \gtrsim 10^4\,\rm{K}$) gas in the SNRs.}
        \label{fig:vorticity_maps}
    \end{figure*}

\section{What fuels the incompressible cascades in supernova-driven turbulence?}\label{sec:vorticity}

    In \autoref{fig:shell2shell} we showed that $\uc$ interactions can transport energy from small scales to large scales, which can then be parasitized upon by low $k$ modes through other flux transfers. Regardless of how efficient this process is, we need some way of turning $\uc$ modes generated by SNe into $\us$ modes to fuel the incompressible cascades. As we derived in \autoref{sec:transfer_functions}, the interaction that creates the $\us \rightarrow \us$ flux transfer is from $\u\cdot\u\cdot\bnabla\otimes\u$. We can expand this into two further terms, $\u\cdot\u\cdot\bnabla\otimes\u = (1/2)\u\cdot\bnabla u^2 - \u\cdot(\u\times\bom)$, where $\u\times\bom$ is the Lamb vector. Hence, by exploring $\bom$ we are able to probe at least part of the turbulence generated via $\u\cdot\bnabla\otimes\u$. The simplest way of determining $\bom$ sources is by directly taking the curl of \autoref{eq:momentum_conservation}. It is,
    \begin{align}
        \label{eq:vorticity}
        \frac{\d{\bom}}{\d{t}} = & \overbrace{-\bom(\bnabla\cdot\u)}^{\text{compression}} + \underbrace{\bom\cdot\bnabla\otimes\u}_{\text{stretching}} + \overbrace{\frac{1}{\rho^2}\bnabla \rho \times \bnabla P}^{\text{baroclinicity}}, 
    \end{align}
    where $\d{}/\d{t}= \partial_t +\u\cdot\bnabla$ is the Lagrangian derivative, $\bnabla\times\bnabla\phi = 0$ by definition, and similarly, by definition, for the point-source supernova term in \autoref{eq:momentum_conservation}. 
    
    The first term $\bom(\bnabla\cdot\u)$ is the vortex compression term, which we associate with the post-shock regions from the supernova detonations. The third term $(1/\rho^2)\bnabla \rho \times \bnabla P$ is the baroclinic term, which is a battery term for $\bom$, generated between misaligned pressure $\bnabla P$ and gas density $\bnabla\rho$ gradients. Misalignment may occur through interacting supernova shells, where the $\bnabla P$ from one shell is not aligned with the $\bnabla\rho$ from a neighboring shell, or even in the simpler case, where there is any obliqueness within a single expanding compressible wave (e.g., a corrugated shock wave or an expanding cooling layer), which is always the case. In fact, any kind of phase mixing may also excite the baroclinic term. This is similar to the Biermann battery effect (only the proportionality constant is different; \citealt{McKee2020}) for generating astrophysical magnetic fields in collisionless cosmic shocks \citep{Biermann1950_battery}. Finally, for an incompressible fluid, $\bnabla\cdot\u = 0$, the second term $\bom\cdot\bnabla\otimes\u$ is the vortex stretching term, which we associate with vorticity structures interacting with a background shear flow. However, our fluid is highly compressible, and hence we must decompose $\bnabla\otimes\u$ into components to extract only the (volume-preserving) stretching terms to make this equivalent to the incompressible vortex stretching operator \citep{Schekochihin2004_dynamo}. Performing the same $\bnabla\otimes\u$ decomposition as in \citet{Beattie2023_bulk_viscosity}, the vorticity equation becomes
    \begin{align}
        \label{eq:vorticity_updated}
        \frac{\d{\bom}}{\d{t}} = & -\frac{2\bom}{3}(\bnabla\cdot\u) + \bom\cdot(\St + \At) + \frac{1}{\rho^2}\bnabla \rho \times \bnabla P, \\
        \St = & \frac{1}{2}\left(\bnabla\otimes\u + [\bnabla\otimes\u]^T \right) - \frac{1}{3}\It\bnabla\cdot\u,\\
        \At = & \frac{1}{2}\left(\bnabla\otimes\u - [\bnabla\otimes\u]^T \right),
    \end{align}
    where $\St$ and $\At$ are the rate of strain and rate of rotation tensors, respectively, and $[\bnabla\otimes\u]^T$ indicates the transpose. It is straightforward to show that $\At$ only changes the unit vector of $\bom$, since the enstrophy equation (for $\omega^2$) gives rise to a term $\bom \otimes \bom : \At$, which is identically zero because $\bom\otimes\bom$ is a symmetric tensor and $\At$ is antisymmetric, completely analogous to the turbulent dynamo process for a magnetic field embedded in a turbulent medium \citep{Beattie2023_bulk_viscosity}. Under this decomposition, the vortex compression term is modified to $-2\bom(\bnabla\cdot\u)/3$, and the stretching term is simply $\sim \bom\cdot\St$, no longer contaminated with compressions, and maintaining the vortex stretching definition as in the $\bnabla\cdot\u = 0$ regime.

    \begin{figure}
        \centering
        \includegraphics[width = \linewidth]{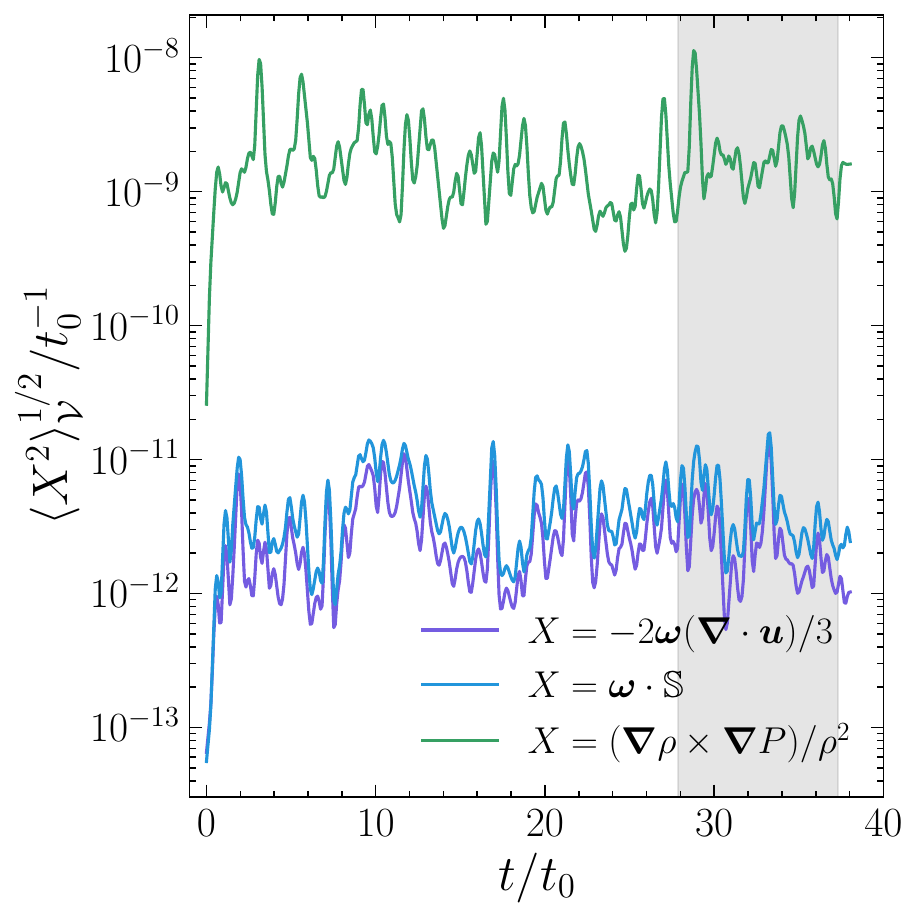}
        \caption{The same terms from \autoref{fig:vorticity_maps} but for the evolution of their rms normalized by $t_0^{-1}$. Throughout the entire evolution of the simulation the baroclinic term (green) dominates over the other two. During the steady state (shaded gray region) it is greater by approximately $\sim 3$ orders of magnitude, making it the leading order term in generating vorticity. Unlike the other two terms, baroclinicity generates vorticity without the need for any seed vorticity, hence can be completely fueled by compressible modes that source strong $\bnabla \rho$ and $\bnabla P$ misalignments through corrugated pressure and density interfaces (like fractal cooling layers, corrugated shocks, etc.) or interacting shock fronts.}
        \label{fig:vort_eq_timeline}
    \end{figure}

    \begin{figure*}
        \centering
        \includegraphics[width=\linewidth]{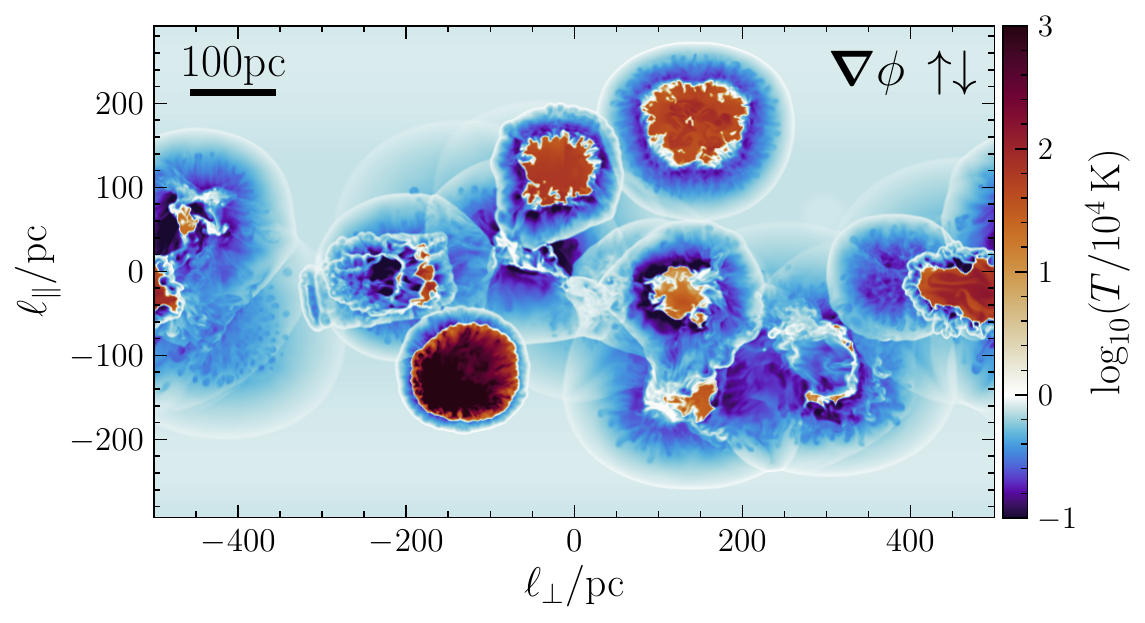}
        \caption{A two-dimensional logarithmic temperature (normalized by $T = 10^4\,\rm{K}$)  slice of a zoom-in of the lower left panel in \autoref{fig:vorticity_maps}, revealing that the baroclinc term, $\propto \bnabla\rho\times\bnabla P$, is the strongest at the fractal cooling layer at $R_{\rm cool}$ between the hot $T > 10^4\,\rm{K}$ plasma inside of the SNRs (shown in red) and the colder $T < 10^4\,\rm{K}$ plasma outside of the SNRs (shown in blue). Because the baroclinic term dominates the total $\bom$ generation (\autoref{fig:vort_eq_timeline}), this suggests that phase mixing through the fractal cooling layer is the strongest source of incompressible turbulence in the SNe-driven medium, and because the incompressible turbulence dominates the total energy in the turbulence (\autoref{fig:spectra}; peaking at $R_{\rm cool}$), phase mixing is potentially the largest source of turbulence, in general.}
        \label{fig:temperature}
    \end{figure*}

    We measure the \autoref{eq:vorticity_updated} compression, stretching and baroclinic terms and show slices of the three-dimensional fields in \autoref{fig:vorticity_maps} and the time-evolution of the first moments in \autoref{fig:vort_eq_timeline}. In \autoref{fig:vorticity_maps} we normalize all $\bom$ terms by the volume integral rms value, revealing the underlying structure of the terms rather than the amplitudes, which we defer to \autoref{fig:vort_eq_timeline}. Firstly, it is clear that there is quite a strong correlation between all three terms, with intense $\bom$ generation in both post-shock regions and at the shock front in the SNe shells. In the early state $t \approx 4t_0$ (left panel), the generation of $\bom$ is concentrated on small scales, where the gas is less dense (see top left-panel in~\autoref{fig:map}, highlighting the stark contrast between the high-dense and low-dense components of the SNRs, potentially separated by the cooling radius; \citealt{Martizzi2015,Martizzi2016}). As the SNRs evolve beyond the Sedov-Taylor stage and into the snowplow stages \citep[e.g.,][]{Martizzi2015}, the shock front becomes a strong source of $\bom$, with all terms (preferentially the $\sim\bom(\bnabla\cdot\u)$ term) showing significant $\bom$ generation around the boundary of the SNRs. By the time the turbulence becomes stationary, the $\bom$ generation spreads through the entire domain, but still with significant concentration in individual SNRs. Undoubtedly, all of the $\bom$ generation in the winds comes in part from the contribution of the inverse cascade discussed in \autoref{ssec:inverse_cascade} and shown in \autoref{fig:Tscs_w_U}. 
    
    The most distinct feature from these slices is that the baroclinic term is strongly enhanced at a particular corrugated surface within the SNRs. This is associated with a cooling layer that develops inside of each SNR, which is well-known to become highly-fractal and corrugated \citep[][where \citet{Lancaster2021_fractal_cooling_layer} has explored similar layers in the context of stellar winds]{Fielding2020_fractal_cooling_layer,Lancaster2021_fractal_cooling_layer,Lancaster2024_cooling_layers}. To show this explicitly, we plot temperature slices at the same $t/t_0$ as in the left column of \autoref{fig:vorticity_maps} in \autoref{fig:temperature}. We specifically zoom in on the SNRs and reveal that indeed, the strong baroclinic source is the cooling layer between the hot $T > 10^4 \, \rm{K}$ that has been heated by the SNe explosion and the colder $T < 10^4 \, \rm{K}$ gas that cools as the SNRs expand. As \citet{Fielding2020_fractal_cooling_layer} explains, these layers are already known to play an important role in facilitating the phase structure of the ISM. But now we also show, based on \autoref{fig:vorticity_maps} and \autoref{fig:temperature}, they also end up being the strongest battery terms for vorticity in our ISM, and at a range of $k$ associated with the fractal nature of the corrugation (e.g., a spectrum of $\us$ modes, $u_s(k) \propto k^\beta$, where $\beta$ is sourced directly from the fractal structure of the layer \footnote{\citet{Fielding2020_fractal_cooling_layer} determined that $\mathcal{A}_{\ell} \propto \ell^{-1/2}$ for cooling layers of this kind, where $\mathcal{A}_{\ell}$ is the area of the layer on length scale $\ell$. As \citet{Fielding2020_fractal_cooling_layer} describes, Koch surfaces have the same $\mathcal{A}_{\ell} \propto \ell^{-1/2}$ scaling, indicating that the surface that produces $\bom$ is a complex, power-law structured field $[\bnabla\rho\times\bnabla P/\rho^2](\k) \sim \k^{-\alpha}$, that would generate an entire spectrum of modes in $\us(\k)$.}, spontaneously generating the incompressible modes from the purely compressible modes driven by SNe detonations.

    We show the rms of each term in \autoref{fig:vort_eq_timeline}, revealing how the magnitude of each term varies as a function of $t/t_0$. In steady state, indicated with the gray band (and even well before then) the ordering of the terms is $(1/\rho^2)\bnabla \rho \times \bnabla P \gg  \bom\cdot\St \gtrsim -2\bom(\bnabla\cdot\u)/3$ in rms. Indeed, the baroclinic term is $\sim 3$ orders of magnitude larger than the other two terms. Hence, by taking the curl of both sides of \autoref{eq:vorticity_updated}, using $\bnabla\times\bnabla\times\u = - \Delta\u$, where $\Delta = \partial_i\partial_i$ is the Laplacian, and by noting $\bom = \bnabla\times\us$, to leading order,
    \begin{align}
        \label{eq:vorticity_approx}
        \frac{\d{\us}}{\d{t}} = & -\Delta^{-1}\bnabla\times\left(\frac{1}{\rho^2}\bnabla \rho \times \bnabla P\right), 
    \end{align}
    in our multiphase ISM, where $\Delta^{-1}$ is the inverse Laplacian. There may be analytical Green's function solutions to \autoref{eq:vorticity_approx}, which we defer for future work that will require understanding local details about the cooling layers. Based on \autoref{fig:vorticity_maps} and \autoref{fig:temperature}, we can confidently say that this is due to the intensely corrugated cooling layers inside the expanding SNRs, which can be seen even in the steady state (last two columns in \autoref{fig:vorticity_maps}). This is the main result from this section. This result is broadly consistent with previous baroclinic studies of SNe driven turbulence \citep{Padoan2016_supernova_driving} and other fluid simulations with non-isothermal EOS \citep{Seta2022_multiphase_dynamo,Mohapatra2022_ICM_driving}, but here we make the result very explicit that it is the baroclinicity directly from the cooling layer embedded within the SNRs themselves that has the largest effect\footnote{This was recently confirmed by \citet{Beattie2025_baroclinicity}.}. Indeed, the cooling layers, which develop when the SNe blast-wave reaches $R>R_{\rm cool}$ are the likely candidates for why the $\us$ modes were extremely energized on $k \approx R_{\rm cool}^{-1}$ in the $\mathcal{P}_{u_{c}}(k)/\P(k)$ ratio in \autoref{fig:spectra_ratio}. 
    
    This effect is a clear departure away from what can be captured in isothermal turbulent boxes, since $P\propto\rho$ isothermality implies that $|\bnabla\rho\times\bnabla P| = 0$, and here we find that it is three orders of magnitude larger than any other of the $\bom$ generation terms, contributing to strongly energizing the $\us$ modes within the gaseous scale-height of the disk. Careful comparisons will need to be made to see if injecting $\us$ modes through large-scale Fourier driving, as is standard practice, can compensate for the deficiency in $\us$ modes that are absent when $|\bnabla\rho\times\bnabla P| = 0$. The top-left panel in \citep[][figure~8]{Federrath2013_universality} suggests that for purely solenoidal driving, it indeed comes close (but potentially on the wrong scales, as indicated in \autoref{fig:spectra_ratio}). Moreover, this may be particularly concerning for simulations that are isothermal and driven with purely compressible modes, since there will be a lack of vorticity, in general, that may not be realized in a true multiphase ISM. 

    To close this section, let us briefly comment on an additional magnetized effect that is deeply tied to the $\bnabla\rho\times\bnabla P$, even though it is not present in the current simulation. As we pointed out in \autoref{sec:vorticity}, the baroclinic vorticity battery sourced from the cooling layer term is proportional to the magnetic Biermann battery term \citep{Biermann1950_battery,Harrison1969_battery_in_early_universe,Kulsrud1997_battery_from_pressure_grad,McKee2020,Martinez2021_biermann_field}, \ie $\partial_t \bm{b} \propto (\bnabla\rho\times\bnabla P)/\rho^2$ if $|\b| = 0$. Indeed, the proportionality leads to a simple relation between $\bm{B}$ and $\bom$, which for a pressure gradient driven battery \citep[e.g.,][]{Kulsrud1997_battery_from_pressure_grad} is,
        \begin{align} \label{eq:b_battery}
            |\bm{B}| = \frac{\bar{m} c}{(1+\chi_i)e}|\bom|,
        \end{align}
    where $\bm{B}$ is the magnetic field, $\bar{m} = \rho/\bar{n}$ is the mean mass of both the neutral and ionized atoms, $\bar{n} = n_i + n_n$ is the total number density, and $\chi_i =n_i/\bar{n}$ is the ionization fraction \citep{McKee2020}. For WIM parameters \citep{Draine2011_ISM_Physics,Beattie2022_ion_alfven_fluctuations}, and using our value for $\omega/t_0^{-1}$ from the baroclinic term in \autoref{fig:vort_eq_timeline} (where $t_0 \sim 3\,\rm{Myr}$, \autoref{tab:sims}), we find
        \begin{align}
             B \approx 10^{-16} \left(\frac{t_0}{3\,\rm{Myr}}\right) \left(\frac{\omega/t_0^{-1}}{10^{-9}}\right)\,\rm{G},
        \end{align}
    providing a significant seed magnetic field over the $k$ modes controlled by the fractal structure of the layer for the ISM which, following the incompressible modes, may ride up the inverse $\us \xrightarrow{\uc}\us$ cascade to larger scales, being further enhanced by the turbulent dynamo \citep{Kriel2022_kinematic_dynamo_scales,Kriel2025_SSD}, and magnetizing the winds and the surrounding medium \citep[e.g.,][]{Tevlin2024_ISM_dynamos_effect_IGM}. This is a strong seed magnetic field compared to the case where the vorticity is sourced from just turbulence alone, where $B \sim 10^{-19}\,\rm{G}$, as calculated in \citet{McKee2020}, and consistent with substituting either $\bom(\bnabla\cdot\u)$ or $\bom\cdot\St$ into \autoref{eq:b_battery}, or generated on larger scales from cosmological shocks and ionization fronts $B \approx 10^{-19} -
    10^{-21}\,\rm{G}$ \citep{Zweibel2013_magnetic_universe}. This is yet another direct consequence of having an extremely strong source of baroclinicity in highly-corrugated cooling layers.
    
\section{Summary, conclusions and limitations}\label{sec:summary}
    In this study, we perform a detailed analysis of the multiphase, hydrodynamical turbulence born from detonations of supernovae in a stratified disk undergoing cooling and heating through a time-dependent chemical network and the adiabatic expansions of the gas. In particular, we focus a lot of our efforts on the transfer of velocity flux through the nonlinear advection term $u_iu_j\partial_ju_i$ in the momentum equation, which defines both the compressible and incompressible mode turbulence cascades. We do this using velocity flux transfer functions, which allow us to compute all three mode interactions in $u_iu_j\partial_ju_i$, split into incompressible $\us$ and compressible $\uc$ modes. In this summary, we will focus on synthesizing the physical picture that has emerged from the analysis of our simulation for how supernova-driven turbulence works in a disk of a galaxy. This will have missing physics, such as rotation, magnetized turbulence and dynamo, and cold plasma phases, but our main goal was always to gather a simple physical picture of supernova-driven turbulence, which we hope can help understand the nature of the turbulence in our own Galaxy, at least in the volume-filling warm phases.

    \begin{figure}
        \centering
        \includegraphics[width=\linewidth]{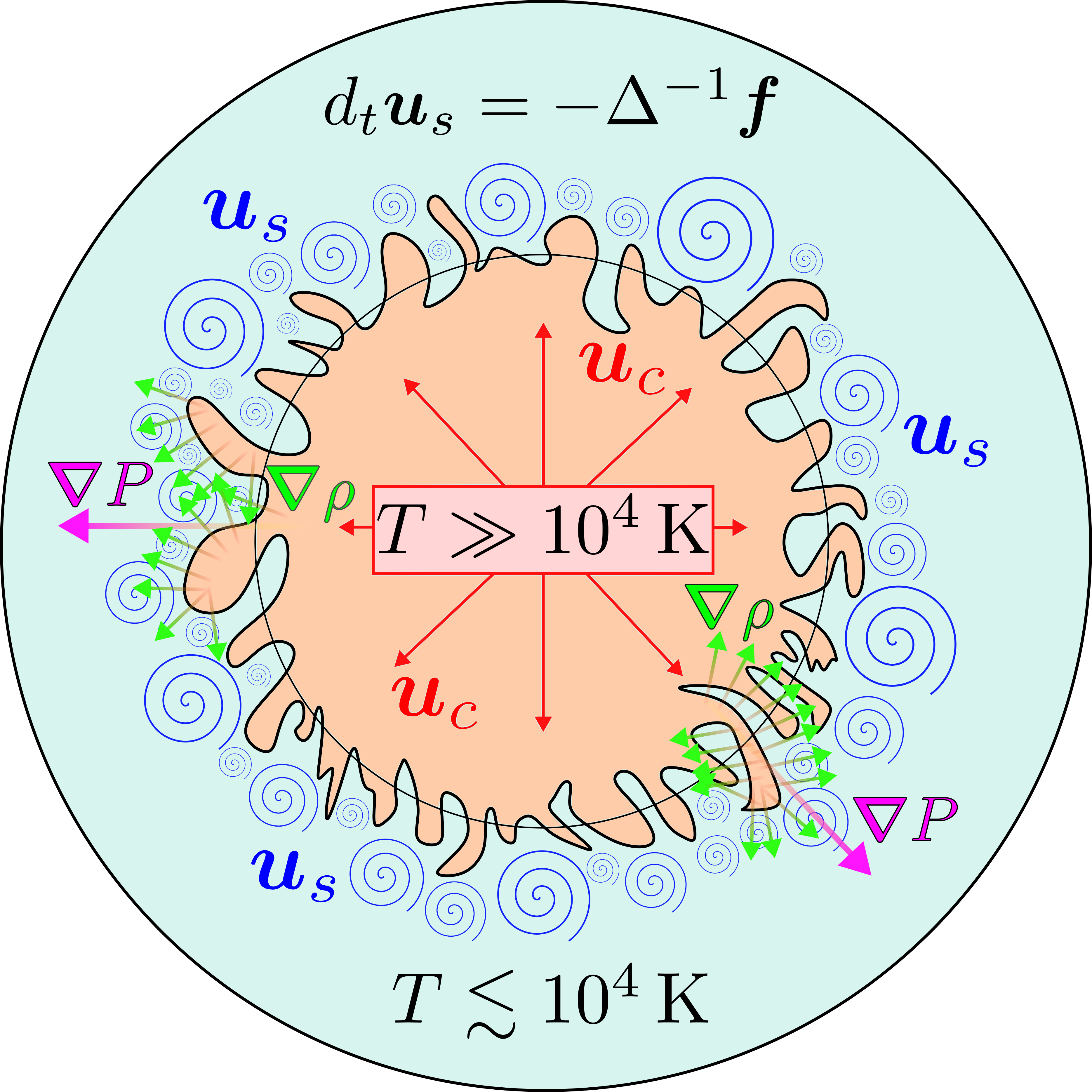}
        \caption{A schematic summarizing the incompressible $\us$ mode (blue) generation through the corrugated cooling layer (black) between the hot (orange) and warm (teal) gas, which are stretched to larger scales by the compressible $\uc$ modes (red) associated with the expanding SNRs. The cooling layer generates a large spectrum of $\us$ modes due to the fractal nature of the layer \citep[see ][for the fractal characterization of such a layer]{Fielding2020_fractal_cooling_layer}. The rate in which the baroclinc term generates $\us$ modes admits to the Green's function solution where $\bm{f} = \bnabla\times(\bnabla\rho\times\bnabla P/\rho^2)$, \autoref{eq:vorticity_approx}, where examples of the $\bnabla P$ for the layer is shown in pink, and the $\bnabla \rho$ or the layer in green.}
        \label{fig:schematic}
    \end{figure}

    \subsection{A supernova-driven turbulence phenomenology}\
        Let us put together everything that we have learnt in this study into a coherent narrative for the nature of SNe-driven turbulence. SNe-driven turbulence injects compressible modes, $\uc$, on the smallest scales in the system, \ie it is compressibly-driven turbulence. As the SNe detonate they heat the gas to $T \gg 10^4\,\rm{K}$ and as they expand into the WIM ($T \sim 10^4\,\rm{K}$) they cool adiabatically, creating corrugated cooling layers at $\ell \approx R_{\rm cool}$, the cooling radius of the SNe. Due to the corrugation, $|\bnabla\rho\times\bnabla P| \gg 1$, which intensely generates incompressible, vortical modes, $\us$, across a broad spectrum of $k$ modes determined by the fractal nature of the layer (see \autoref{fig:schematic}). These modes energize the scales below the gaseous scale height, $\ell_0$, peaking at $ \approx R_{\rm cool}$, such that almost 90\% of the energy is in $\us$ on these scales. On scales beyond $R_{\rm cool}$, the expanding SNRs facilitate an inverse cascade via the three-mode interaction, $\us \xrightarrow{\uc} \us$, that brings $\us$ modes up to scales well beyond $\ell_0$, e.g., into the galactic winds and throughout the disk. At the same time, the post-shock regions have relatively local, direct cascades, $\us \xrightarrow{\us} \us$, that transport velocity flux down to smaller scales. On the large scales, the $\us \xrightarrow{\us} \us$ cascade is highly non-local, with $\us$ modes more or less coupling across all scales, meaning energy from $\us$ is deposited directly from the large scales ($1\,\rm{kpc}$) to the small scales ($10\,\rm{pc}$), without undergoing a \citet{Kolmogorov1941}-style cascade. The compressible modes from the expanding SNRs interact with one another through the three-mode interaction $\uc \xrightarrow{\uc} \uc$, which gives rise to a relatively non-local direct transfer of energy from large to small scales, perhaps resembling the non-local \citet{Burgers1948}-type turbulence. However, there is a $\uc$ cascade that is very local across all scales through the $\uc \xrightarrow{\us} \uc$ interaction, perhaps scattering the $\uc$ modes down the $\us$-dominated galactic winds and post-shocked regions. By this time, fully developed turbulent spectra have been established in the medium, albeit with significant time variability associated with the strong SNe detonation events. The spectra are $\mathcal{P}_{u_c}(k) \sim k^{-2}$ for the compressible modes and $\mathcal{P}_{u_s}(k) \sim k^{-3/2}$ for the incompressible. Over the range of scales for which these power laws hold, the $\uc$ modes are undergoing both local $\uc \xrightarrow{\us} \uc$ and non-local $\uc \xrightarrow{\uc} \uc$ cascades, whereas the $\us$ modes are undergoing $\us \xrightarrow{\uc} \us$ inverse and $\us \xrightarrow{\us} \us$ non-local cascades. Hence, neither of the incompressible cascades in SNe-driven turbulence are consistent with the \cite{Kolmogorov1941} phenomenology of turbulence. 
        
        Finally, it is worth highlighting that in the phenomenology we have described, whether it be by bringing energy from the small-to-large-scales via an inverse cascade, or generating vorticity via the baroclinic term, the compressible modes play a critical and vital role in a SNe-driven medium, and are not passive in any of the processes, which has been the conventional wisdom of the magnetohydrodynamic ISM community \citep{Lithwick2001_compressibleMHD}. Indeed, our results suggest that the multiphase ISM turbulence ecosystem relies critically on both the incompressible and compressible modes playing different but very important roles in creating and maintaining the turbulent fluctuations and cascades. 

        We now itemize some of the key results from this study:
        \begin{itemize}
            \item We drive a stratified ISM into stationarity with SNe detonations happening at the grid scale, resulting in $\M = 1.75 \pm 0.05$ (\autoref{fig:mach}). The SNe and our time-dependent cooling network give rise to a multiphase ISM, with a stable phase at $T \approx 10^4\,\rm{K}$ (\autoref{fig:phase}). Regardless, the gas is able to cool to $\lesssim 10^2\,\rm{K}$ through the adiabatic expansion of the disk, as shown with the blue adiabat in \autoref{fig:phase}. Based on the velocity dispersion, $\Exp{u^2}{\V}^{1/2} = 28 \pm 6 \,\rm{km\,s^{-1}}$, and gaseous scale-height, $\ell_0 = 85 \pm 6 \,\rm{pc}$, we calculate a turbulent turnover time on $\ell_0$ as $t_0 = 2.9 \,\rm{Myr}$.
            \item We decompose the velocity field into incompressible $\us$ and compressible $\uc$ modes using the Helmholtz decomposition, and with this directly calculate cylindrically ($\ell_\parallel, \ell_{\perp}$, with respect to the gravitational potential gradient, $\bnabla\phi$) summed energy spectra $\P(k)$ for each type of mode (\autoref{fig:spectra}). We find $\mathcal{P}_{u_s}(k) \sim k^{-3/2}$ and $\mathcal{P}_{u_c}(k) \sim k^{-2}$, regardless of the direction, indicating that the turbulence is rather isotropic. By computing the ratio between the spectra we find that the turbulence is everywhere dominated by $\us$ modes, and particularly on modes $k > \ell^{-1}_0$, \ie on scales below the scale-height, peaking at the estimates SNe cooling radius $R_{\rm cool}$. On these scales, $\us$ is more dominant than what even local turbulent boxes driven completely with solenoidal modes can achieve. We show that the correlation scales of the turbulence are $\sim 4\ell_0$, indicating that the turbulence is correlated well into the galactic winds.
            \item Using spectral transfer functions for the $\uc$ and $\us$ kinetic flux we investigate four different cascades in two different directions generated from the $u_iu_j\partial_ju_i$ nonlinearity (\autoref{eq:Tccc}-\autoref{eq:Tsss}), shown via shell-to-shell transfers in \autoref{fig:shell2shell}. Similar to the power spectrum, there is not a significant difference between the two directions. The $\uc \xrightarrow{\us} \uc$ is a highly-local, direct cascade, $\uc \xrightarrow{\uc} \uc$ is a more non-local, direct cascade, potentially due to a mechanism like \citet{Burgers1948}-type turbulence. $\us \xrightarrow{\us} \us$ is a highly non-local, direct cascade, specifically at scales where $\mathcal{P}_{u_s}(k) \sim k^{-3/2}$ and $\us \xrightarrow{\uc} \us$ is a mixture of inverse cascade on the scales, where $\mathcal{P}_{u_s}(k) \sim k^{-3/2}$ and a direct cascade at smaller scales. The cascades are roughly split at the SNe cooling radius. The cascade varies significantly in total velocity flux, length and even direction when sampled across multiple realizations in the steady state (\autoref{fig:vel_disp_snapshotsID}).
            \item By visualizing the velocity flux everywhere in space (\autoref{fig:Tscs_w_U}) we reveal that there are simultaneous inverse and direct cascades happening everywhere in the $\us \xrightarrow{\uc} \us$ transfer of velocity flux. Specifically, we identify strong sources of inverse cascades as expanding SNRs that advect $\us$ modes to lower $k$, larger $\ell$, energizing $\uc$ modes in the winds, beyond the gaseous scale height of the disk. 
            \item We further show that the dominant source of incompressible modes in SNe-driven turbulence is from baroclinicity in fractal cooling layers in the SNRs (\autoref{fig:vorticity_maps} and \autoref{fig:vort_eq_timeline}). This is the corrugated layer between the hot, $T \gtrsim 10^4\,\rm{K}$ SNR interior and warm $T \sim 10^4\,\rm{K}$ surrounding ISM, shown very clearly in \autoref{fig:temperature}. We hypothesize that the cooling layer generates the $\us$ modes that are then taken to the large scales through the  $\us \xrightarrow{\uc} \us$ inverse cascade mechanism, which contributes to feeding $\us$ modes into the winds and also the direct cascades. We use the measured $\omega/t_0^{-1}$ and WIM plasma parameters to directly estimate the Biermann field that the cooling layer would generate (\autoref{eq:b_battery}), which is $B \sim 10^{-16}\,\rm{G}$ -- strong seed magnetic field that could contribute to magnetizing the galaxy and surrounding medium when coupled to other processes, like the turbulent dynamo.
        \end{itemize}

\subsection{Limitations of the numerical model}
    In \autoref{sec:sims} and in the current section we discuss the limitations of the simulations presented in this study. 
    
    \paragraph{Background shear profile}
    Firstly, we want to stress that our models are simplistic descriptions of galactic SNe-driven turbulence, with the aim of understanding the nature of SNe-driven turbulence in a stratified disk. Unlike our Galaxy, the disk in our model does not have a velocity and hence shear profile \citep{Huang2016_MW_rotation_curve}. However, this means we are more easily able to probe vorticity generation directly from the SNe, which is desirable for our study.
    
    \paragraph{Stable cold phase gas}
    Further, our cooling function, $\Lambda$, is mostly truncated at $T \approx 10^4\,\rm{K}$, as we show in \autoref{fig:cooling}. This means we do not form cold gas through $\Lambda$ (e.g., through far-infrared fine-structure lines, low-$J$ molecular transitions, or dust). Instead, the $T \lesssim 10^4\,\rm{K}$ gas, which is the bulk of the gas in the disk, settling to $T \approx 4\times10^3\,\rm{K}$, creates a WNM phase analogue through macroscopic processes (e.g., adiabatic expansion) rather than through $\Lambda$. This mechanism significantly contributes to the formation of colder gas in an unstable WNM, which is very strongly influenced by the nature of the turbulence itself (e.g., in volume-filling factor), as \citet{Hu2025_multiphaseISM} shows in their Figure~2, and as \citet{Connor2025} shows by modeling the sound speed spectrum. 

    In a medium with an even colder molecular ($T \approx 10\,\rm{K}$) phase, the cold gas will condense, even in the absence of self-gravity \citep{Fielding2022_ISM_plasmoids}. However, as \citet{Fielding2022_ISM_plasmoids} shows, the turbulent kinetic energy spectrum does not significantly differ in a simulation in which this phase is formed via cooling, versus an isothermal approximation (their Figure~4, top panels), with the turbulence organizing itself in $k$-space in a universal manner. The cold clumps that form from condensing ISM plasmas (or gases) are typically of order 0.1-10\,$\rm{pc}$ \citep{Planck2011_cold_clumps}, which, in our simulations, are at the numerical diffusion limit for our uniform grids or less. Even though it is unlikely that the cold phase gas will change the spectral slopes, if the galaxy models are able to produce a molecular cold phase, there will be more phase interfaces to create $\bnabla P \times \bnabla \rho$, which promotes incompressible turbulence generation, as we showed in \autoref{sec:vorticity}. This means the $\bom$ generation (and hence the Biermann field) are only  lower limits of what one may expect from phase mixing in SNRs alone. With extremely high-resolution simulations of single SNe, \citet{Guo2024_high_res_SN} showed that the internal structure of a SNR can host many phases (indeed, many pioneering ideas about the multiphase ISM were developed specifically in terms of SNe forming all phases of the ISM; \citealt{McKee1977_SN_ISM}). Rigorous analysis on local simulations such as the ones in \citet{Guo2024_high_res_SN} may reveal in detail how a fully resolved bi- or tri-stable ISM (locally within an SNR) generated by SNe can be batteries for both vorticity and magnetism. The expectation is that stronger batteries than those in our study will be present.

    \paragraph{Self-gravity, star formation and SNe prescription}
    Along with the omission of the cold phase, our simulations do not include self-gravity, which would be particularly important for the cold phase component and being able to self-consistently for star-cluster particles to be used as SNe detonation coordinates, as modeled in other simulations \citep[e.g.,][]{Rosdahl2017_SNe_feedback}. Our SNe prescription, first used in \citet{Martizzi2016}, is elementary in this sense. We uniformly seed SNe through roughly a gaseous scale height in the disk. This is in the aim of performing simple, controlled numerical experiments to understand the overall flow of velocity flux, without the need for performing a full star formation and feedback simulation. We present a more detailed study on the effect of the SNe seeding and galactic model (parameterized by both the static potential \autoref{eq:grav_pot}, and SNe detonation rate $\dot{n}_{\rm SNe}$) in a forthcoming study (Connor, Beattie, et.~al, in prep.).

    \paragraph{Magnetic fields and cosmic rays}
    A significant omission to the numerical model is the lack of magnetic fields and cosmic rays, two important components of the pressure budget of a star-forming galaxy \citep{Ruszkowski2023_CRs_in_galaxies_review}. Both magnetic fields and cosmic rays directly change the nature of the turbulence. By comparing \citealt{Federrath2020} with \citealt{Beattie2025_nature_astro}, we see that the subsonic spectrum changes from $k^{-5/3}$ to $k^{-3/2}$ for the velocity with the addition of a magnetic field, meaning that the magnetic field allows for the sequestering of velocity modes on trans-to-sub-sonic scales more efficiently with a magnetic field. \citet{Sampson2025_CRMHD_heating} showed that CRs can be constantly reaccelerated through parasitic interactions with the $\u_c$ modes, decaying the $\u_c$ spectrum on all scales. \citet{Bustard2022_turbulent_reacceleration} showed a similar effect for the $\sqrt{\rho}\u$ spectrum. Furthermore, cosmic rays can facilitate large-scale pressure gradients that constantly accelerate winds and plasma out of the Galaxy \citep{Booth2013_CRwinds,Ruszkowski2023_CRs_in_galaxies_review}, which will facilitate stronger anisotropies between the perpendicular and parallel transfer functions in \autoref{fig:shell2shell}.
    
    Furthermore, even in the absence of a strong magnetic field, a dynamo, both small-scale \citep{Brandenburg2005_astro_dynamos,Federrath2016_dynamo,Kriel2022_kinematic_dynamo_scales,Beattie2023_growth_or_decay,Kriel2025_SSD,Gent2021_supernova_turbulence_and_dynamo}, and large-scale (e.g., stochastic $\alpha$, even in the absence of large-scale rotation; \citealt{Rincon2019_dynamo_theories}) may be excited. The small-scale dynamo will be saturated almost immediately to $\ekin \approx \emag$, given the large Reynolds number of the warm and cold ISM and the kinematic growth rates scaling like $\gamma \approx \rm{Re}^{1/2}$ \citep{Schekochihin2004_dynamo}. This will make the turbulent magnetic field dynamically important, independent of its initial seed field \citep{Beattie2023_growth_or_decay}. Our simulations neglect these effects, which will change the nature of the energy flux transfers by magnetizing the turbulent fluctuations \citep{Grete2017_shell_models_for_CMHD,Beattie2025_nature_astro}. 

    \paragraph{Kinetic plasma effects in SNRs}
    Finally, we should highlight that SNe shocks, the type that drives the turbulence in our simulations, are known to be collisionless, in that the shock thickness becomes of the order of the ion skin depth, $d_i$ \citep[e.g.,][]{Raymond2023_collisionless_SNe_shocks}. This gives rise to a number of effects that are not captured by a fluid model, like the Weibel instability \citep{Zhou2021_weibel,Zhou2024_weibel_to_dynamo}, or the details of the shock heating on the underlying plasma species \citep{Ghavamian2013_eletron_ion_temps_SNe}. The Weibel instability can efficiently seed a magnetic field at $d_i \sim 10^5-10^8\,\rm{cm}$ in the ISM \citep{Ferriere2020_reynolds_numbers_for_ism}, quickly establishing a field that becomes comparable to the thermal energy on $d_i$ \citep{Huntington2015_Weibel_in_the_lab}. Further simulations, utilizing a generalized Ohm's law that includes $-\bnabla P_e/n_e$, the electron pressure and density, will be required to understand how strong the true Biermann field (not just estimated, as in \autoref{sec:vorticity}) is compared to estimates of the Weibel fields in the SNRs.

\section*{\textbf{Acknowledgments}}
    We thank Philipp Grete for the useful comments that allowed us to find a bug in our original transfer functions. We also thank Mark Krumholz for the helpful comments about the warm ISM phases. We thank Isabelle Connor for the productive discussions about the simulation setup. J.~R.~B.~acknowledges financial support from the Australian National University, via the Deakin PhD and Dean's Higher Degree Research (theoretical physics) Scholarships and the Australian Government via the Australian Government Research Training Program Fee-Offset Scholarship and the Australian Capital Territory Government funded Fulbright scholarship, which facilitated this collaboration and project. J.~R.~B. further acknowledges the support from NSF Award 2206756, and compute allocations rrg-ripperda from the Digital Research Alliance of Canada, as well as high-performance computing resources provided by the Leibniz Rechenzentrum and the Gauss Center for Supercomputing grant~pn76gi~pr73fi and pn76ga. A.~N.~K. and E.~R.-R. acknowledge support by the Heising-Simons Foundation, the Danish National Research Foundation (grant No. DNRF132) and the NSF (grant Nos. AST-2206243, AST-1911206, and AST-1852393).
    C.~F.~acknowledges funding provided by the Australian Research Council (Discovery Project grants DP230102280 and DP250101526), and the Australia-Germany Joint Research Cooperation Scheme (UA-DAAD). C.~F.~further acknowledges high-performance computing resources provided by the Leibniz Rechenzentrum and the Gauss Centre for Supercomputing (grant~pr32lo), the Australian National Computational Infrastructure (grant~ek9) and the Pawsey Supercomputing Centre (project~pawsey0810) in the framework of the National Computational Merit Allocation Scheme and the ANU Merit Allocation Scheme.

\facilities{This research was undertaken with the assistance of resources from the National Computational Infrastructure (NCI Australia), an NCRIS enabled capability supported by the Australian Government.}

\software{Data analysis and visualization software used in this study: \textsc{C++} \citep{Stroustrup2013}, \textsc{numpy} \citep{Oliphant2006,numpy2020}, \textsc{numba}, \citep{numba:2015}, \textsc{matplotlib} \citep{Hunter2007}, \textsc{cython} \citep{Behnel2011}, \textsc{visit} \citep{Childs2012}, \textsc{scipy} \citep{Virtanen2020},
\textsc{scikit-image} \citep{vanderWalts2014}, \textsc{cmasher} \citep{Velden2020_cmasher}, 
\textsc{yt} \citep{yt}, \textsc{pandas} \citep{pandas}, \textsc{joblib}\citep{joblib}} 
    
\FloatBarrier

\appendix

\section{Isotropic power spectrum comparisons}\label{app:spect_comparison}
\begin{figure}
        \centering
        \includegraphics[width=0.5\linewidth]{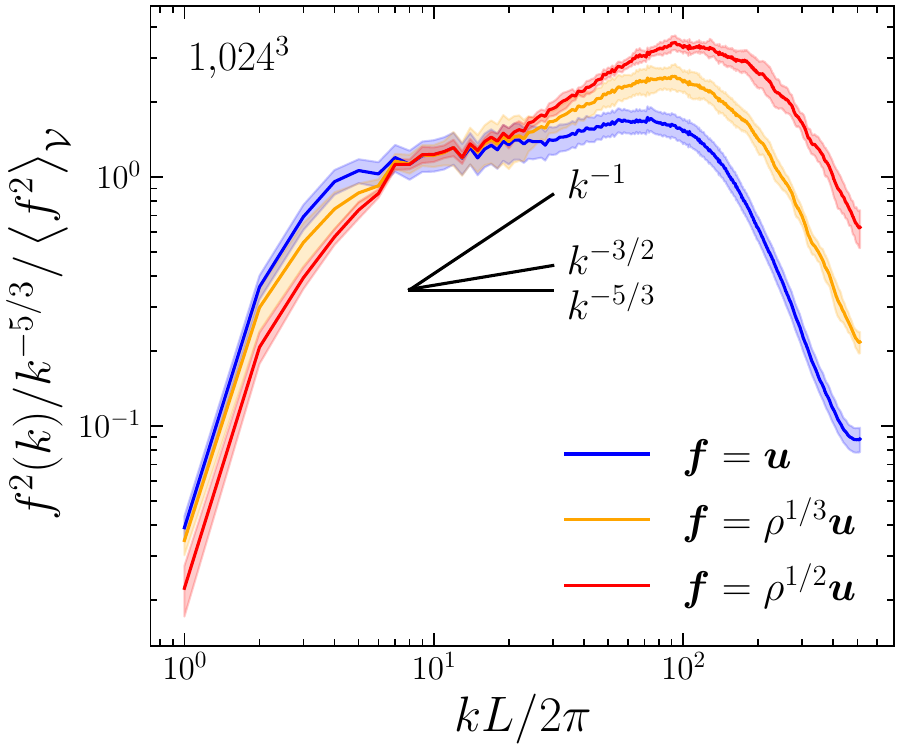}
        \caption{The three most commonly reported velocity and density-weighted velocity power spectrum for the $1,\!024^3$ simulation, averaged over the statistically steady state shown in \autoref{eq:mach_number}. Each spectrum is isotropically integrated. The pure velocity spectrum $\bm{f} = \bm{u}$ (same as main text) is shown in blue, the compressible energy flux-corrected spectrum $\bm{f}=\rho^{1/3}\bm{u}$ \citep{Kritsuk2007} in orange, and the kinetic energy spectrum, with $\bm{f} = \rho^{1/2}\bm{u}$, in red.}
        \label{fig:isotropic_spectrum}
    \end{figure}

    In \autoref{fig:isotropic_spectrum} we show the isotropically integrated power spectrum of three quantities that have been previously reported and compared in supersonic turbulence box studies \citep[][specifically, see Appendix~1 in \citealt{Federrath2010_solendoidal_versus_compressive}]{Kritsuk2007,Federrath2010_solendoidal_versus_compressive,Federrath2013_universality,Grete2017_shell_models_for_CMHD,Fielding2022_ISM_plasmoids}. All spectra are averaged over the stationary state, and compensated by the fiducial $k^{-5/3}$ \citet{Kolmogorov1941} expectation as well as the integral energy. The first spectrum is the velocity spectrum, where $\bm{f} = \bm{u}$, as in the main study. As we showed previously in \autoref{fig:spectra}, this spectrum shows a universal power law across all $k$ modes, regardless of the disk and wind geometry or thermal phase structure. As we indicated previously, it is shallower than the $k^{-5/3}$ expectation, with a roughly $k^{-3/2}$ power law (see \autoref{tab:powerspec}), which can be indicative of a reduction in the nonlinearities that give rise to the turbulence \citep{Boldyrev2005_MHD_spectrum,Beattie2025_scale_dependent_alignment}, potentially due to the inverse fluxes that we show in \autoref{fig:shell2shell}. 
    
    The next spectrum we show is the compressible flux-corrected velocity spectrum, $\bm{f} = \rho^{1/3}\bm{u}$ (orange). This quantity is motivated from the fact that $\varepsilon \sim \rho u^3/\ell = {\rm{const.}} \implies \rho^{1/3}u \propto \ell^{1/3}$, which can be Fourier transformed to give $(\rho^{1/3}u)^2(k) \propto k^{-5/3}$, as discussed in detail in \citet{Federrath2013_universality}. \citet{Federrath2013_universality} found that $(\rho^{1/3}u)^2(k)$ lacked universality when comparing different modes of turbulent driving in isothermal supersonic turbulence, with a range between $k^{-5/3}$ and $k^{-19/9}$. Unlike $(\rho^{1/3}u)^2(k)$, \citet{Federrath2013_universality} found that the $u^2(k)$ spectrum was relatively more universal across all simulations. Our $(\rho^{1/3}u)^2(k)$ is shallower than $k^{-5/3}$, close to $k^{-1}$ on small scales. This is most likely due to cooling in the disk creating a significant amount high-density clumps at small scales, which increases the power at high-$k$. Indeed, \citet{Kim2005} shows that this is exactly how the $\rho$ spectrum behaves for increasing $\M$, asymptotic towards $\rho^2(k) \propto k^{0}$, as $\M \rightarrow \infty$.  

    The final spectrum we investigate is the kinetic energy spectrum $\rho u^2(k)$, where $\bm{f} = \rho^{1/2}\bm{u}$. Compared to $(\rho^{1/3}u)^2(k)$, this puts even more weight on $\rho$, resulting in an even shallower spectrum on the high-$k$ modes, now following very closely $k^{-1}$. The key takeaway is, as in \citet{Federrath2013_universality}, that the velocity power spectrum shows significantly more universality than the other spectra, demonstrating that the turbulent velocity modes connect the turbulence between the hot winds and the warm disk in a self-similar manner, which we explore with our transfer functions in \autoref{sec:transfer_functions} in more detail.

    \begin{figure}
        \centering
        \includegraphics[width = \linewidth]{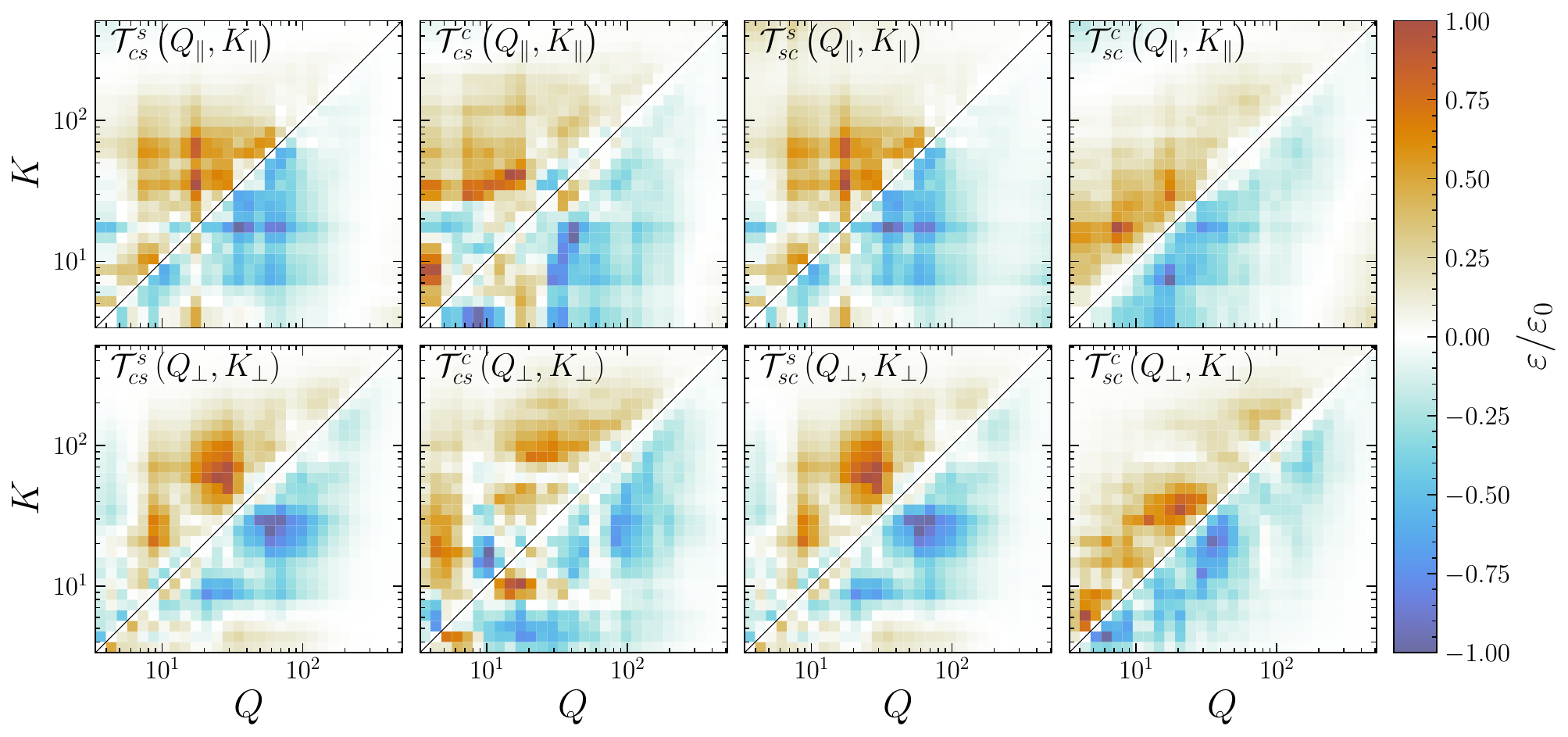}
        \caption{Same as \autoref{fig:shell2shell} but for mixed mode shell-to-shell transfers, $\uc \leftrightarrow \us$, \autoref{eq:all_TF_mixed_1}-\autoref{eq:all_TF_mixed_2}.}
        \label{fig:mixed_shell2shell}
    \end{figure}

\section{Mixed mode shell-to-shell transfers functions}\label{app:mixed-modes}
    We compute a total of 20 transfer functions, as discussed in \autoref{sec:transfer_functions}. The main text included only like-mode transfers, $\us \rightarrow \us$, $\uc \rightarrow \uc$, which probe the velocity flux directly related to the spectrum we show in \autoref{fig:spectra}. However, there are the mixed mode combinations (visualized in the center of \autoref{fig:schematic}) that we did not show in the main text, $\us \leftrightarrow \uc$. These describe the flux transfers between $\us$ and $\uc$, relating to the exchange of velocity flux between the two spectra. One can also think of these transfer functions as probing the amount of $\uc$ modes turning into $\us$ modes, and vice-versa. The transfer functions that describe these processes are
    \begin{align}
        & \quad\quad\quad\quad\quad\quad \us \xrightarrow{\uc} \uc, \quad\quad\quad\quad\quad\quad\quad\quad\quad\quad\quad\quad\quad\quad\quad \uc \xrightarrow{\uc} \us, \nonumber \\
        \T_{sc}^c(Q,K) &= - \int\dthree{\bell}\; \u^K_c\otimes \uc : \bnabla \otimes \u^Q_s,  \quad\quad
        \T_{cs}^c(Q,K) = - \int\dthree{\bell}\; \u^K_s\otimes \uc : \bnabla \otimes \u^Q_c, \label{eq:all_TF_mixed_1} \\
        & \quad\quad\quad\quad\quad\quad \us \xrightarrow{\us} \uc, \quad\quad\quad\quad\quad\quad\quad\quad\quad\quad\quad\quad\quad\quad\quad \uc \xrightarrow{\us} \us, \nonumber \\
        \T_{sc}^s(Q,K) &= - \int\dthree{\bell}\; \u^K_c\otimes \us : \bnabla \otimes \u^Q_s, \quad\quad \T_{cs}^s(Q,K)= - \int\dthree{\bell}\; \u^K_s\otimes \us : \bnabla \otimes \u^Q_c. \label{eq:all_TF_mixed_2}
    \end{align}
    We show the shell-to-shell transfers in \autoref{fig:mixed_shell2shell}, organized in a similar way as our cascade shell-to-shell transfers \autoref{fig:shell2shell}, and transformed into purely antisymmetric fluxes as described in \autoref{app_eq:anti_sym_transfers}. As was the case in the cascade transfers, there is no significant difference between $\ell_\perp$ and $\ell_{\parallel}$, so we focus our discussion on each mixed transfer. Firstly, there are no strict cascades in any of these transfers (local and diagonally dominated). Both $\T^s_{cs}$ and $\T^s_{sc}$ show that $\uc$ modes donate energy to $\us$ modes across quite a broad range of $k$ when mediated by either $\uc$ or $\us$. The same is true for the $\T^c_{cs}$ transfer, has $\varepsilon > 0$ on all $k$ modes, but especially at low-$k$, showing that $\uc \rightarrow \us$ on large scales when mediated by an additional $\uc$ mode. What we learn is that there are many flux exchanges between the two modes on all scales.

\section{Cross scale flux and total flux amplitude of cascade transfers}\label{app:cross}
    In \autoref{subsec:shell2shell} we analyzed the energy transfer between individual $k$ mode shells, $Q$ and $K$. However, following \citet{Grete2017_shell_models_for_CMHD}, here we consider the cross-scale energy transfer $\Pi(k)$, and the total flux amplitude, $\Pi_0$, of each transfer function. They are defined
    \begin{equation}
        \Pi(k) = \sum_{Q \leq k} \sum_{K > k} \T_{ii}^{j}(Q,K), \quad\quad   \Pi_{0} = \sum_{k_{\rm inertial}} \Pi(k),
        \label{eq:cross_transfer}
    \end{equation}
    which is the total velocity flux from all $Q$ shells smaller than $k$, to all $K$ shell modes larger than $k$. We use this statistic to define an inertial range, $k_{\rm inertial}$, where $\varepsilon \approx \rm{const.}$ (very approximately), 'a la Kolmogorov, and then define the total cross-scale transfer, $\Pi_0$, by simply summing over $\Pi(k)$ in the inertial range $k_{\rm inertial}$. We use $\Pi_0$ to normalize all transfers (both cross and shell-to-shell; \autoref{sec:transfer_functions}) throughout the study, and use the $k_{\rm inertial}$ derived to indicate where the inertial range is for \autoref{fig:spectra} and \autoref{sec:power_spectra}.

\begin{figure}
        \centering
        \includegraphics[width = \linewidth]{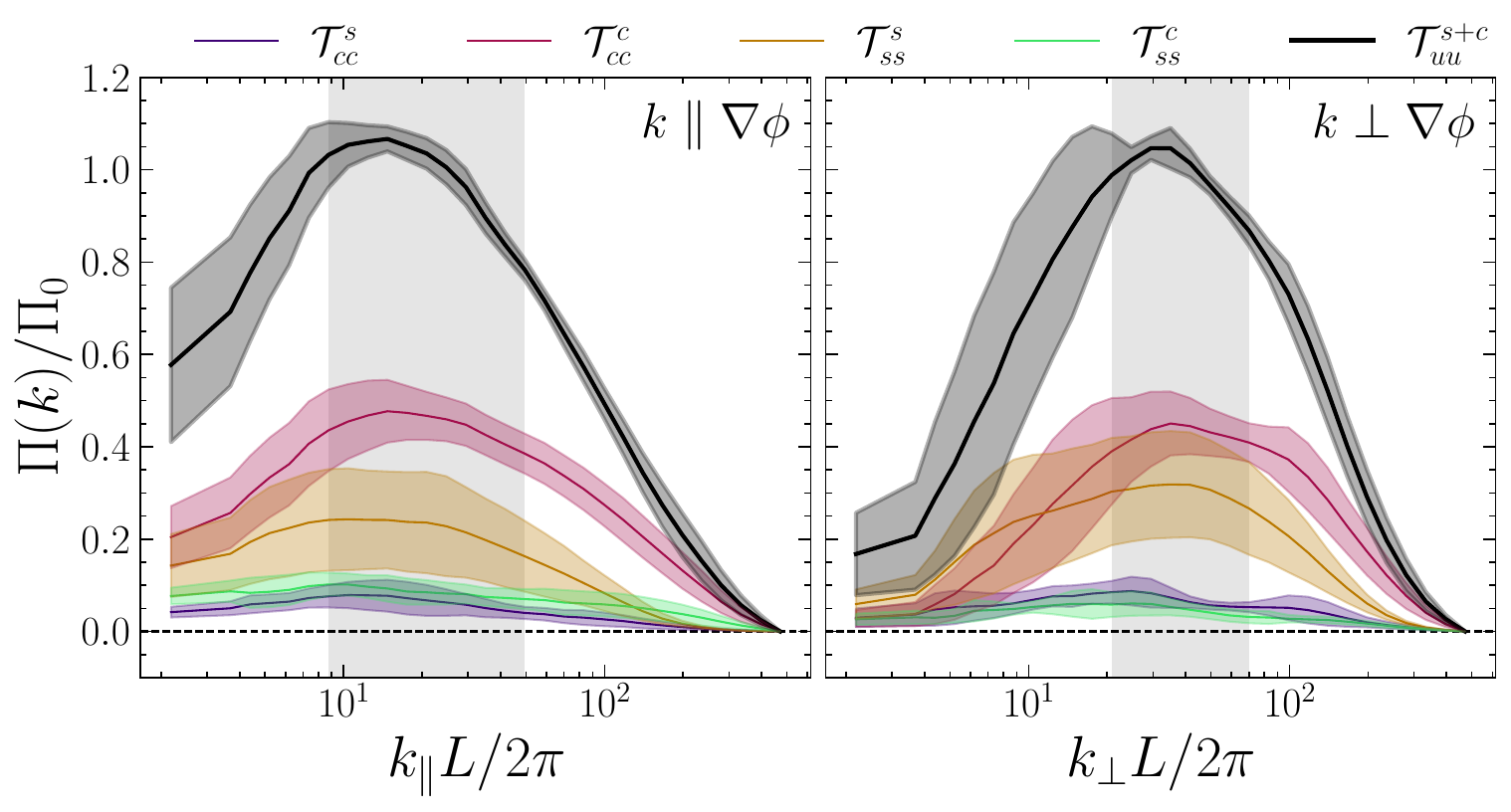}
        \caption{Cross scale transfer, $\Pi(k)/\Pi_0$ (\autoref{eq:cross_transfer}), for like-mode ($\uc \rightarrow \uc$ or $\us \rightarrow \us$; \autoref{eq:Tccc} to \autoref{eq:Tsss}) and total mode interactions, normalized by the cross scale transfer in the $\sim$ inertial range, $\Pi_0$ (shaded black region, based on the classical assumption for \citet{Kolmogorov1941}-type turbulence, $\Pi(k)/\Pi_0 = \text{const.}$) along ($\kpar$; top panel) and across ($\kperp$; bottom panel) the $\bnabla\phi$. Colored, shaded bands on each $\Pi(k)/\Pi_0$ curve denote the $1\sigma$ time variation. The flux from three like-mode interactions (either all $\uc$ or all $\us$) dominate the flux, and show the closest behavior to the expected $\Pi(k)/\Pi_0 = \text{const.}$ across a small range of wavemodes in the simulations.}
        \label{fig:crossscale}
    \end{figure}

\begin{figure}
        \centering
        \includegraphics[width = \linewidth]{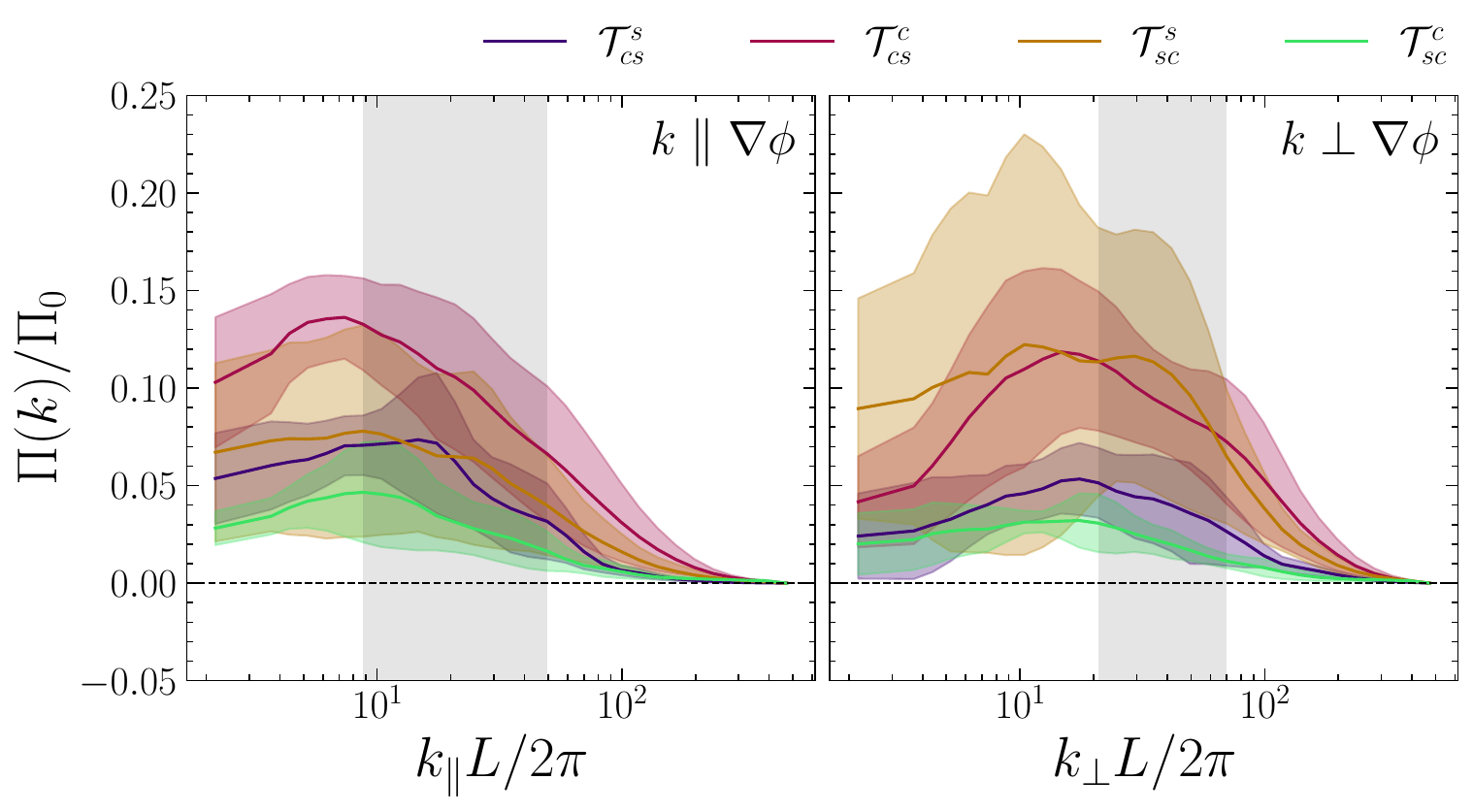}
        \caption{The same as figure \autoref{fig:crossscale} but for mixed mode cross scale transfers, $\uc \rightarrow \us$ and $\us \rightarrow \uc$.}
        \label{fig:mixed_cross_scale}
    \end{figure}

    We show the cross-scale transfer function for the cascade mode transfers in \autoref{fig:crossscale} and mixed mode transfers in \autoref{fig:mixed_cross_scale}. Firstly, we note that the velocity flux does not become completely constant at any range of $k$. However, there is a small range of $k$ in the isolated transfers, $\T_{cc}^c$ and $\T_{ss}^s$ that is constant within the $1\sigma$ fluctuations. We indicated where this region is with the gray bands, and this is the region we call the inertial range. Note that the $\T_{cc}^c$ and $\T_{ss}^s$ fluxes dominate the total velocity flux, showing that these two cascades (both quite non-local transfers, see \autoref{fig:shell2shell}) are the most efficient at moving energy through the different modes. The mixed mode transfers shown in \autoref{fig:mixed_cross_scale} have a factor of two smaller flux. Hence, the $\T_{cc}^c$ and $\T_{ss}^s$ cascades are significantly more efficient at pushing energy down the cascade than the mode interactions between the cascades are at exchanging energy.

\begin{figure}
        \centering
        \includegraphics[width=0.8\linewidth]{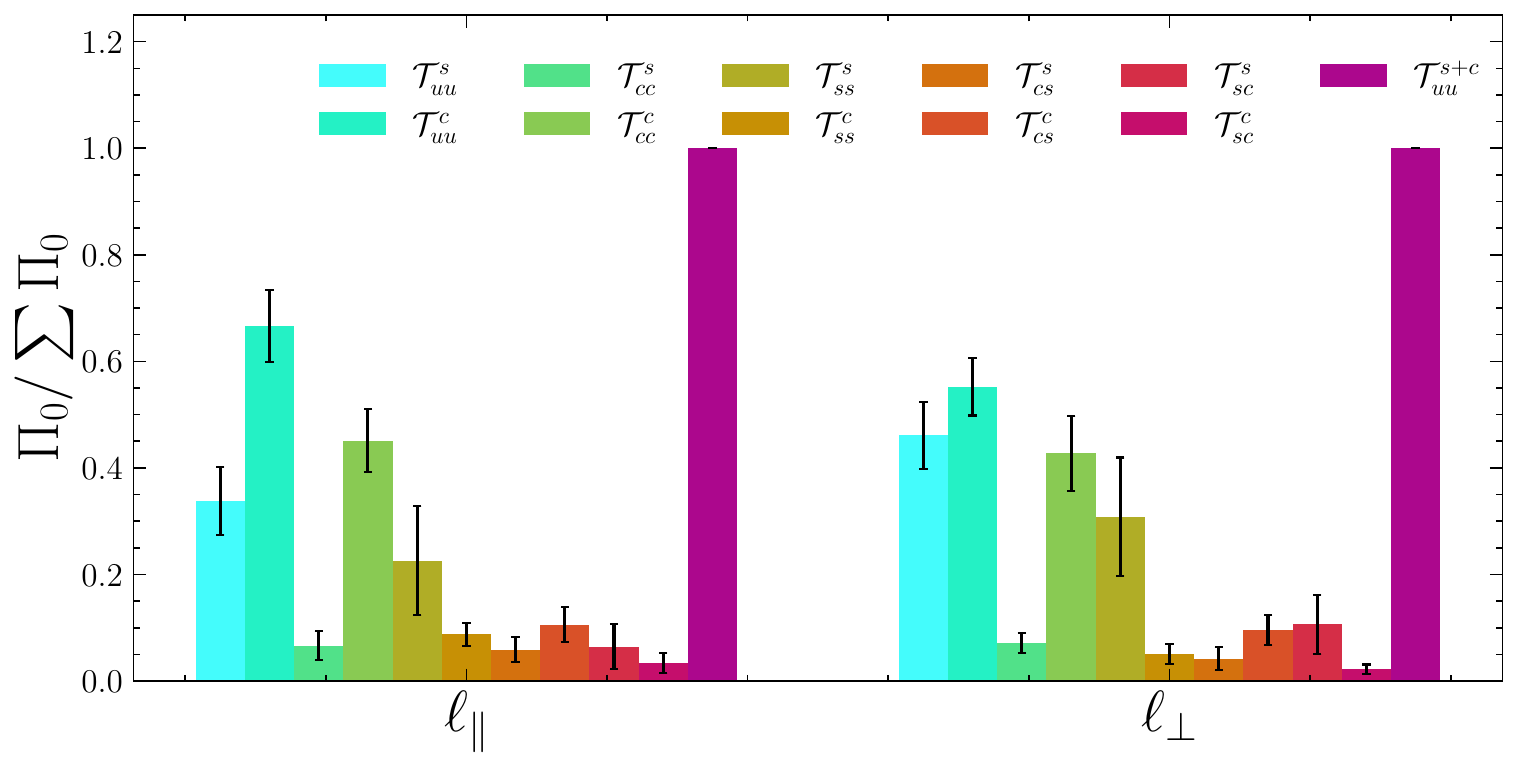}
        \caption{The integrated velocity flux contributions from each of the transfers integrated in the range of scales in the inertial range $k_{\rm inertial}$ (gray bands in \autoref{fig:crossscale} and \autoref{fig:mixed_cross_scale}). All transfers are normalized by the velocity flux of the total transfer function $\T^{u}_{uu}$ (last bar), such that the $\T^{s}_{uu} + \T^{c}_{uu} = 1$ (first two bars) and then the remaining eight transfers, $\T^s_{cc} + \T^c_{cc} + \T^{s}_{ss} + \T^{c}_{ss} + \T^{s}_{cs} + \T^{c}_{cs} + \T^{s}_{sc} + \T^{c}_{sc} = 1$. The cascade flux $\T^c_{cc}$ dominates the flux, followed by  $\T^s_{ss}$.}
        \label{fig:total_integrated_flux}
    \end{figure}

    Finally, in \autoref{fig:total_integrated_flux} we plot the total summed velocity flux in the inertial range indicated by \autoref{fig:crossscale}. We normalized all the fluxes by the total across all transfers, such that the first two columns sum to one, $\T^{s}_{uu} + \T^{c}_{uu} = 1$, and so do the next eight columns, $\T^s_{cc} + \T^c_{cc} + \T^{s}_{ss} + \T^{c}_{ss} + \T^{s}_{cs} + \T^{c}_{cs} + \T^{s}_{sc} + \T^{c}_{sc} = 1$. The $\uc$ mode-mediated transfers have the largest flux, which is dominated by the $\T^c_{cc}$ transfer.

\section{Antisymmetric property of Helmholtz decomposed transfer functions}\label{app:antisymmetry}
    In general, transfer functions should have the antisymmetric property,
    \begin{align}
        \T(Q,K) = -\T(K,Q),
    \end{align}
    and in this section we shall explore this property for the Helmholtz decomposed transfers, focusing, as we have done for the entire study, just on the $\u\otimes\u:\bnabla\otimes\u$ nonlinearity. Consider the classical turbulence cascade term that we use to explore the cascade in \autoref{sec:transfer_functions},
    \begin{align}
        -\int_{\V}\d{\V}\; u^K_i u_j \partial_j u^Q_i = \int_{\V}\d{\V}\; \partial_j (u^K_i u_j u^Q_i) + \int_{\V}\d{\V}\; u^K_i u^Q_i \partial_j u_j+ \int_{\V}\d{\V}\; u_j u^Q_i \partial_j u^K_i,
    \end{align}
    from the product rule, written in tensorial form. Using the divergence theorem,
    \begin{align}
        \int_{\V}\d{\V}\;\partial_j (u^K_i u_j u^Q_i) = \oiint_{\partial\mathcal{V}} u^K_i u_j u^Q_i\d{\partial\mathcal{V}_j} = 0,
    \end{align}
    if the domain is periodic. In our case, we have only doubly-periodic boundaries, so this surface integral will not be exactly zero. However, based on the antisymmetry that is preserved in the incompressible mediated transfer functions (see \autoref{fig:shell2shell}), seemingly this has a very small effect. Then,
    \begin{align}
        -\int_{\V}\d{\V}\; u^K_i u_j \partial_j u^Q_i = \int_{\V}\d{\V}\; u^K_i u^Q_i \partial_j u_j+ \int_{\V}\d{\V}\; u_j u^Q_i \partial_j u^K_i.
    \end{align}
    Now consider the compressible $u_i^c$ and incompressible $u_i^s$ mode decomposition that we perform throughout \autoref{sec:transfer_functions}. Without loss of generality, if we consider only the mediating mode, for the compressible mode we have
    \begin{align}\label{eq:comp_tranasfer_expanded}
        -\int_{\V}\d{\V}\; u^K_i u_j^c \partial_j u^Q_i = \int_{\V}\d{\V}\; u^K_i u^Q_i \partial_j u_j^c + \int_{\V}\d{\V}\; u_j^c u^Q_i \partial_j u^K_i,
    \end{align}
    and for incompressible,
    \begin{align}
        -\int_{\V}\d{\V}\; u^K_i u_j^s \partial_j u^Q_i= \int_{\V}\d{\V}\; u_j^s u^Q_i \partial_j u^K_i.
    \end{align}
    This means that transfer functions mediated by $u_i^s$ are antisymmetric with themselves, as is expected from regular incompressible transfer function theory \citep{Alexakis2005_shell_to_shell}, but transfer functions mediated by $u_i^c$ are not. From \autoref{eq:comp_tranasfer_expanded}, they are instead antisymmetric with the total velocity transfer function,
    \begin{align}
        \T^c(Q,K) = -\int\dthree{\bell}\; \left[\u^K\otimes\left( \uc:\bnabla\otimes\u^Q + \u^Q:(\bnabla\cdot\uc)\It\right)\right], \label{app_eq:anti_sym_transfers}
    \end{align}
    which we do not investigate in this study, due to the fact that the antisymmetric component of a transfer function we are already computing does not contain any further information. However, this does explain why all of the $\T^c(Q,K)$ transfer functions are not antisymmetric with themselves, whereas the $\T^s(Q,K)$ transfers are. 

    Antisymmetric transfers are at the heart of cascade transfers. Hence, we perform one final step to the $\T^c(Q,K)$ transfer functions to ensure antisymmetry, where we redefine then such that
    \begin{align}
        \T^c(Q,K) \leftarrow \frac{1}{2}\left(\T^c(Q,K) + \T^c(K,Q)\right),
    \end{align}
    isolating only the antisymmetric transfers in $\T^c(Q,K)$.

\section{Kinetic helicity}\label{app:kinetic_hel}
    The kinetic helicity, $\mathcal{H}_{\rm kin} = \Exp{\bom\cdot\u}{\V}$, describes the volume-averaged asymmetry in the left versus right eigenmodes of $\us$ in the fluid \citep[clockwise orientated $\bom_+$ versus anticlockwise orientated $\bom_-$; ][]{Alexakis2017_helically_decomposed,Plunian2020_inverse_cascade}. If $|\Exp{\bom\cdot\u}{\V}| \gtrsim 0$ then we expect there to be an excess of left or right modes, which, when interacting, $\bom_\pm \rightarrow \bom_\pm$, have, on average, a flux from small to large scales, $\varepsilon < 0$. Hence, $\bom_+ \rightarrow \bom_+$ or $\bom_- \rightarrow \bom_-$ velocity mode interactions can give rise to an inverse cascade, as demonstrated utilizing $\bom_\pm$ transfer functions in three-dimensional hydrodynamic turbulence driven with net kinetic helicity \citep{Alexakis2017_helically_decomposed,Plunian2020_inverse_cascade}. In our study, we show that the $\us \rightarrow \us$ mediated by $\uc$ interaction can also provide a mechanism for inverse transfer. Therefore, a natural question is the nature of the handedness of $\us \rightarrow \us$ transfers during this interaction. We leave the details of this for a future study, where we intend to track both eigenmodes, but we provide $\mathcal{H}_{\rm kin}$ as a function of $t/t_0$ for the $512^3$ simulation to at least understand if the supernova-driven turbulence becomes net helical (note that $\mathcal{H}_{\rm kin}$ is no longer an invariant in compressible hydrodynamics, so small helicity fluctuations could indeed grow). 
    
\begin{figure}
    \centering
    \includegraphics[width=0.8\linewidth]{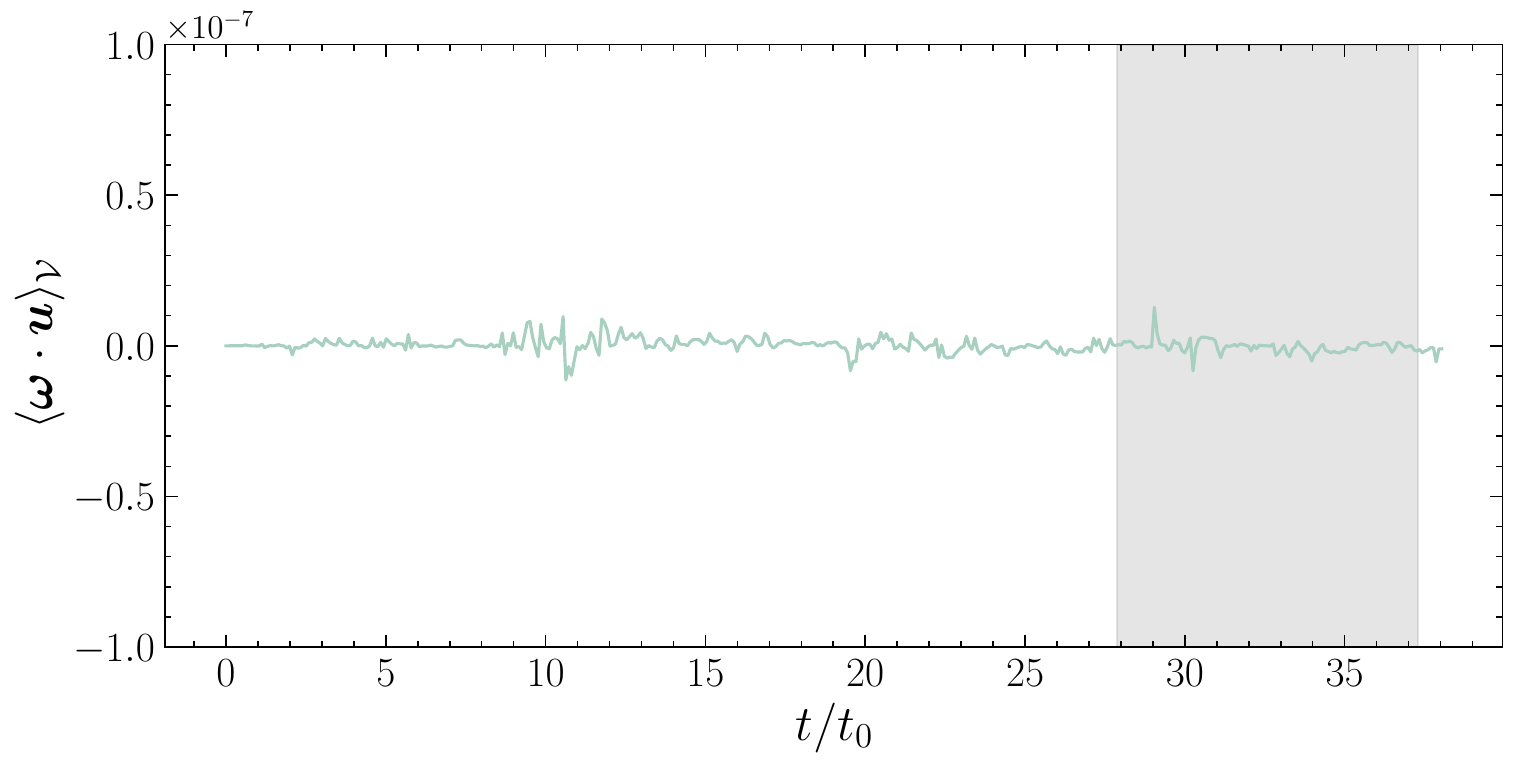}
    \caption{The kinetic helicity, $\mathcal{H}_{\rm kin} = \Exp{\bom\cdot\u}{\V}$, as a function of $t/t_0$ for the $512^3$ simulation, with the gray band showing the stationary state for the supernova-driven turbulence (see \autoref{fig:mach}). Both in the non-stationary and stationary states the turbulence is non-helical to single-precision, $\mathcal{H}_{\rm kin} \approx 0$, meaning that the inverse transfer that we observe does not require net kinetic helicity. Hence, the inverse cascade we measure in \autoref{sec:transfer_functions} is a completely different mechanism compared to homochiral-mediated inverse cascade in 3D turbulence \citep{Plunian2020_inverse_cascade}.}
    \label{fig:kinetic_helicity}
\end{figure}

\begin{figure}
    \centering
    \includegraphics[width=\linewidth]{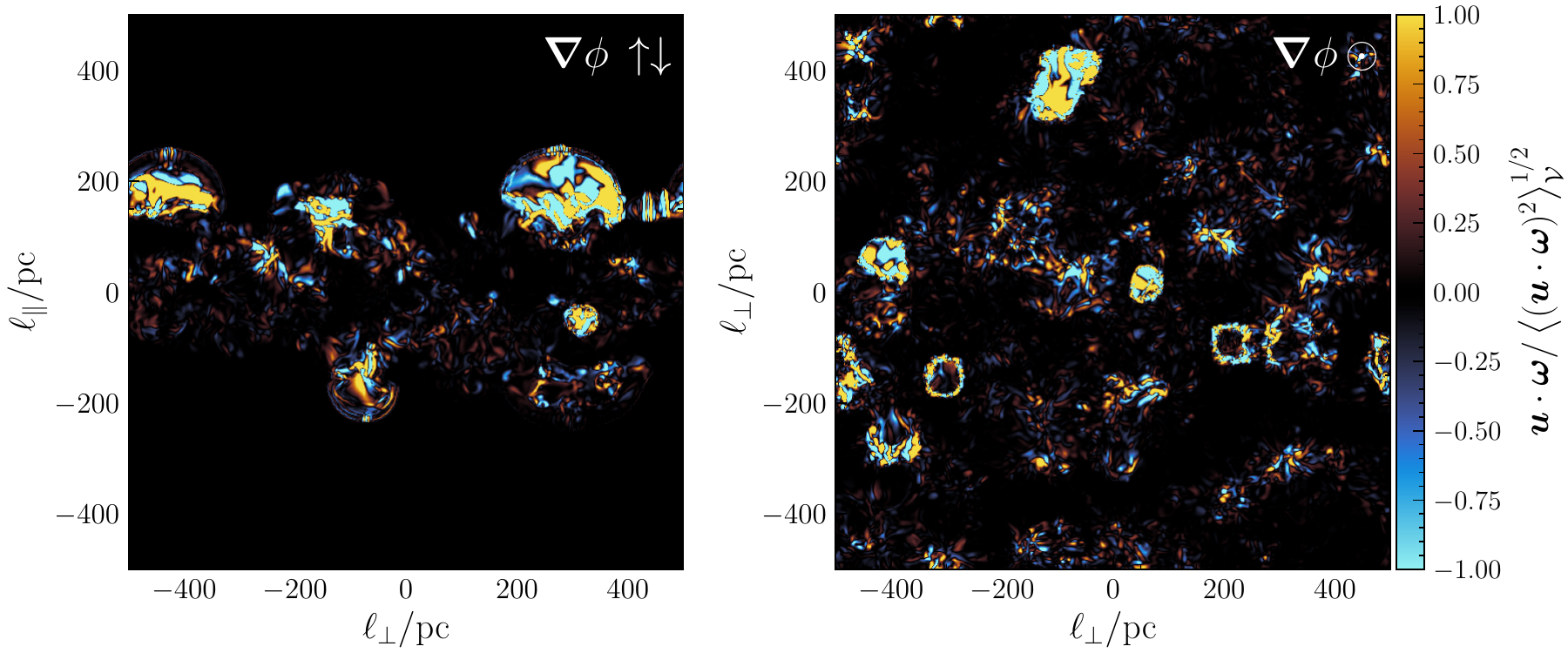}
    \caption{The point-wise kinetic helicity $\bom\cdot\u$ normalized by the rms for a two-dimensional slice along $\bnabla\phi$ (left) and perpendicular to $\bnabla\phi$ (right). The largest $\bom\cdot\u$ fluctuations, both in amplitude and size scale, trace the expanding remnants, showing how, even in the absence of global shear and with $\mathcal{H}_{\rm kin} \approx 0$, strong local kinetic helicity fluctuations are generated through detonating SNe.}
    \label{fig:kinetic_helicity_map}
\end{figure}

    We show the kinetic helicity in \autoref{fig:kinetic_helicity}, where for all time the supernova-driven turbulence is completely non-helical (in volume average) $|\Exp{\bom\cdot\u}{\V}| \approx 0$, at least to single precision. We also show a two-dimensional slice of the point-wise kinetic helicity $\bom\cdot\u$ normalized by the rms, \autoref{fig:kinetic_helicity_map}, with the same annotations as in \autoref{fig:temperature}. This is early in the evolution of the simulation, just to easily visualize what is happening in and around the SNRs, but no conclusions change in the steady state. What we observe are strong $\bom\cdot\u$ fluctuations, localized around the SNRs. The signed fluctuations become comparable to the size of the typical size of the SNRs, significantly larger than the background fluctuations in the disk. Furthermore, they are bright, indicating that the fluctuations contain strong alignment between $\u$ and $\bom$. This suggests, like $\us \xrightarrow{\uc} \us$ interactions, the $\uc$ modes generated in the SNRs take strong $\bom\cdot\u$ fluctuations to large scales. As previously discussed in \citet{2018Kapyla_helicity}, this is important for large-scale, stochastic $\alpha$ dynamos \citep{Rincon2019_dynamo_theories}. So even though $\mathcal{H}_{\rm kin} \approx 0$, the fluctuations in $\bom\cdot\u$ may still be important for both the ISM dynamo and the handed interactions in the turbulence. 

\section{Convergence test} \label{app:convergence}
    In order to test the convergence of our results, we calculate the total velocity power spectra, $\P(k)$ for all three resolutions (\texttt{MW\_256}, \texttt{MW\_512}, \texttt{MW\_1024}; see \autoref{tab:sims}) runs included in this study, shown in \autoref{fig:spectra}. As in \autoref{sec:power_spectra}, we normalize each spectrum by its respective integral, \ie $\langle u^2 \rangle_{\V}$, by Parseval's theorem. For both $\kpar$ and $\kperp$ wave modes, the spectra at $256^3$ are approximately converged for $k \leq 20$. As the resolution increases, $\P(k)$ extends to higher and higher $k$-modes, and looks reasonably similar, just with more power at higher $k$. The correlation scales, in the form $(\ell_{\rm cor,\parallel}, \ell_{\rm cor,\perp})$, for each of the $\P(k)$ are as follows, $\texttt{MW\_1024} \approx (4.1, 3.5)\ell_0$, $\texttt{MW\_512} \approx (4.1, 3.4)\ell_0$, $\texttt{MW\_256} = (4.1,3.6)\ell_0$, where $\ell_0 \approx 85\,\rm{pc}$ is the measured gaseous scale height.
    
\begin{figure}
    \centering
    \includegraphics[width=\linewidth]{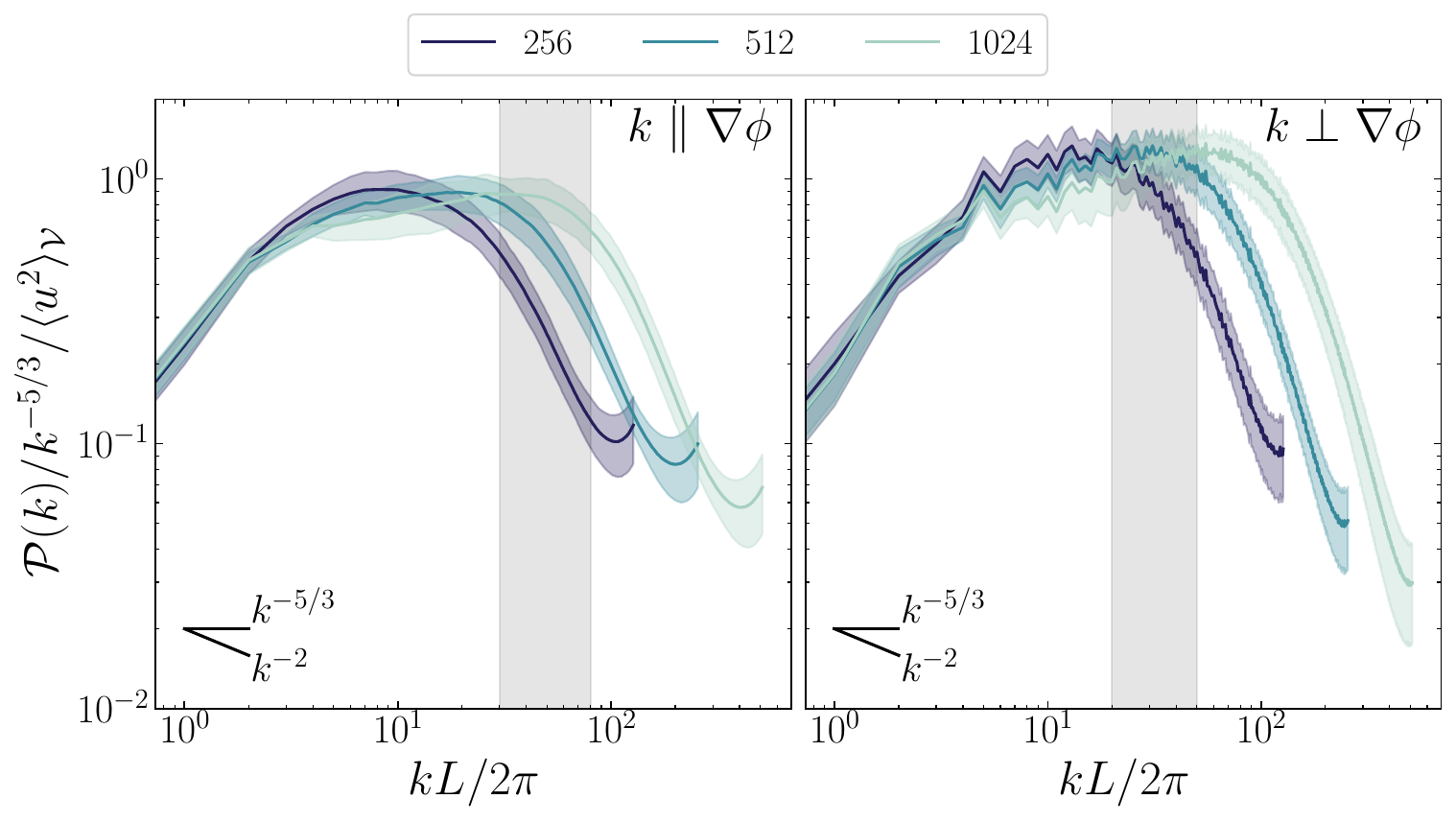}
    \caption{Cylindrically integrated velocity power spectra, $\P(k)$, as a function of wavenumbers parallel to the static gravitational potential, \autoref{eq:grav_pot} ($\kpar = |k_z|$; left column) and perpendicular to it ($\kperp = \sqrt{k_x^2 + k_y^2}$; right column). Each color represents a different resolution, going from linear resolution $N_{\rm grid} = 256$ (blue) to $N_{\rm grid} = 1024$ (light green). We indicate a gray band to show where the cascade is in each spectrum, based on the flux results in \autoref{sec:transfer_functions}. Each spectrum is normalized by the integral square velocity, $\Exp{u^2}{\V}$ and time-averaged in the stationary regime (see \autoref{fig:mach}) so that intense intermittent events from supernova detonations do not have a strong influence on the spectra.}
    \label{fig:P_resolution}
\end{figure}

\bibliography{Mar24,james_bib}
\bibliographystyle{aasjournal}

\end{document}